\documentclass[%
 reprint,
 amsmath,amssymb,
 aps,
]{revtex4-2}

\usepackage{graphicx}
\usepackage{dcolumn}
\usepackage{bm}
\usepackage[utf8]{inputenc}
\usepackage{graphicx}
\usepackage{dcolumn}
\usepackage{bm}
\usepackage{bbold}
\usepackage{physics}
\usepackage{mathtools} 
\usepackage{multirow}
\usepackage{hyperref}
\usepackage{makecell}

\usepackage[normalem]{ulem}
\usepackage{xcolor}

\usepackage{soul,xcolor}
\newcommand\s{\bgroup\markoverwith{\textcolor{magenta}{\rule[0.5ex]{2pt}{0.4pt}}}\ULon}

\newcommand\R{\bgroup\markoverwith{\textcolor{blue}{\rule[0.5ex]{2pt}{0.4pt}}}\ULon}

\begin{document}

\title{Diffusive Epidemic Process with quenched disorder}
\author{Valentin Anfray$^1$}
\email{valentin.anfray@gmail.com}
\author{Hong-Yan Shih$^{1,2}$}%
\email{hongyan@as.edu.tw}
\affiliation{%
 $^1$Institute of Physics, Academia Sinica, Taipei 115201, Taiwan\\
 $^2$Physics Division, National Center for Theoretical Sciences, Taipei 106319, Taiwan
}%

\begin{abstract}
Epidemic spreading often occurs in spatially heterogeneous environments, yet how quenched heterogeneity reshapes its onset and critical dynamics remains poorly understood. The diffusive epidemic process, a minimal reaction–diffusion model whose absorbing-state transition is controlled by the relative diffusion of healthy and infected species, provides a natural setting for this question. Using a new single-seed algorithm that effectively simulate infinite systems for the infected individuals, we find that effective global diffusion rates can be used to predict disorder relevance and we identify two distinct infinite-disorder fixed points. Notably, we find that disorder in diffusion rates is qualitatively different from that in reaction rates as it can even induce a total suppression of the active phase, a phenomenon not observed with other types of disorder.These results establish mobility disorder as a distinct route by which quenched heterogeneity qualitatively reorganizes spreading dynamics, with implications for systems ranging from cell polarity to epidemic propagation in heterogeneous media.
\end{abstract}

\maketitle

Epidemic-type models provide a minimal framework for studying how self-sustaining activity spreads through space, with applications ranging from infectious disease and malware propagation to microbial colonization and species invasion \cite{kermack_contribution_1997, mack_biotic_2000, wang_epidemic_2003, bettencourt_power_2006, keeling_modeling_2008}. Their central organizing feature is the epidemic threshold, which separates an active phase of persistent spreading from an absorbing phase in which activity eventually dies out. Determining this threshold and the critical behavior is essential for understanding spreading dynamics and informing control or mitigation strategies in epidemic and biological systems.

\begin{figure}
    \centering
    \includegraphics[width=\linewidth]{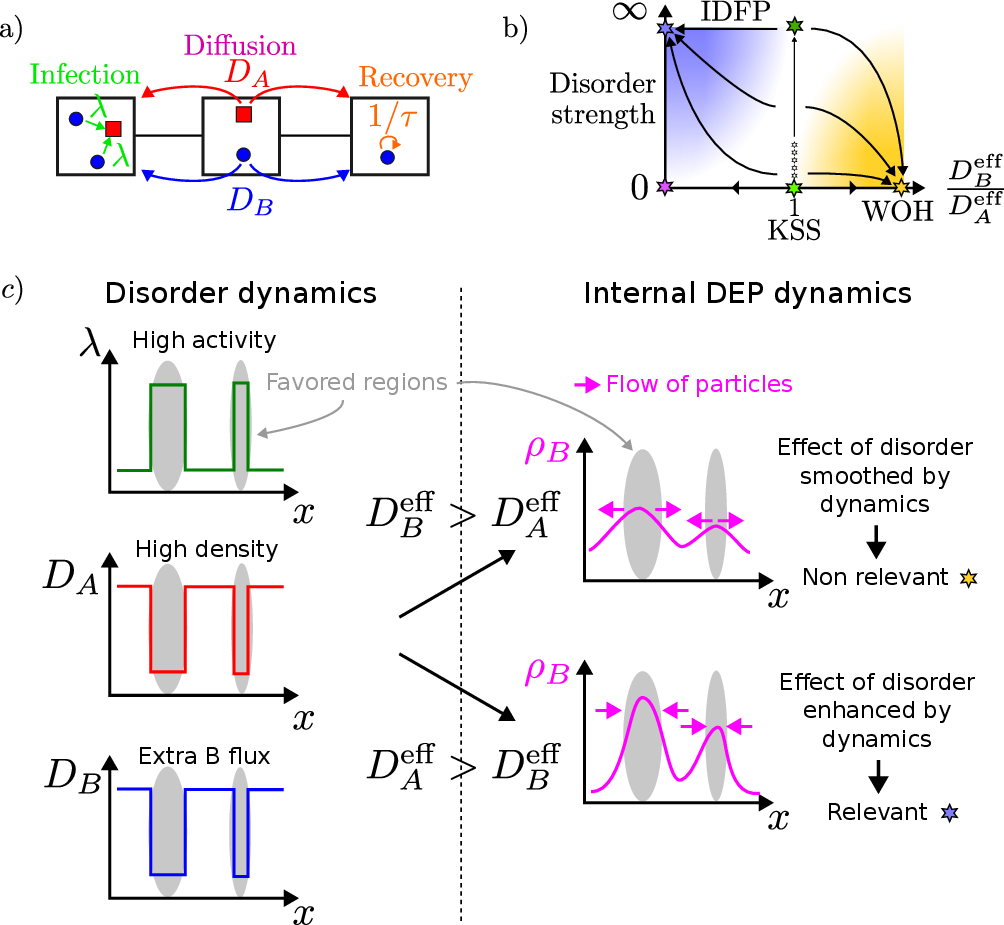}
    \caption{(a) Schematic of the Diffusive Epidemic Process (DEP). \( A \) (red) and \( B \) (blue) particles diffuse with rates \( D_A \) and \( D_B \); \( B \) infects \( A \) at rate \( \lambda \) and recovers at rate \( 1/\tau \).  
    (b) Potential renormalization Group (RG) flow as a function of disorder strength and \( D_B^{\text{eff}}/D_A^{\text{eff}} \) (\( D_A^{\text{eff}},D_B>0 \)). Pure fixed points appear for \( D_A>D_B \) (pink), \( D_A=D_B \) (green) \cite{kree_effects_1989}, and \( D_A<D_B \) (yellow) \cite{van_wijland_wilson_1998}. For \( D_B^{\text{eff}}<D_A^{\text{eff}} \), disorder drives an infinite-disorder fixed point (IDFP, blue); for \( D_A^{\text{eff}}=D_B^{\text{eff}} \), critical properties vary slowly with disorder strength up to another IDFP.
    (c) Introducing disorder in \( \lambda(x) \), \( D_A(x) \), or \( D_B(x) \) creates activity-favored regions, where $B$ particles tend to stay longer,  prior to the DEP dynamics. Depending on the relative difference between the effective rates \( D_A^{\text{eff}} \) and \( D_B^{\text{eff}} \), these activity-favored regions are either enhanced or diminished by the internal dynamics. This behavior is consistent with the Harris criterion and allows one to predict when disorder is relevant.}
    \label{fig:DEP_Universality_and_Disorder}
\end{figure}

The diffusive epidemic process (DEP) \cite{kree_effects_1989, van_wijland_wilson_1998} is one significant class of these models \cite{hanski_ecology_2010, colizza_reactiondiffusion_2007, colizza_epidemic_2008}, involving two types of particles: healthy ($A$) and infected ($B$), which diffuse with rates $D_A$ and $D_B$ respectively. The system evolves through a reaction-diffusion process with onsite infection events, \(A+B \to 2B\), occurring at rate \(\lambda\), and spontaneous recovery, \(B \to A\), occurring at rate \(1/\tau\). Throughout the dynamics the total particle number, and hence the total density \(\rho\), is conserved (Fig.~\ref{fig:DEP_Universality_and_Disorder}a). DEP exhibits an absorbing phase transition into a state containing only healthy particles. Within a mean-field approximation, the critical point is located at \(\lambda_{\mathrm{MF}} = (\rho \tau)^{-1}\), and the resulting critical behavior is independent of diffusion \cite{van_wijland_wilson_1998}. In contrast, the critical behavior bellow the upper-critical dimension is known to fall into three distinct universality classes, determined by the relative diffusion rates of the two species \footnote{The case $D_A=0$ featuring infinitely many absorbing states is not discussed here.}. While for the cases of $D_A=D_B$ and $D_B>D_A$ inconsistencies are found between the critical exponents predicted by the field theory \cite{kree_effects_1989, van_wijland_wilson_1998} and estimated by the numerical simulations \cite{de_freitas_critical_2000, fulco_critical_2001, maia_diffusive_2007, filho_critical_2010, da_silva_critical_2013, polovnikov_subdiffusive_2022, dickman_nature_2008, alencar_two-dimensional_2023}, the case of $D_A>D_B$ presents a significant paradox: analytical predictions suggests a discontinuous transition \cite{van_wijland_wilson_1998}, whereas numerical simulations revealed a continuous one \cite{fulco_critical_2001, maia_diffusive_2007, filho_critical_2010, da_silva_critical_2013, polovnikov_subdiffusive_2022, dickman_nature_2008, alencar_two-dimensional_2023}. A strong-coupling regime with subdiffusive spreading has been suggested to explain why renormalization group analyses fail to predict the observed critical behavior \cite{polovnikov_subdiffusive_2022}.

Beyond this fundamental interest in statistical physics, the DEP also serves as a minimal framework for epidemic spreading and pattern formation, including cell polarization \cite{altschuler_spontaneous_2008, brauns_phase-space_2020}, where spatially asymmetric intracellular protein distributions emerge and organize cell morphology and function.
However, spatial heterogeneity is ubiquitous in real environments  \cite{anderson_absence_1958, mezard_spin_2004,pickett_landscape_1995} including the intracellular space \cite{xiang_single-molecule_2020, huang_cytoplasmic_2022}, and there is growing evidence that heterogeneity and individual-level variation can strongly shape epidemic dynamics \cite{tkachenko_stochastic_2021,heltberg_spatial_2022,berestycki_epidemic_2023}.
Determining how quenched disorder modifies the critical behavior of the DEP can provide a rigorous baseline for interpreting more complex mechanisms and offers testable predictions for spreading in heterogeneous media.

Moreover, the explicit dependence of the DEP critical properties on the diffusion rates raises an important question for real-world systems with intrinsic inhomogenities: can the impact of spatial quenched disorder—such as its relevance at criticality—be predicted?

In this letter, we investigate quenched spatial disorder in the DEP using extensive Gillespie-type algorithm \cite{gillespie_exact_1977, gibson_efficient_2000, gillespie_stochastic_2007} with an enhanced scheme that emulates an infinite system size, thereby eliminating finite-size effects. We demonstrate that the relevance of random-diffusion disorder can be predicted by introducing effective diffusion rates.
We further show that, when predicted to be relevant, strong disorder may give rise to infinite-disorder fixed points (IDFPs) \cite{hooyberghs_absorbing_2004, vojta_infinite-randomness_2009, vojta_monte_2012,buono_slow_2013, juhasz_rare-region_2012} characterized by an activated dynamical scaling \cite{fisher_critical_1995} and off-critical Griffiths singularities exhibiting nonuniversal power laws which arise from long-lived locally supercritical rare regions \cite{griffiths_nonanalytic_1969,vojta_criticality_2014}.
Moreover, sufficiently strong disorder in the healthy-species diffusion rate can either eliminate the active phase or, when activity persists, drive a crossover to an IDFP that is distinct from those induced by infection or recovery disorder. In the latter case, diffusion heterogeneity generates pronounced high- and low-density regions where the infectious dynamics share similarities with the disordered contact process.
These results reveal that disorder in the diffusion rates can be qualitatively different from disorder in the reaction rates. This suggests that quenched transport variation is an important, yet often overlooked, factor in determining the behavior of nonequilibrium transitions, with broad implications for the robustness of patterns in cell polarity and the persistence of epidemics in patchy environments.

\textit{Observables.}---To characterize activity spreading, we monitor the survival probability \(P_s(t)\), the total number of $B$ particles \(N_B(t)\), and their mean-square displacement \(R^2(t)\) for surviving samples \cite{grassberger_reggeon_1979, lubeck_universal_2004, henkel_non-equilibrium_2008}. At criticality, these scale as:
\begin{equation*}
    P_s(t)\sim t^{-\delta}, \quad N_B(t) \sim t^{\Theta}, \quad R^2(t)\sim t^{2/z}
\end{equation*}

where $\delta  = \beta'/\nu_{\parallel}$, $z = \nu_{\parallel}/\nu_{\perp}$ is the dynamical exponent and $\Theta = d/z - \alpha - \delta$ is the slip exponent ($\alpha = \beta/\nu_{\parallel}$).  For DEP, \(\Theta\) and \(\delta\) are non-universal, depending on initial particle distributions \cite{van_wijland_wilson_1998}; here we consistently employ Poissonian initial conditions. Corrections to scaling are addressed by computing the effective exponents \( \delta(t), \Theta(t), 2/z(t) \) via local slopes such as \( -\delta(t) = \frac{\partial \ln P_s(t)}{\partial \ln t} \), which are then extrapolated to \( t \to \infty \) \cite{riedel_effective_1974, grassberger_directed_1989, hinrichsen_nonequilibrium_2000}. 

To ensure particle conservation without tracking the entire lattice, we develop a local-update algorithm [Sec.~I~C~\cite{supp}] restricted to the active region. This effectively models an infinite system up to $T_{\max}$; our pure-case critical exponents align with recent numerical results [Table~S1~\cite{supp}].

\textit{Harris criterion}---Predictions for quenched heterogeneity are often guided by the Harris criterion \cite{harris_effect_1974, igloi_strong_2005}, which states that disorder is relevant when $d\nu_{\perp}<2$. In the one-dimensional DEP, random-mass disorder—which couples to the square of the order parameter in field theory \cite{vojta_phases_2013} and corresponds here to disorder in the infection rate $\lambda$ or recovery rate $1/\tau$—is relevant for $D_A>D_B$ and marginal for $D_A\leq D_B$ \cite{polovnikov_subdiffusive_2022, van_wijland_wilson_1998} \footnote{There is no consensus between numerical and field-theoretic analyses. In particular, for \(D_A > D_B\), a predicted discontinuous transition is not observed numerically. See Sec.~II~C~\cite{supp}.}. However, for random-diffusion disorder ($D_A$ or $D_B$), applying the Harris criterion is ambiguous when regions with $D_A(x) \gtrless D_B(x)$ are mixed as the value of $\nu_{\perp}$ to use is unclear. Furthermore, the behavior of locally favored regions cannot be extrapolated from pure-system properties alone. While a reduction in $D_A$ typically expands the absorbing phase in homogeneous systems, the introduction of spatial heterogeneities in $D_A$ produces the opposite effect: regions with lower $D_A$ accumulate a higher local density $\rho_A(x)$, where infection is locally favored. This accumulation acts as an effective spatial disordered infection rate, $\lambda(x) = \lambda [ 1 + \Delta \rho(x)/\rho ]$ prior to the infection (e.g. at $t=0$) or when $\rho_B\approx 0$ (e.g. close to the critical point), suggesting that the Harris criterion could remain applicable.

\textit{Infection-rate disorder}---To test the Harris criterion across the three different universality classes, we introduce heterogeneities in the infection rate ($\lambda$-disorder). Following Refs.~\cite{hooyberghs_absorbing_2004, dickison_monte_2005}, we consider a binary probability distribution for \(\lambda\) at site \(i\):
\begin{equation}
        \mathcal{P}(\lambda_i) = (1-p_{\lambda})\delta(\lambda_i - \lambda) + p_{\lambda}\delta(\lambda_i - c\lambda),
        \label{eq:DisorderDistributionLambda}
\end{equation}
where $c, p_{\lambda} \in [0, 1]$ represent the impurity strength and concentration, respectively.

\begin{figure}
    \centering
    \includegraphics[width=0.49\columnwidth]{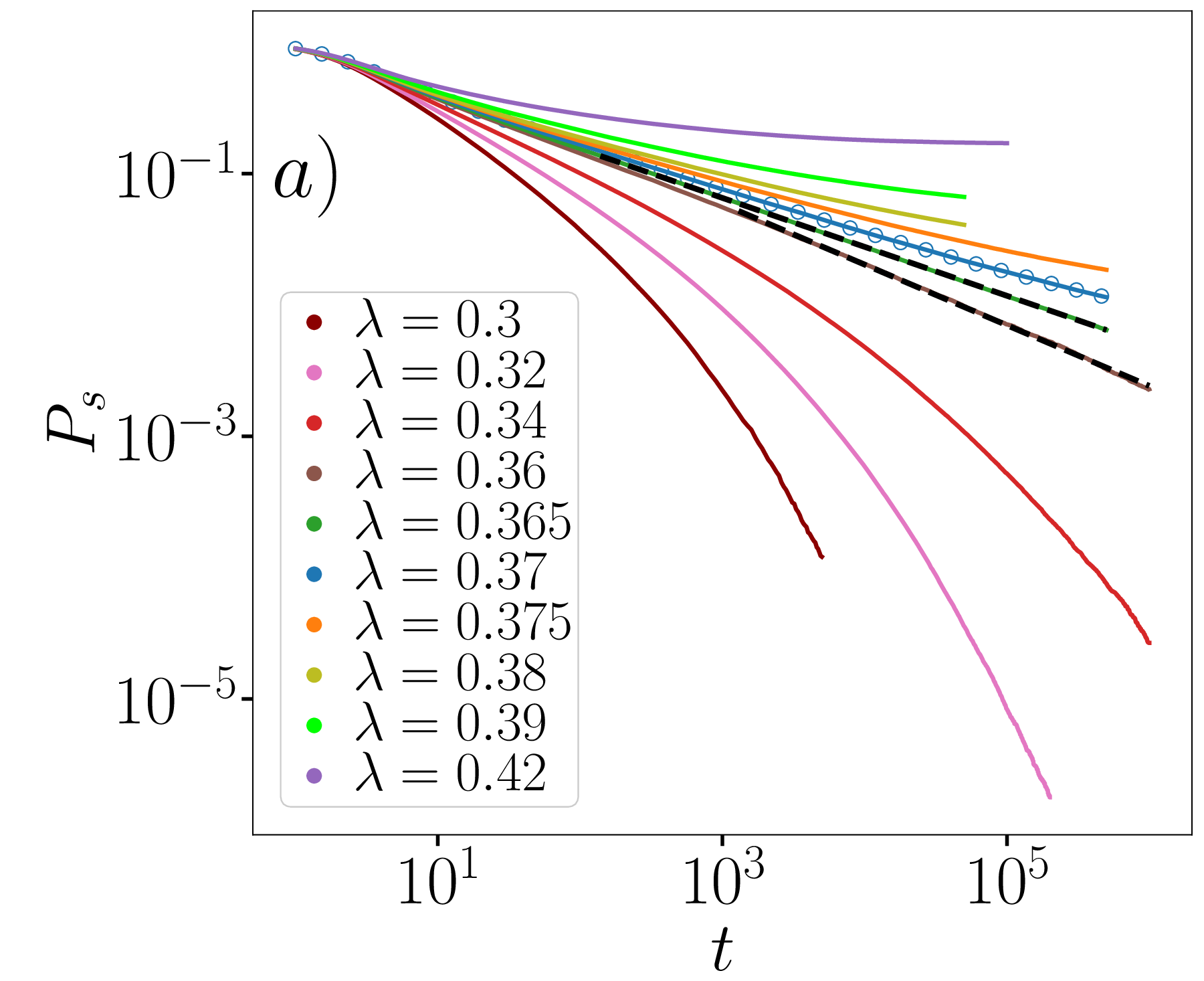}
    \includegraphics[width=0.49\columnwidth]{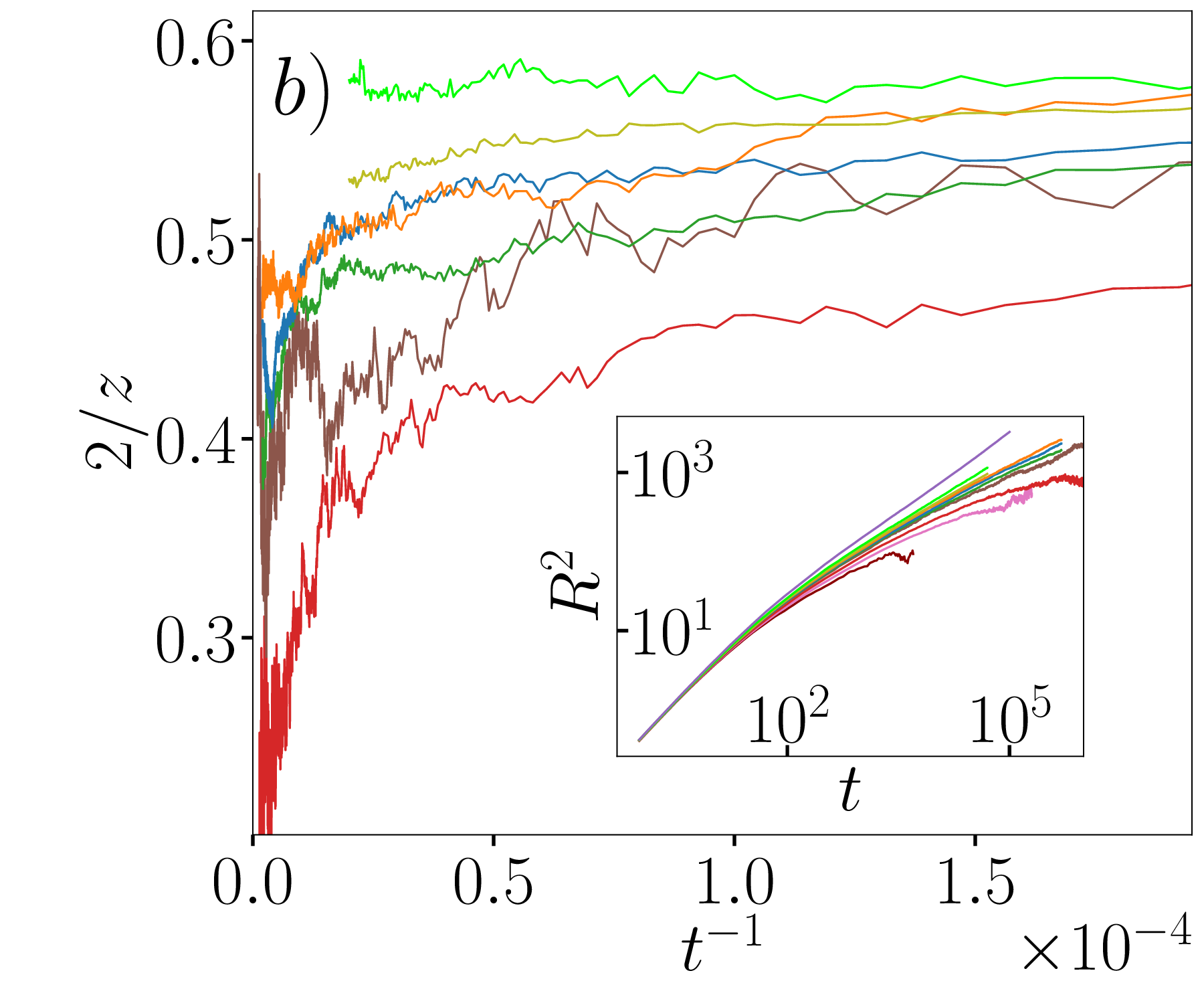}
    \caption{Effects of $\lambda$-disorder for $1=D_A > D_B=0.375$ ($p_{\lambda}=0.3, c=0.2$). (a) Log-log evolution of the survival probability $P_s(t)$. A power-law decay (black dotted lines) is observed for different $\lambda$, whereas the estimated critical point (circles, blue line) exhibits upward curvature. (b) Evolution of the effective dynamical exponent $2/z(t)$. The pronounced decrease toward zero reflects the activated, slower-than-power-law growth of the mean-square displacement $R^2(t)$ (inset).}
    \label{fig:LambdaDisorderDblDa}
\end{figure}

We focus first on the regime $D_A > D_B$, which is biologically relevant to systems, as in cell polarity, where signaling molecules diffuse faster in the cytoplasm than on the membrane \cite{altschuler_spontaneous_2008, brauns_phase-space_2020}.
We study the evolution of the survival probability \(P_s(t)\) [Fig.~\ref{fig:LambdaDisorderDblDa}a)] and find that at the estimated critical point [see Fig.~S11~\cite{supp}], the decay of \(P_s(t)\) deviates from a power law. 
Below criticality, $P_s(t)$ exhibits power-law decay (black dotted lines) throughout the accessible timescale, while the mean-square displacement $R^2(t)$ shows an unusually large, time-dependent dynamical exponent $z(t)$ indicative of sub-diffusive, slow dynamics [Fig.~\ref{fig:LambdaDisorderDblDa}b)]. 
These observations are inconsistent with a conventional fixed point but are compatible with an infinite-disorder fixed point (IDFP) \cite{fisher_critical_1995, igloi_strong_2005, igloi_strong_2018}, surrounded by Griffiths phases with power-law singularities, featuring activated scaling \cite{hooyberghs_strong_2003, vojta_critical_2005}:  
\begin{gather}
    P_s(t) \sim [\ln(t/t_0)]^{-\overline{\delta}}, \quad N_B(t)\sim [\ln(t/t_0)]^{\overline{\Theta}}, \notag \\
    R^2(t) \sim [\ln(t/t_0)]^{2/\psi},
    \label{eq:IDFP}
\end{gather}
where the tunneling exponent \(\psi\) and the non-universal microscopic scale $t_0$ relate correlation lengths \(\xi_{\perp}\) and \(\xi_{\parallel}\) via $\ln(\xi_{\parallel}/t_0) \sim \xi_{\perp}^{\psi}$. 

Conversely, for $D_B > D_A$, our simulations recover pure-system critical properties, confirming the predicted marginal irrelevance of disorder [Fig.~S10~\cite{supp}]. The case $D_A = D_B$ is more subtle: exponents $\beta$ and $\nu_\perp$ remain unchanged, yet $z$ increases with disorder strength. This suggests a weak-universality scenario [Figs.~S8--S9~\cite{supp}]. An IDFP might further emerge only above a disorder threshold, similar to predictions in the random quantum clock chain \cite{senthil_critical_1996}.

\textit{Diffusion-rate disorder}---Predicting the effects of diffusion heterogeneities is more complex because regions with $D_A(x) \gtrless D_B(x)$ may coexist. To assess disorder relevance, we define an effective rate $D_X^{\text{eff}}$ corresponding to the long-time diffusivity. For a binary distribution 
\begin{equation}
    \mathcal{P}(D_{X,i} ) = p_X\delta(D_{X,i}-D^0_{X}) + (1-p_X)\delta(D_{X,i}-D^1_{X}), 
    \label{eq:DisorderDistributionDiffusion}
\end{equation}
where $D_X^1 > D_X^0$ and $p_X$ is the probability of a slow site, the effective rate satisfies $1/D_X^{\text{eff}} = p_X/D_X^0 + (1-p_X)/D_X^1$. We heuristically use the relative difference between $D_A^{\text{eff}}$ and $D_B^{\text{eff}}$ to predict diffusion disorder relevance.

\begin{figure}
    \centering
    \includegraphics[width=0.7\columnwidth]{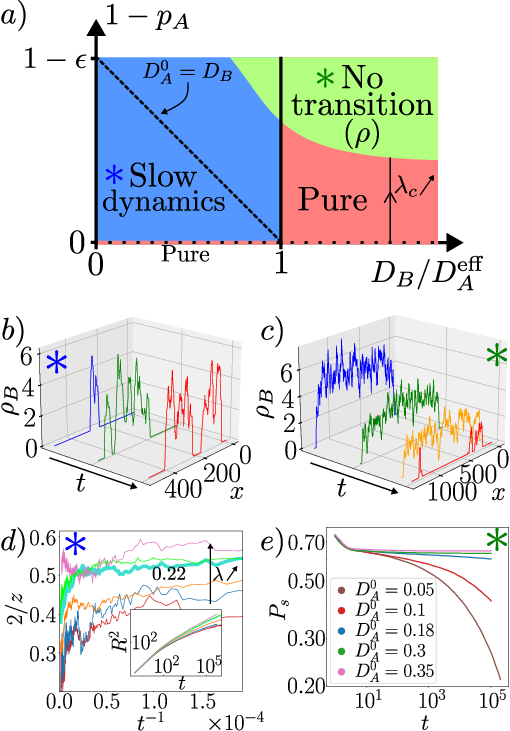}
    \caption{Phase transitions in $D_A$-disordered systems with fixed $D_A^1=2000$. 
    (a) Schematic phase diagram in the ($1-p_A, D_B/D_A^{\text{eff}}$) plane ($\epsilon > 0$ is a small positive value). Blue: slow-dynamics region, defined by a continuously decaying critical exponent $2/z(t)$. Red: pure critical properties. Green: suppressed active phase with density-dependent boundaries. 
    (b) Spatial profiles of the $B$-particle density $\rho_B(x)$ (averaged over 20 neighboring sites) at several times near criticality ($\lambda_c \approx 0.22, D_A^0 > D_B$). Infection survives in rare regions (red curve) and spreads slowly. 
    (c) Same as (b) for $D_A^0 < D_B$ and $\lambda=50$. The spreading of the infected cluster is arrested by extended low-density gaps (blue curve). Within these trapped regions, the density equilibrates; since $D_B > D_A^0$, the density progressively decays (yellow curve) toward spontaneous recovery, precluding a stable active phase (red curve precedes extinction).
    (d) Time evolution of the effective dynamical exponent $2/z(t)$. Continuous decay at the estimated critical point, $\lambda_c \approx 0.22$, is compatible with an IDFP. Other parameters: $p_A=0.5, D_A^0=1.4, D_B=1$. 
    (e) Evolution of the survival probability $P_s(t)$. Despite the large $\lambda=50$, which would typically yield a constant $P_s(t)$ at late times, the observed decay for sufficiently small $D_A^0$, confirms the suppression of the active phase. Other parameters: $p_A=0.5, D_B=1$.}
    \label{fig:DaDisorder}
\end{figure}

To test this, we first consider fixed $D_B$ with random $D_{A,i}$. In the extreme limit $D_A^1 \gg D_A^0, D_B$, simulations confirm Harris relevance based on $D_A^{\text{eff}}$ [Fig.~\ref{fig:DaDisorder}(a)]. For $D_A^{\text{eff}} > D_B$, critical properties differ from the pure case, yielding a time-decreasing effective exponent $2/z(t)$ (denoted ``slow-dynamics'') characteristic of an IDFP [Fig.~\ref{fig:DaDisorder}(d)]. This relevance is theoretically supported by an analytical upper bound on $2/z$ derived for $D_A^1 \to \infty$, which decays to 0 as $p_A$ decreases [Eq.~S16,~\cite{supp}]. This implies that critical exponents must differ from pure values at least for sufficiently small $p_A$, and that the disordered $z$ must either depend on $p_A$ or become infinite, as expected at an IDFP. This ``slow-dynamics'' regime is not restricted to the large-fluctuation limit; it is also observed for small disorder variance ($D_A^1 \approx D_A^0$) [Fig.~S13,~\cite{supp}].

The most striking consequence of strong heterogeneities is the emergence of a regime lacking a stable active phase [Fig.~\ref{fig:DaDisorder}(a)], also compatible with the disorder-dependent analytical bound suggesting that critical properties shift with $p_A$. Here, $A$ particles concentrate in $D_A^0$ regions, forcing the infection to bridge low-density gaps. This behavior is governed by the ratio $D_B/D_A^0$: for $D_B > D_A^0$, outward particle fluxes drive a progressive decay in average density [Fig.~\ref{fig:DaDisorder}(c)], suppressing survival even at large $\lambda$. In this case, rare density fluctuations leaving a single particle per site have sufficient time to occur, allowing spontaneous recovery independent of $\lambda$ [Fig.~\ref{fig:DaDisorder}(e)]. Conversely, if $D_B < D_A^0$, inward fluxes sustain long-lived infected clusters [Fig.~\ref{fig:DaDisorder}(b)], preserving the absorbing phase transition.

Finally, $D_B$-disorder appears comparatively weaker, and the disorder strength is more difficult to tune [Sec.~IV~D,~\cite{supp}]. To accelerate the expected crossover from the pure to the disordered fixed point, we introduce perfectly correlated disorder over $X=10$ sites. We observe the effective exponent $2/z(t)$ at criticality decreasing continuously with $t$ to values below the pure case for $D_B^{\text{eff}} < D_A$ [Fig.~S20,~\cite{supp}], consistent with activated scaling. However, these correlations introduce a mesoscopic time scale that limits the accessible asymptotic range. While we cannot exclude that true asymptotics differ from those observed up to $T_{\max}=10^7$, a scenario where only $D_B$-disorder is irrelevant---in contrast to $\lambda$- and $D_A$-disorder---would contradict expectations and necessitate a new theoretical framework.

\begin{figure}
    \centering
    \includegraphics[width=0.95\columnwidth]{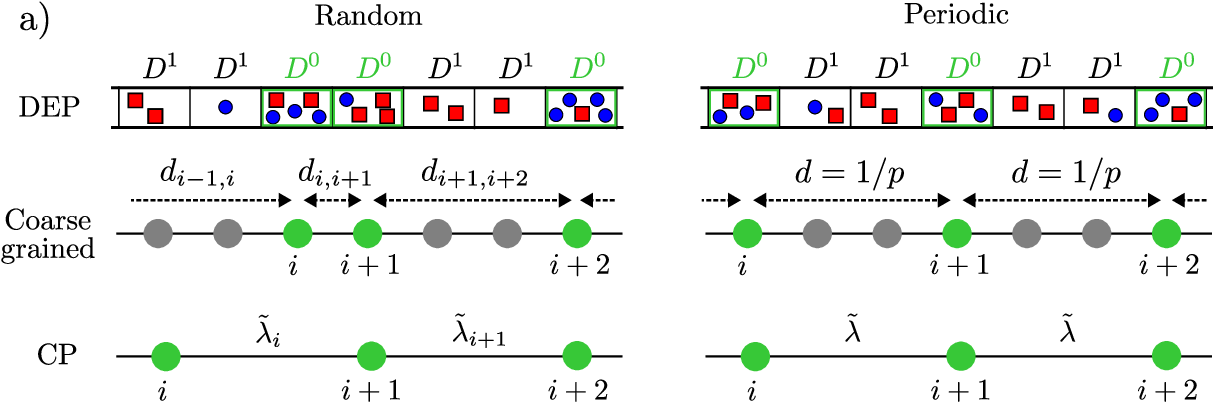}
    \includegraphics[width=0.45\columnwidth]{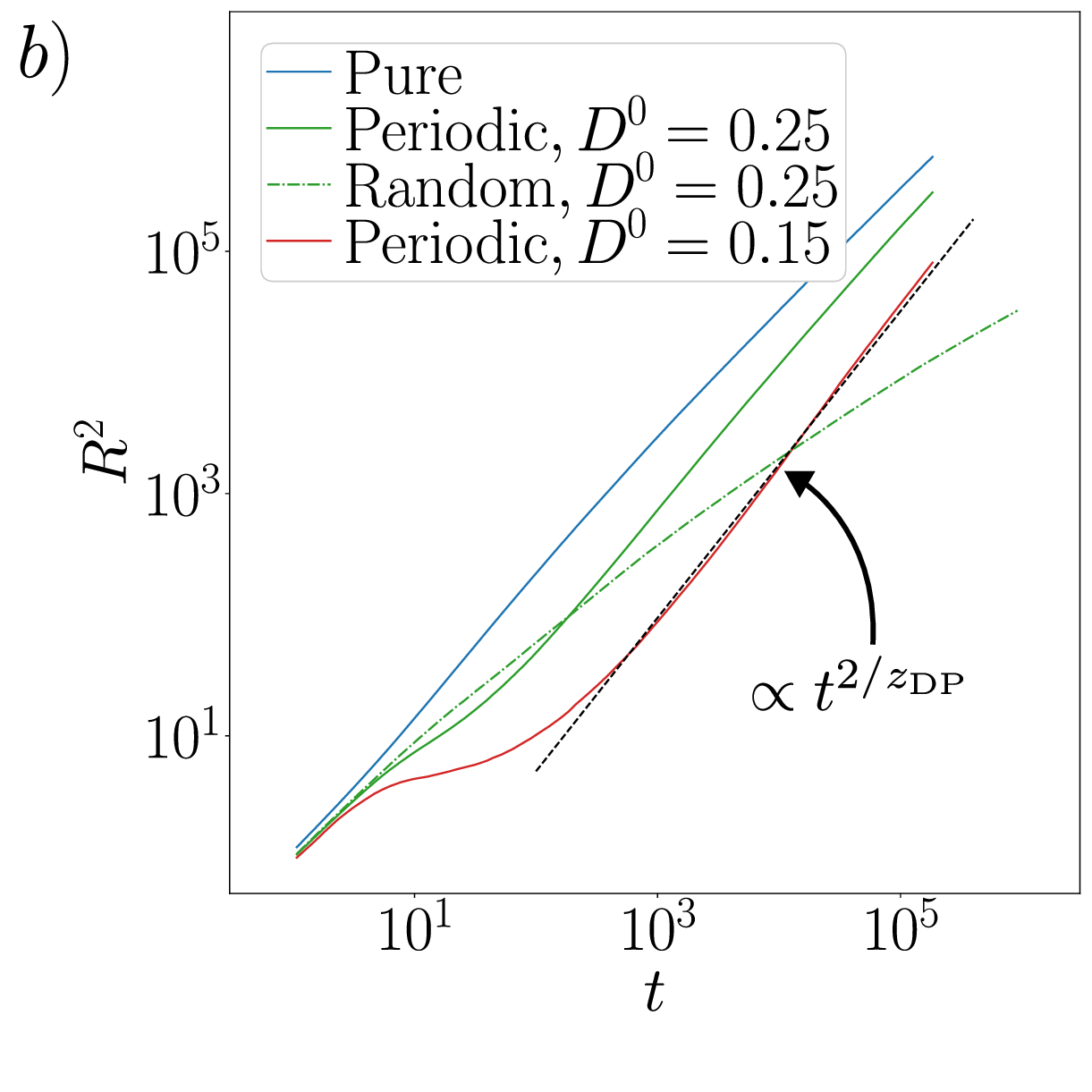}
    \includegraphics[width=0.45\columnwidth]{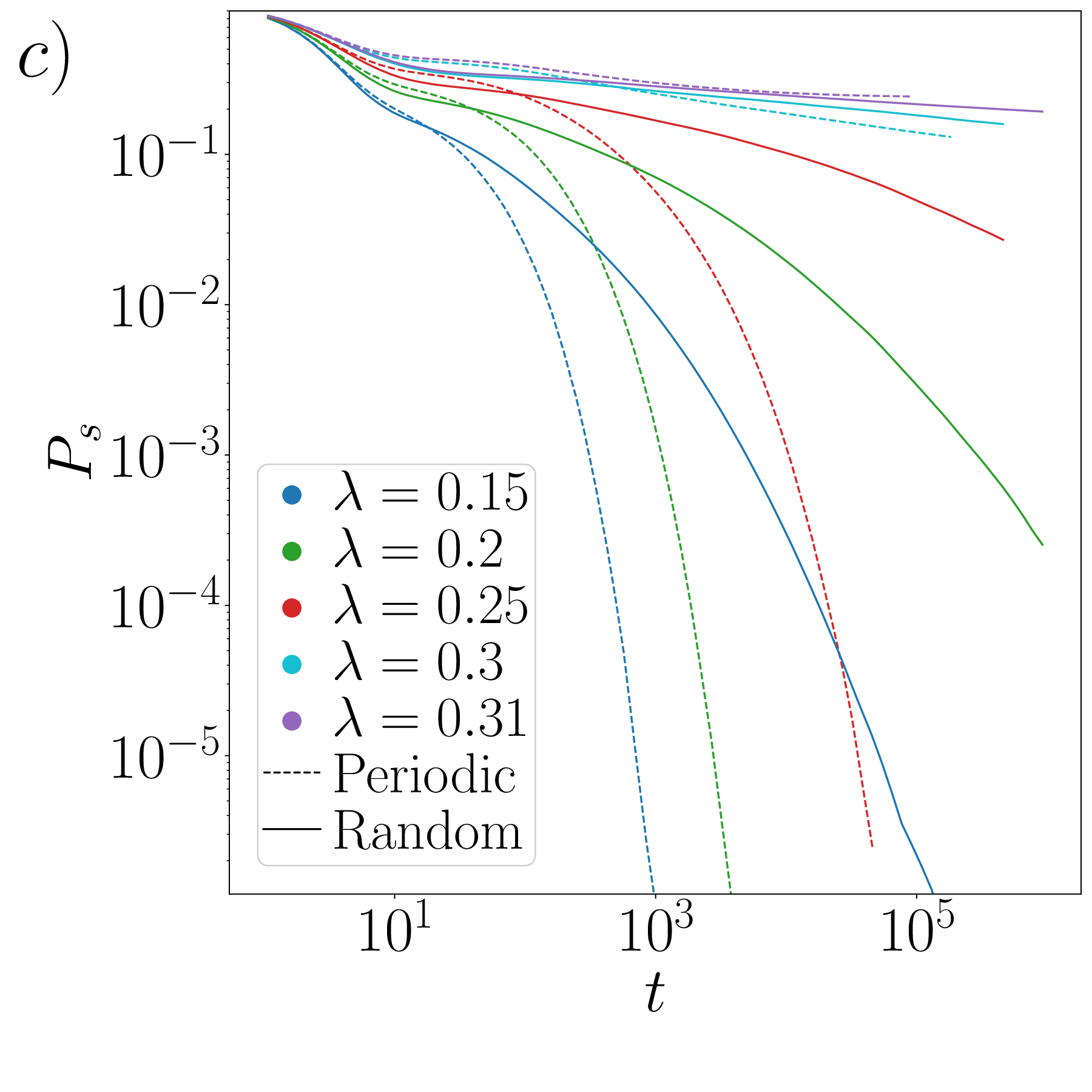}
     \caption{(a) Schematic of perfectly correlated disorder ($D_{A,i}= D_{B,i} = D^{0/1}$). Sites with $D^0$ are either randomly distributed (left) or periodically spaced (right). The coarse-grained mapping retains only the state of $D^0$ sites and their mutual distances $d_{i,j}$, mapping the system to a contact process with either a random (left) or uniform (right) effective infection rate $\tilde{\lambda}$. (b) Mean-square displacement $R^2(t)$ at criticality for various $D^0$ ($p=0.1, D^1=1$) under periodic disorder, illustrating different transient growth profiles. Randomizing the position of $D^0$ sites leads to significantly slower, activated growth (dotted green line). (c) Survival probability $P_s(t)$ for periodic (dotted) and random (solid) disorder ($D^0=0.25, D^1=1, p=0.1$). The slower decay for random disorder is attributed to persistent activity within rare clusters of adjacent $D^0$ sites. }
    \label{fig:correlatedDisorder}
\end{figure}

\textit{Correlated diffusion disorder.}---We examine perfectly correlated rates $D_{A,i} = D_{B,i} = D_i$ using the binary distribution of Eq.~\ref{eq:DisorderDistributionDiffusion}. This configuration maintains $D^{\text{eff}}_A = D^{\text{eff}}_B$ while producing a stationary, spatially heterogeneous density profile. To interpret the resulting dynamics, we propose a coarse-grained mapping to a contact process dominated by high-density (slow $D^0$) sites, treating fast $D^1$ sites as a background that effectively modifies the inter-site infection rate based on distance [Fig.~\ref{fig:correlatedDisorder}(a)]. Under this mapping, periodic spacing of $D^0$ sites corresponds to a pure DP process, whereas random spacing mimics a disordered DP.

While periodic disorder asymptotically belongs to the pure DEP class, highlighting the limits of the coarse-grained mapping, the transient dynamics are significantly altered [Fig.~\ref{fig:correlatedDisorder}(b)]. Decreasing $D^0$ enhances particle localization, introducing a mesoscopic timescale and a plateau in $R^2(t)$ before it eventually accelerates toward $2/z_{\text{DP}}$ scaling. In contrast, randomly spaced slow sites result in much slower, activated growth. This distinction is further reflected in the survival probability $P_s(t)$ [Fig.~\ref{fig:correlatedDisorder}(c)], where random disorder exhibits a slower decay in the absorbing phase due to activity persistence in rare $D^0$ clusters—a hallmark of a Griffiths phase. This picture clarifies how disorder strength fundamentally reshapes the dynamics even in the marginal $D_A=D_B$ regime. However, unambiguously inferring the asymptotic critical properties remains challenging, as the existence of long-lived transients may obscure the true late-time limit.

\begin{figure}
    \centering
    \includegraphics[width=0.7\linewidth]{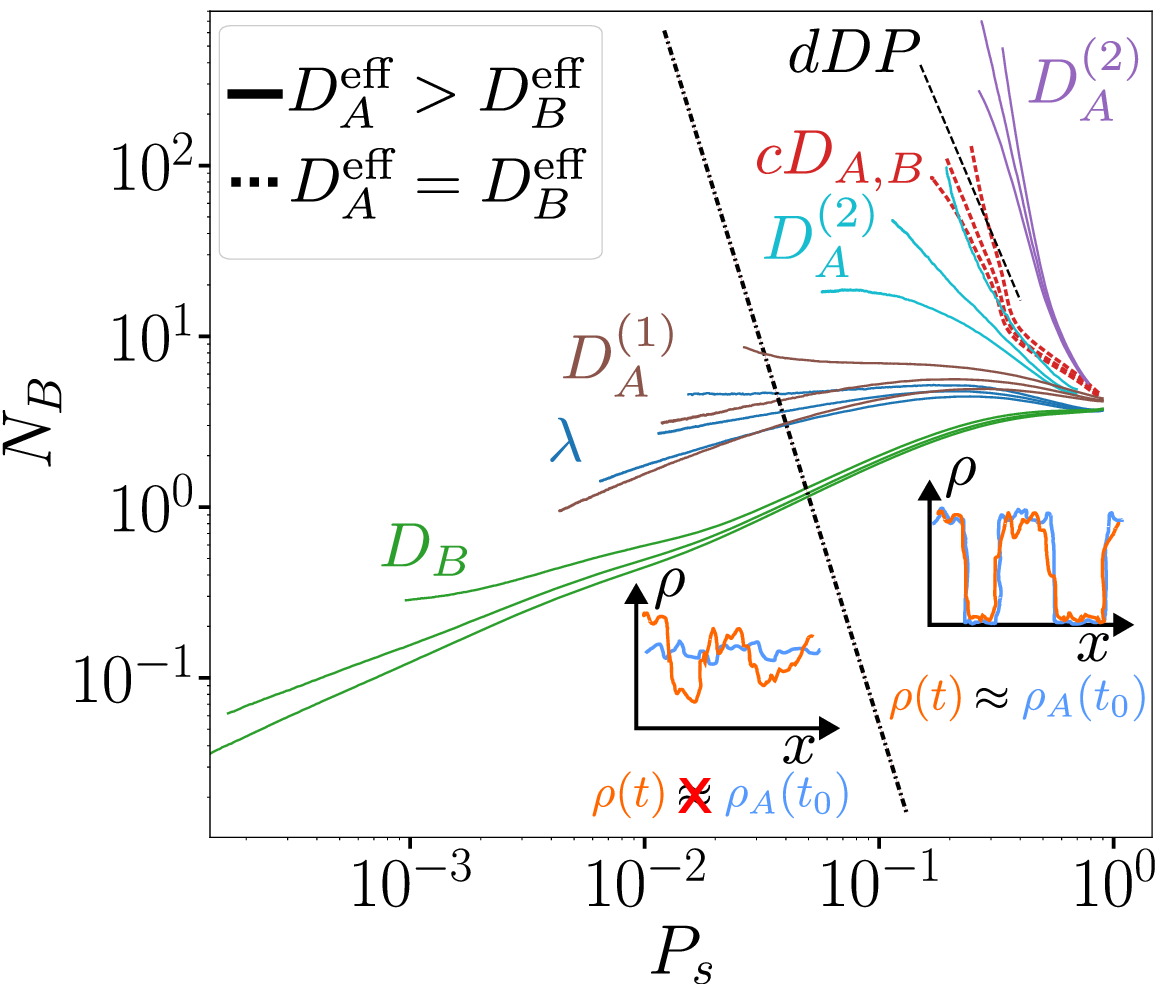}
    \caption{$P_s$ versus $N_B$ at criticality for $D_A^{\text{eff}} > D_B^{\text{eff}}$ under varying quenched disorder. Colors: green ($D_B$ block), blue ($\lambda$-disorder), brown ($D_A$ weak variance), cyan/purple ($D_A$ strong variance for $D_A^0 > D_B$ and $D_A^0 < D_B$, respectively). Correlated disorder (red) with $D_A^{\text{eff}} = D_B^{\text{eff}}$ is included as a reference. The black dotted line represents the slope $-\overline{\Theta}/\overline{\delta}$ of the disordered contact process. Two distinct regimes emerge: a positive-slope class (resembling pure DEP) and a negative-slope class (analogous to the disordered contact process). The transition of $D_A$-disorder across both regimes highlights the non-trivial impact of diffusion variance on scaling. Insets: Qualitative evolution of total density after infection; the first class shows $\rho(t) \neq \rho_A(t_0)$, whereas the second shows $\rho(t) \approx \rho_A(t_0) \, \forall t$.
}
    \label{fig:ComparisonNbPs_ptcrit_differentsModels}
\end{figure}

\textit{Comparison of IDFPs.}---We assess the ratio of critical exponents $\overline{\Theta}/\overline{\delta}$ at the estimated critical point $\lambda_c$ for the previously discussed disorder distributions satisfying $D_A^{\mathrm{eff}}>D_B^{\mathrm{eff}}$. By including data from a range of $\lambda$ values near $\lambda_c$, we quantify the uncertainty in this slope determination [Fig.~\ref{fig:ComparisonNbPs_ptcrit_differentsModels}].

The results reveal two clearly distinct universality classes. The first, characterized by a positive slope, encompasses $\lambda$-disorder, block-$D_B$ disorder, and $D_A$-disorder with weak variance ($D_A^1/D_A^0 \simeq 1$). The second, exhibiting a negative slope, includes $D_A$-disorder with strong variance ($D_A^1/D_A^0 \gg 1$), correlated disorder with $D_A^{\text{eff}} = D_B^{\text{eff}}$, and the disordered contact process.

This separation indicates that the microscopic distribution $\mathcal{P}(D_{A,i})$ can fundamentally alter critical behavior, suggesting either a crossover between distinct IDFPs or the emergence of a disorder-dependent continuum of fixed points. We hypothesize that when $D_A^1/D_A^0 \gg 1$, the system separates into regions of extreme density contrast. Here, the specific microscopic dynamics of DEP are suppressed, and the density profile $\rho(x)$ becomes enslaved to the initial distribution $\rho_A(x_0)$, mirroring the correlated diffusion case. Further work is required to determine whether the observed similarities with the disordered contact process—currently identified through the ratio of critical exponents—extend to the full scaling set. In that case, it would further enlarge this super-universal class with a process with a conserved quantity \cite{senthil_critical_1996, hooyberghs_strong_2003, kang_superuniversality_2020, anfray_numerical_2021}.

\textit{Discussion.}---Despite the sensitivity of critical properties to relative diffusion rates, we show for the one-dimensional DEP that the relevance of quenched disorder is unambiguously predicted by the Harris criterion using effective macroscopic rates. Our simulations confirm that weak disorder is relevant for $D_A^{\mathrm{eff}} > D_B^{\mathrm{eff}}$, revealing signatures of an infinite-disorder fixed point (IDFP), distinct from that of the disordered contact process \cite{dickison_monte_2005}, and presence of Griffiths phases. Such slow dynamics could have important consequences in cell-polarity systems, where quenched heterogeneity in molecule transport, turnover, or scaffold abundance can create rare favorable cortical sites \cite{xiang_single-molecule_2020, smigiel_protein_2022, huang_cytoplasmic_2022, garner_vast_2023}. These sites may dictate pattern selection, producing broad polarization latencies and mediating slow or intermittent competition between competing domains. The slow dynamics are also reminiscent of smoldering-like persistence, in which infection or activity can appear to fade globally while surviving for long times in rare favorable spatial regions \cite{bejon_stable_2010,santos_definition_2019,dom_housing_2026,noe_mapping_2018}.

To model stationary, spatially heterogeneous populations, we examined correlated spatial rates ($D_B(x)=D_A(x)$). Strong density fluctuations act as relevant perturbations potentially driving the system toward an IDFP. While Griffiths effects are documented in disordered contact processes \cite{hooyberghs_absorbing_2004, vojta_infinite-randomness_2009, vojta_monte_2012, buono_slow_2013, juhasz_rare-region_2012}, the role of static density variations remains largely unexplored. Whether the same IDFP governs this regime is an open question. 

Extending this analysis to higher dimensions or complex networks would test the robustness of these findings and facilitate comparisons with connectivity disorder, which is irrelevant in the two dimensional DEP \cite{alencar_two-dimensional_2023}. Finally, in many nonequilibrium models, diffusion does not alter the critical behavior \cite{dickman_universality_1989, jensen_time-dependent_1993}, and spatial variations of the diffusion rates are typically expected to be irrelevant by naive power counting. The DEP provides a definitive counterexample: random-diffusion disorder modifies critical scaling and triggers phenomena fundamentally inaccessible to random-mass disorder. This suggests that spatially varying transport processes can significantly influence absorbing-state phase transitions.

\nocite{chayes_finite-size_1986, eisenberg_expectation_2008, feller_introduction_1968, norden_distribution_1982, park_high-precision_2013}

\textit{Acknowledgement}---We gratefully acknowledge valuable discussions with Fr\'ed\'eric van Wijland. This work was supported by grants from the National Science and Technology Council, Taiwan (Grant No. NSTC 111-2112-M-001-027-MY3 and 114-2112-M-001-062) and Academia Sinica Career Development Award (Project No. AS-CDA-114-M02). VA acknowledges support from Academia Sinica Postdoctoral Scholar Program.

\bibliography{references3}

\end{document}


\title{Supplementary material}

\author{Valentin Anfray$^1$}
\email{valentin.anfray@gmail.com}
\author{Hong-Yan Shih$^{1,2}$}%
\email{hongyan@as.edu.tw}
\affiliation{%
 $^1$Institute of Physics, Academia Sinica, Taipei 115201, Taiwan\\
 $^2$Physics Division, National Center for Theoretical Sciences, Taipei 106319, Taiwan
}%

\maketitle

This supplementary material details the simulation implementation and methods for estimating critical properties. We present supporting data and figures for the critical exponents of the three pure universality classes of the Diffusive Epidemic Process (DEP), along with the impact of quenched disorder on the contamination rate in each case. Additional data on disorder effects in \( D_A \) and \( D_B \) are provided, as well as an analytical study of a simplified model explaining the influence of \( D_A \)-disorder, particularly the slow growth of the mean-square displacement and the absence of an active phase. Finally, we include additional data for cases where on-site diffusion rates are perfectly correlated and equal, and further discuss cases where they are not.

\tableofcontents

\newpage 

\section{Numerical details}

\subsection{Simulation setup}

We employ the Gillespie algorithm \cite{gillespie_exact_1977, gillespie_stochastic_2007} to simulate our system. To explicitly account for spatial heterogeneity, we compute the propensity at each site and use it to determine the time of the next reaction. We utilize the next reaction method \cite{gibson_efficient_2000}, which is particularly well-suited to our model, as an event occurring at one site can, at worst, modify the propensity of one of its neighboring sites.

Our simulations primarily begin with an initially infected seed to investigate its spreading behavior \cite{grassberger_reggeon_1979, henkel_non-equilibrium_2008}. For each disorder configuration, we typically perform 32 or 64 runs, placing the initial seed randomly within a range of 1000 sites and simulating up to a maximum time $T_{\max}$. The results are then averaged over multiple disorder configurations, typically ranging from $10^3$ to $10^6$.

We fix the total density at $\rho = 5$ and the recovery rate at $\tau = 1$, using the contamination rate $\lambda$ as the control parameter. The diffusion rates $D_A$ and $D_B$ are varied to explore the three different universality classes of the DEP. Initially, the number of $A$ particles per site follows a Poisson distribution, whose mean depends on both the density $\rho$ and the diffusion rates $D_{A,i}$, corresponding to the steady state in the absence of $B$ particles. Using the binary distribution from Eq.~(4) in the main text, rewritten here for completeness:
\begin{equation}
    \mathcal{P}(D_{X,i}) = p_X\delta(D_{X,i}-D^0_{X}) + (1-p_X)\delta(D_{X,i}-D^1_{X}),
    \label{eq:DisorderDistributionDiffusion}
\end{equation}
the initial average density for sites with the diffusion rate $D_A^0$ is given by:
\begin{equation}
    \rho_0 = \frac{\rho/D_A^0}{(1-p_A)/D_A^1 + p_A/D_A^0},
\end{equation}
and otherwise by $(\rho - p_A \rho_0)/(1 - p_A)$, where $p_A$ denotes the probability of a site having $D_{A,i} = D_A^0$.

At the position of the infected seed, all $A$ particles are converted into $B$ particles. Simulations continue until $t = T_{\max}$ or until no $B$ particles remain. Notably, since we simulate effectively infinite system sizes, we do not impose any constraints on the total number of particles $N_{\text{tot}}$. This contrasts with typical simulations on finite systems of size $L$ where the density $\rho = N_{\text{tot}}/L$ is explicitly enforced.

\subsection{Method to estimate the critical properties}
\label{subsection:estimateCriticalProperties}

To estimate the position of the critical point, $\lambda_c$, and the critical exponents, we calculate three effective exponents, \( \delta(t) \), \( \Theta(t) \), and \( 2/z(t) \), which relate to the time variations of \( P_s \), \( N_B \), and \( R^2 \), respectively, as described in Eq.~(1) of the main text. For instance, $-\delta(t) = \frac{\partial \ln P_s(t)}{\partial \ln t}$ \cite{riedel_effective_1974} is determined numerically by\( -\delta(t) = \frac{\ln[P_s(t)/P_s(t/b)]}{\ln b}\) \cite{hinrichsen_nonequilibrium_2000}. We use $b = 4$, as lower values introduce greater noise and require more samples. Due to scaling corrections, we expect \(\delta(t) = \delta + \frac{a}{t^{\omega}} + \cdots\)\cite{grassberger_directed_1989}. Since precise estimation of the critical exponents is not the primary goal, we plot the effective exponents versus $1/t$ to extrapolate as $t \to \infty$. To estimate the leading correction $\omega$, larger maximum times and more samples would be necessary \cite{park_high-precision_2013}.

For the correlation length exponent \( \nu_{\parallel} \), we plot \( N_B t^{-\Theta} \) versus \( t \Delta^{\nu_{\parallel}} \), where \( \Delta = \frac{|\lambda - \lambda_c|}{\lambda_c} \), and adjust only \( \nu_{\parallel} \) to achieve a satisfactory data collapse, with the other parameters \( \lambda_c \) and \( \Theta \) being previously estimated. This estimate does not account for corrections and strongly depends on the previously determined \(\lambda_c\) and \( \Theta \), which explains the larger systematic uncertainty observed for \( \nu_{\parallel} \).

In the following sections, to extract the critical properties, we use the effective critical exponent \( \Theta(t) \) to identify the position of the critical point, as it is empirically the most sensitive indicator \cite{henkel_non-equilibrium_2008}. Away from the critical point, the effective exponent should either curve upward or downward. In some cases, we also plot \( P_s \) versus \( N_B \) to estimate the position of the critical point, as \( P_s \sim N_B^{-\delta/\Theta} \) holds at both a conventional critical point and an infinite disorder fixed point \cite{vojta_infinite-randomness_2009}. Once the position of the critical point is estimated, we use \( \delta(t) \) and \( 2/z(t) \) at this value to estimate \( \delta \) and \( 2/z \). If sufficient data are available outside the critical point, we estimate \( \nu_{\parallel} \) via data collapse.

\subsection{Infinite system size simulation}

Due to particle conservation, even when starting with a seed of infected particles, it is necessary to track the positions of all other particles. This contrasts with the contact process, where only the positions of infected sites are required to evolve the system. To overcome this difficulty, we exploit the fact that $A$ particles undergo free diffusion, meaning that the trajectory of any individual $A$ particle is independent of its surroundings unless it encounters an infected particle.

To simulate effectively infinite systems, we partition particles into two subsystems: $S_{AB}$ which contain all the $B$ particles and the relevant $A$ particles (defined in the following) where we perform exact Gillespie simulations and $S_A$ where only simple diffusion is performed. In principle, $S_{AB}$ is infinite, but in practice, we can evolve its size with $t$ such that it remains finite up to $T_{\max}$. The goal of our algorithm is to transfer particles between the two systems such that the dynamics of the $B$ particles remain identical to those in an infinite system while keeping the number of particles in $S_{AB}$ as small as possible. The simulation time $t$ is defined by the time in $S_{AB}$, where the exact dynamics take place. We denote $x^B_{\min}$ and $x^B_{\max}$ as the leftmost and rightmost positions of the infected particles, respectively.

We introduce a free parameter, $T_{\text{upt}}$, representing the maximum time interval between updates of $S_A$ and $S_{AB}$. During this time, we are interested in the maximum distance the infection front might travel. Since this is a complex question, we instead define another free parameter, $l_{B,\max}$, which represents the maximum expected distance. Additionally, we need to determine which $A$ particles from $S_A$ and $S_{AB}$ could occupy the region $[x^B_{\min} - l_{B,\max}, x^B_{\max} + l_{B,\max}]$ during $t \in [t_i, t_i + T_{\text{upt}}]$. To estimate this, we use the first hitting time probability, $F_1(L_j, t)$, which gives the probability that a single particle starting at $x=L_j$ will reach a target at $x=0$ by time $T_{\text{upt}}$. Given a diffusion coefficient $D$, it is expressed as:

\begin{equation}
    F_1(L_j,T_{\text{upt}}) = 1-\textrm{erf}(L_j/\sqrt{4DT_{\text{upt}}}),
\end{equation}
with $\textrm{erf}(x)$ the error function.
As a result, the probability that all particles starting at $x=0$ will remain at a position $x<L_j$ for all times up to $t+T_{\text{upt}}$ is:

\begin{equation}
    K_j(L_j,t) = \prod_k \textrm{erf}\left[L_j/\sqrt{4D(t+T_{\text{upt}}-T_k)}\right]^{N_k}.
    \label{eq:ProbNoHitting}
\end{equation}

where $N_k$ is the number of particles updated at time $T_k$ located at site $j$. Here, we take into account that site $j$ might be occupied by many $A$ particles with different last update times $T_k$.

We also introduce a probability threshold, $p_{\text{out}}$, which defines the set of $A$ particles in $S_A$ that should be included in $S_{AB}$ if the probability of reaching the region $[x^B_{\min} - l_{B,\max}, x^B_{\max} + l_{B,\max}]$ by time $t + T_{\text{upt}}$ is below $p_{\text{out}}$. Specifically, we define the leftmost site, $x_{\text{left}}$, as the first position where $\prod_{j=0}^4 K_{x_{\text{left}}+j}(L_{x_{\text{left}}+j},t) > p_{\text{out}}$, where $L_{x_{\text{left}}+j}$ is the distance between site $x_{\text{left}}+j$ and $x^B_{\min} - l_{B,\max}$. A similar approach is used to determine the rightmost site to include in $S_{AB}$.

Using exact simulation, we evolve all $A$ particles in $S_A$ initially within the interval $[x_{\text{left}}, x_{\text{right}}]$ up to time $t$. Next, we transfer the $A$ particles in $S_A$ that might reach the interval $[x^B_{\min} - l_{B,\max}, x^B_{\max} + l_{B,\max}]$ by time $t + T_{\text{upt}}$ into $S_{AB}$. Conversely, we move the $A$ particles from $S_{AB}$ to $S_A$ if they are not likely to reach the interval $[x^B_{\min} - l_{B,\max}, x^B_{\max} + l_{B,\max}]$ by time $t + T_{\text{upt}}$.

\begin{figure}
    \centering
    \includegraphics[width=0.5\linewidth]{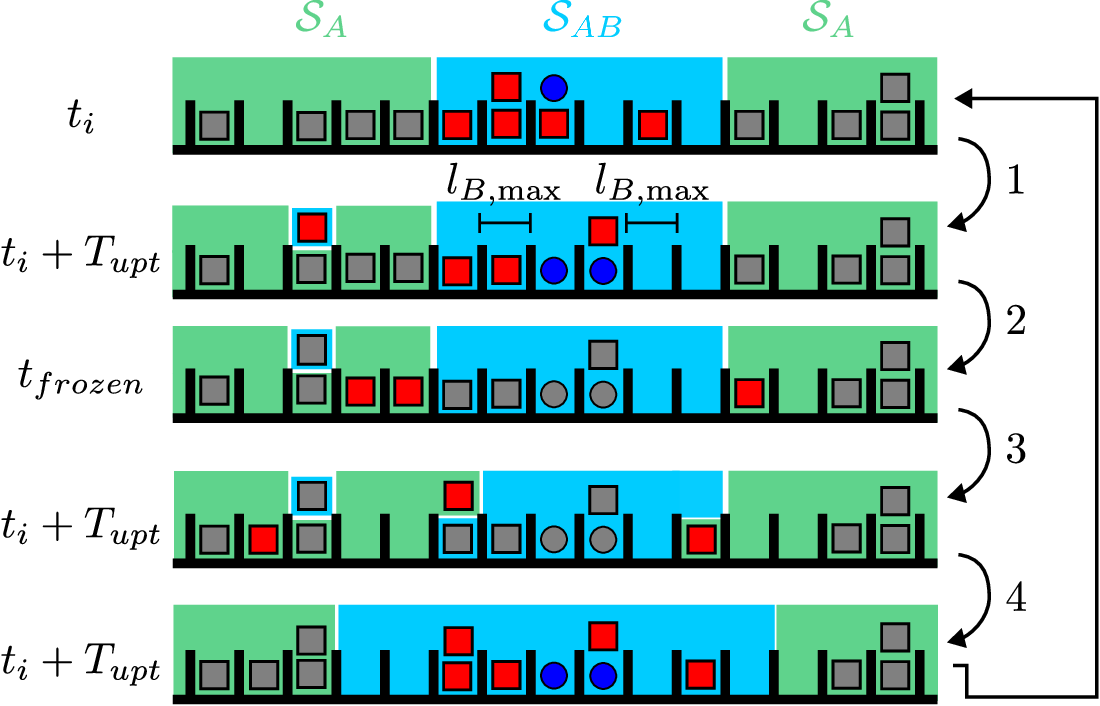}
    \caption{Schematic representation of the main steps of the infinite-size algorithm, see text for details. Note: each healthy particle $k$ in $\mathcal{S}_A$ has its own $t^k_{\text{frozen}}$, not represented here for clarity.}
    \label{fig:UnbseedAlgo}
\end{figure}

The steps are summarized in Figure \ref{fig:UnbseedAlgo}. Suppose we start at time $t_i$ with particles belonging to either $S_A$ or $S_{AB}$:

\begin{enumerate}
    \item Perform Gillespie simulation on $S_{AB}$ up to $t_{i} + T_{\text{upt}}$.
    \item Use Eq.~\eqref{eq:ProbNoHitting} to identify $A$ particles that could interact with $B$ particles between $t_i+ T_{\text{upt}}$ and $t_i + 2T_{\text{upt}}$, based on their most recent update time, $t_{\text{frozen}}$.
    \item Perform single-particle diffusion to update the positions of those particles up to time $t_i+T_{\text{upt}}$.
    \item Use Eq.~\eqref{eq:ProbNoHitting} again to refine the list of particles that could encounter a $B$ particle during $t_i+T_{\text{upt}}$ to $t_i + 2T_{\text{upt}}$, and update the systems $S_A$ and $S_{AB}$ accordingly.
    \item Return to step 1.
\end{enumerate}

We have monitored the accuracy of this procedure by tracking the number of free-diffusing $A$ particles (in $S_A$) during step 3 that reach the position of a $B$ particle (in $S_{AB}$), indicating potential missed interactions.

\begin{figure}
    \centering
    \includegraphics[width=0.4\linewidth]{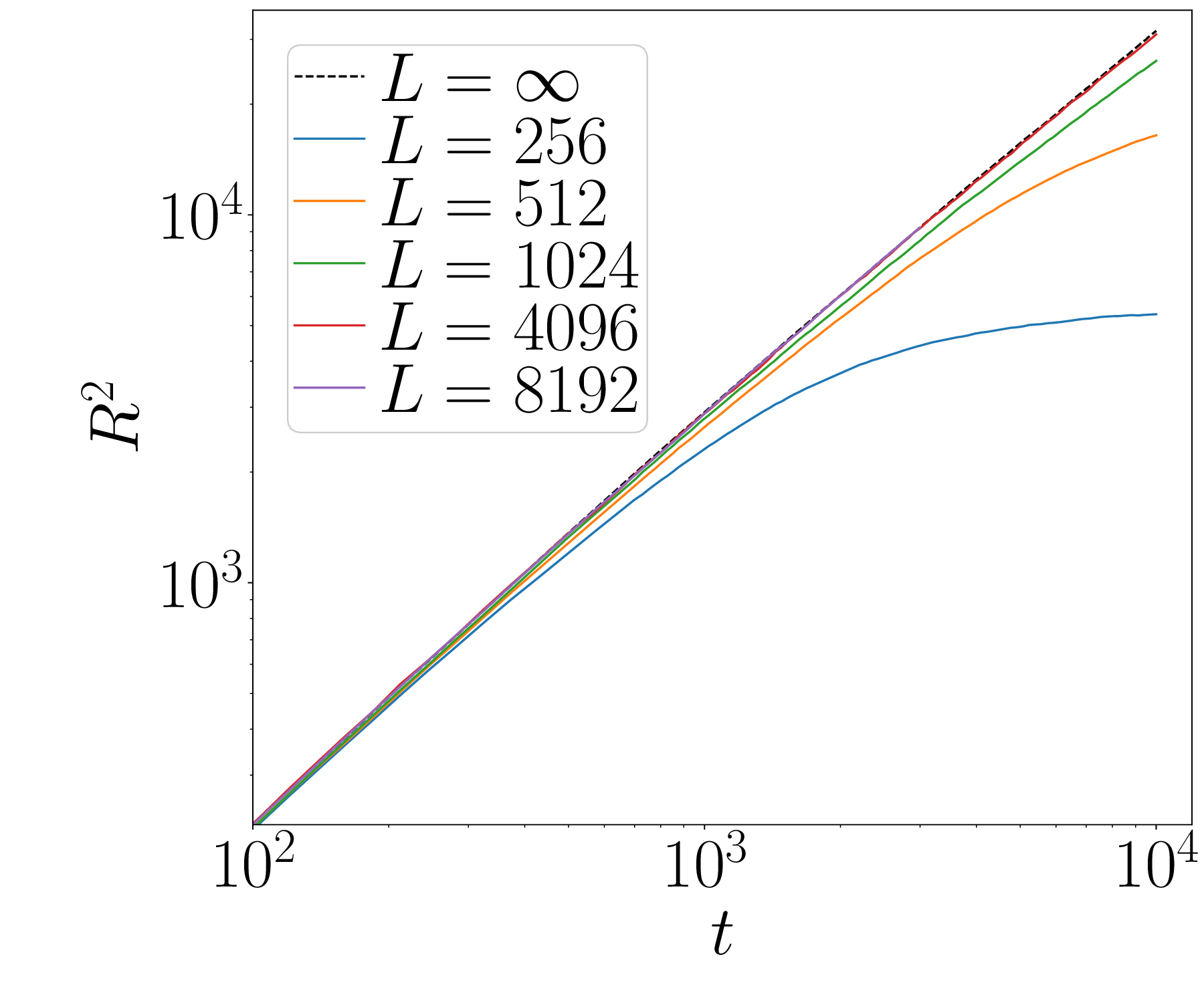}
    \caption{Mean-square displacement \(R^2\) close to the critical point starting from an infected seed for various system sizes \(L\), including the case \(L = \infty\) using our algorithm. Parameters: \(D_A = D_B = 1\), \(\lambda = 0.349\), \(\rho = 5\), and \(\tau = 1\).}
    \label{fig:Comparaison_Unbseed_Seed_Rsquare}
\end{figure}

In Figure \ref{fig:Comparaison_Unbseed_Seed_Rsquare}, we compare $R^2(t)$ for different system sizes $L$ and for our algorithm indicated by $L=\infty$. Finite-size effects become apparent as $R^2$ saturates after some time. For $L=256$, saturation occurs around $t=10^4$, but deviations from infinite system behavior appear earlier, around $t = 300$. Similar behavior is observed for other system sizes. 

Extracting critical exponents from the slope becomes challenging due to these finite-size effects, which our infinite-size algorithm avoids. However, the effective system size grows with time, limiting us to $T_{\max} \sim 10^5/10^6$ (depending on $D_A, D_B$), typically corresponding to $L \sim 10^5$. While simulating longer times is possible, it would require averaging over fewer independent runs. Nevertheless, our infinite-size algorithm remains faster than finite-size simulations, at least until the latter deviate due to finite-size effects. To obtain the data in Figure \ref{fig:Comparaison_Unbseed_Seed_Rsquare} for $T_{\max} = 10^4$, our algorithm is 0.5, 0.9, 1.6, and 8 times faster than the usual finite-size simulations for $L=256, 512, 1024, 4096$, respectively after averaging over $10^5$ samples. This speedup arises from two factors: first, we avoid simulating particles that are not yet relevant to the spreading dynamics, and second, simulating single-particle diffusion is faster than using spatial Gillespie methods. We have typically set the parameters to $p_{\text{out}}=0.995, T_{\text{upt}}=40, l_{B,\max}=70$. 

It is important to note that although our algorithm simulates a system of an effective size $L$, the results are not identical to those obtained from a finite system of the same size $L$. In finite systems, the total number of particles is fixed, a constraint absent in our algorithm, which aims to simulate an infinite-size system without local particle number restrictions. Consequently, fluctuations differ, potentially affecting the dynamics. This particle number constraint is particularly evident when starting from a Poisson initial distribution. Finite size effects appear sooner than expected based solely on the addition of new sites at the end of the chain and originate from the impossibility of realizing an initial state with a true Poisson distribution in a finite system. Our algorithm is free from this additional finite size effect.

\section{Pure critical properties}

In the next subsections we detail the results obtained in the three different pure cases depending on the sign $D_A-D_B$. We summarize our estimates of the critical exponents for the three universality classes in the Table~\ref{tab:exponents1D} which we compare with the other numerical estimates found in the literature.

\begin{table}[h]
\centering
\setlength\extrarowheight{-3pt}
\begin{tabular}{|c|c|c|c|c|c|c|}
\hline
& $\Theta$ & $\delta=\beta' / \nu_\parallel$ & $\beta / \nu_\bot$ & $\nu_\bot$ & $z$ & Ref. \\ \hline
\multirow{7}{*}{$D_A=D_B$} 
& -&-&$0.197(2)$  & $2.21(5)$ & - & \cite{de_freitas_critical_2000} \\ \cline{2-7} 
& -&-&$0.226(20)$ & - & - & \cite{fulco_critical_2001} \\ \cline{2-7} 
& -&-&$0.192(4)$ & $2.0(2)$ & $2.02(4)$ & \cite{maia_diffusive_2007} \\ \cline{2-7} 
& -&-&-  & $2.04(3)$ & $1.98(4)$ & \cite{filho_critical_2010} \\ \cline{2-7} 
& 0.32(2)&0.087 &$0.175(15)$  & $2.03(24)$ & $2.00(4)$ & \cite{polovnikov_subdiffusive_2022} \\ \cline{2-7} 
&\textbf{ 0.32(1)}&\textbf{ 0.090(5)} & \textbf{0.173(24)}  &  \textbf{2.03(10)} & \textbf{2.02(3)} & - \\ \cline{2-7} 
&-&- &\textit{0.3125} & \textit{2} & \textit{2} & \cite{kree_effects_1989, van_wijland_wilson_1998} \\ \hline

\multirow{6}{*}{$D_A>D_B$} 
& -&-&$0.336(15)$ & - & - & \cite{fulco_critical_2001} \\ \cline{2-7} 
& -&-&$0.404(10)$ & $2.3(3)$ & $2.01(4)$ & \cite{maia_diffusive_2007} \\ \cline{2-7} 
& -&-&-  & $2.02(3)$ & $1.97(4)$ & \cite{filho_critical_2010} \\ \cline{2-7} 
& -&-&$0.34(4)$  & $2.00(16)$ & $2.00(4)$ & \cite{da_silva_critical_2013} \\ \cline{2-7} 
& -0.38(4) & 0.66(3) & - & $1.3(2)$ & $3.0(1)$ & \cite{polovnikov_subdiffusive_2022} \\ \cline{2-7} 
& \textbf{-0.32(2)} & \textbf{0.58(2)} & \textbf{0.30(8)}  & \textbf{1.41(9)} & \textbf{2.70(7)} & - \\ \cline{2-7} 
& \multicolumn{5}{c|}{\textit{first order}} & \cite{van_wijland_wilson_1998} \\ \hline

\multirow{5}{*}{$D_A<D_B$} 
& -&-&$0.165(22)$ & - & - & \cite{fulco_critical_2001} \\ \cline{2-7} 
& -&-&$0.113(8)$ & $1.77(3)$ & $1.6(2)$ & \cite{maia_diffusive_2007} \\ \cline{2-7} 
& -&-&- &  $2.00(16)$ & $1.99(16)$ & \cite{filho_critical_2010} \\ \cline{2-7} 
& \textbf{0.44(1)} & \textbf{0.11(1)} & \textbf{0.10(3)}  &\textbf{ 2.01(12)} & \textbf{1.64(2)} & - \\ \cline{2-7} 
& -&-&\textit{1 / 2} & \textit{2} & \textit{2} & \cite{van_wijland_wilson_1998} \\ \hline
\end{tabular}
\caption{Critical exponents of the pure unidimensionnal DEP from numerical simulations (in bold are the results of this paper) and field theory (italic). Our estimates of $\beta/\nu_{\perp}$ comes from the hyperscaling relation $\Theta z = d-\beta/\nu_{\perp} - z\delta$ with $d$ the dimensionality of the system. }
\label{tab:exponents1D}
\end{table}

\subsection{$D_A=D_B$}

\begin{figure}[h]
    \centering
    \includegraphics[width=0.34\textwidth]{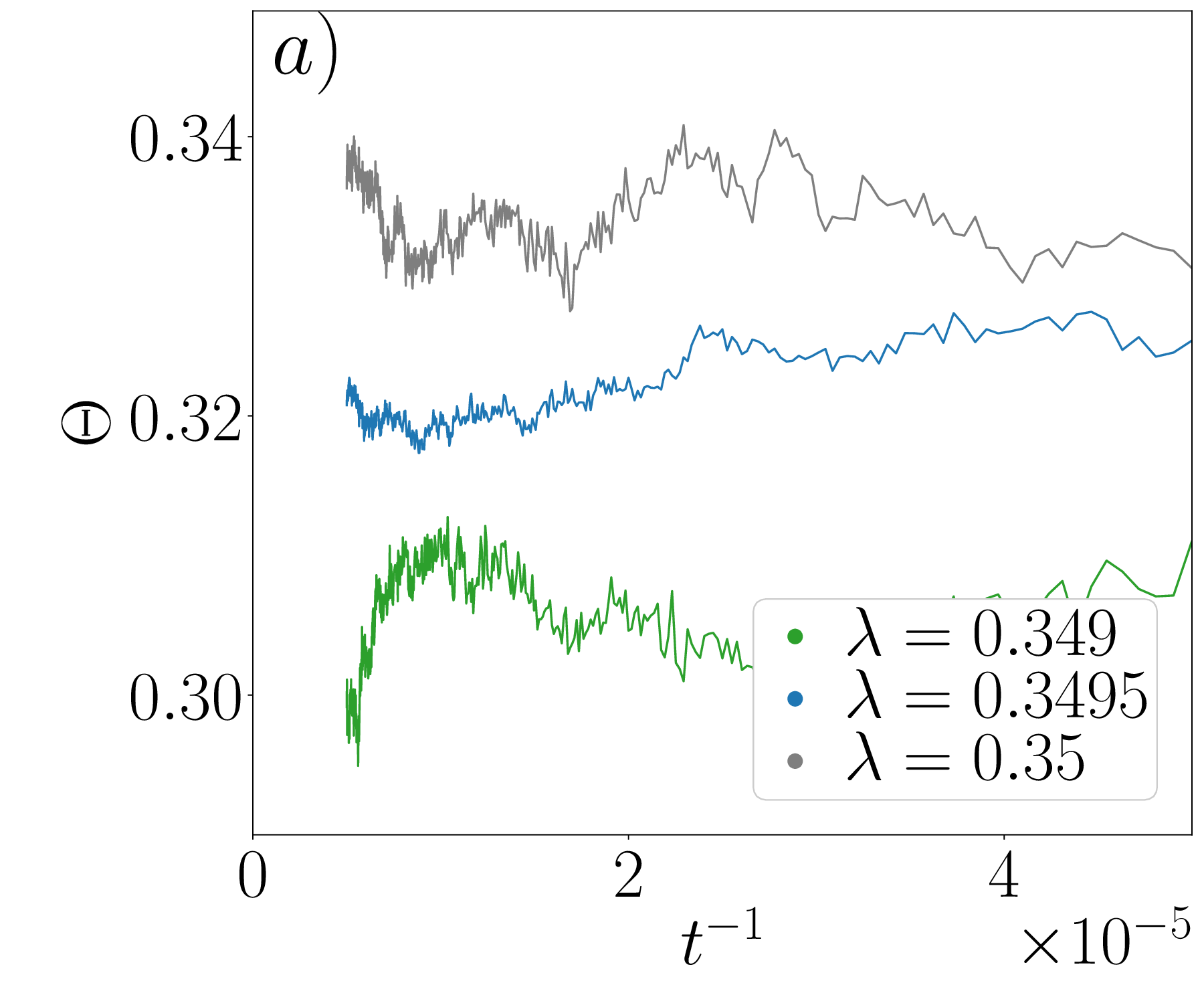}
    \includegraphics[width=0.34\textwidth]{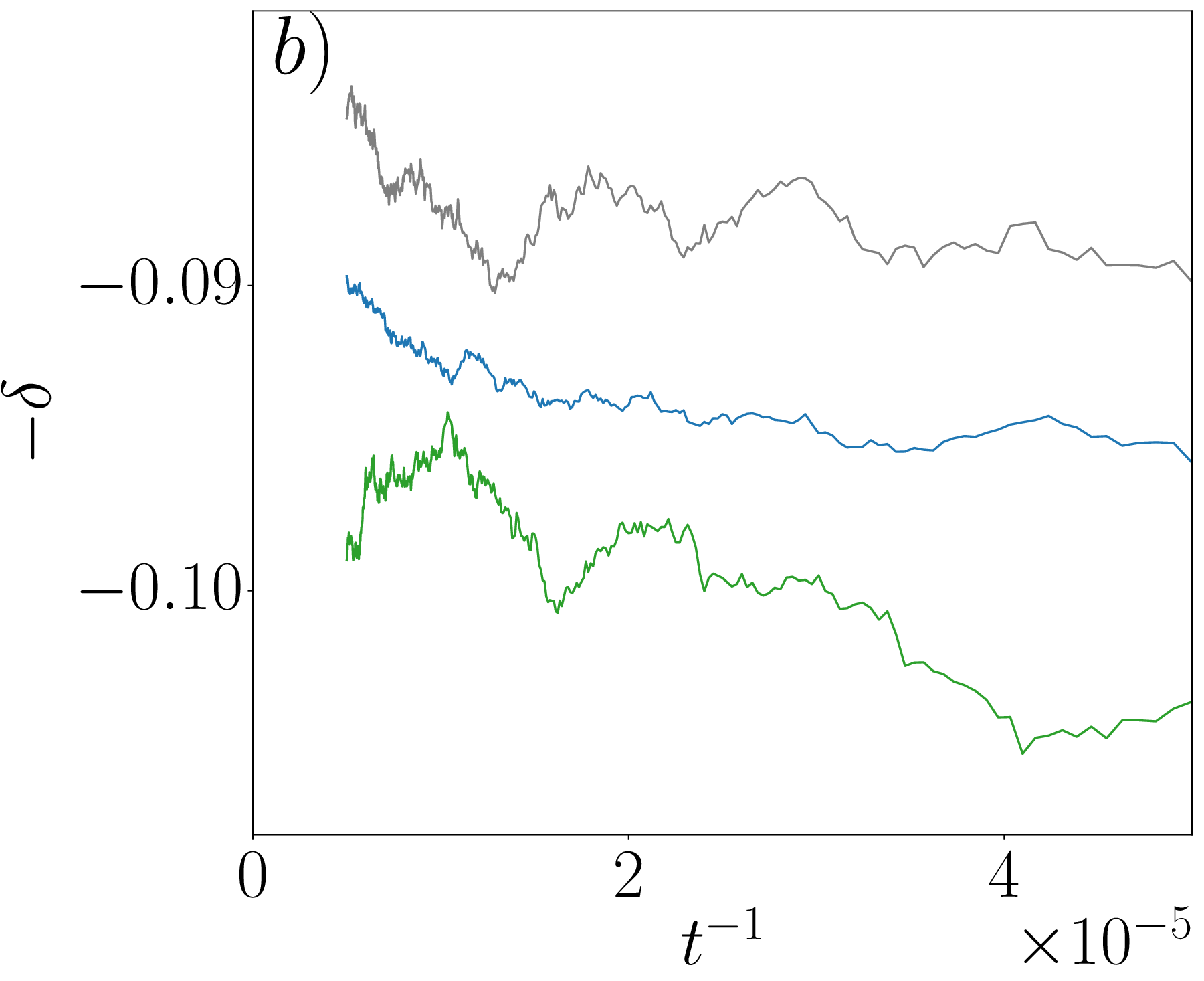}
    \includegraphics[width=0.34\textwidth]{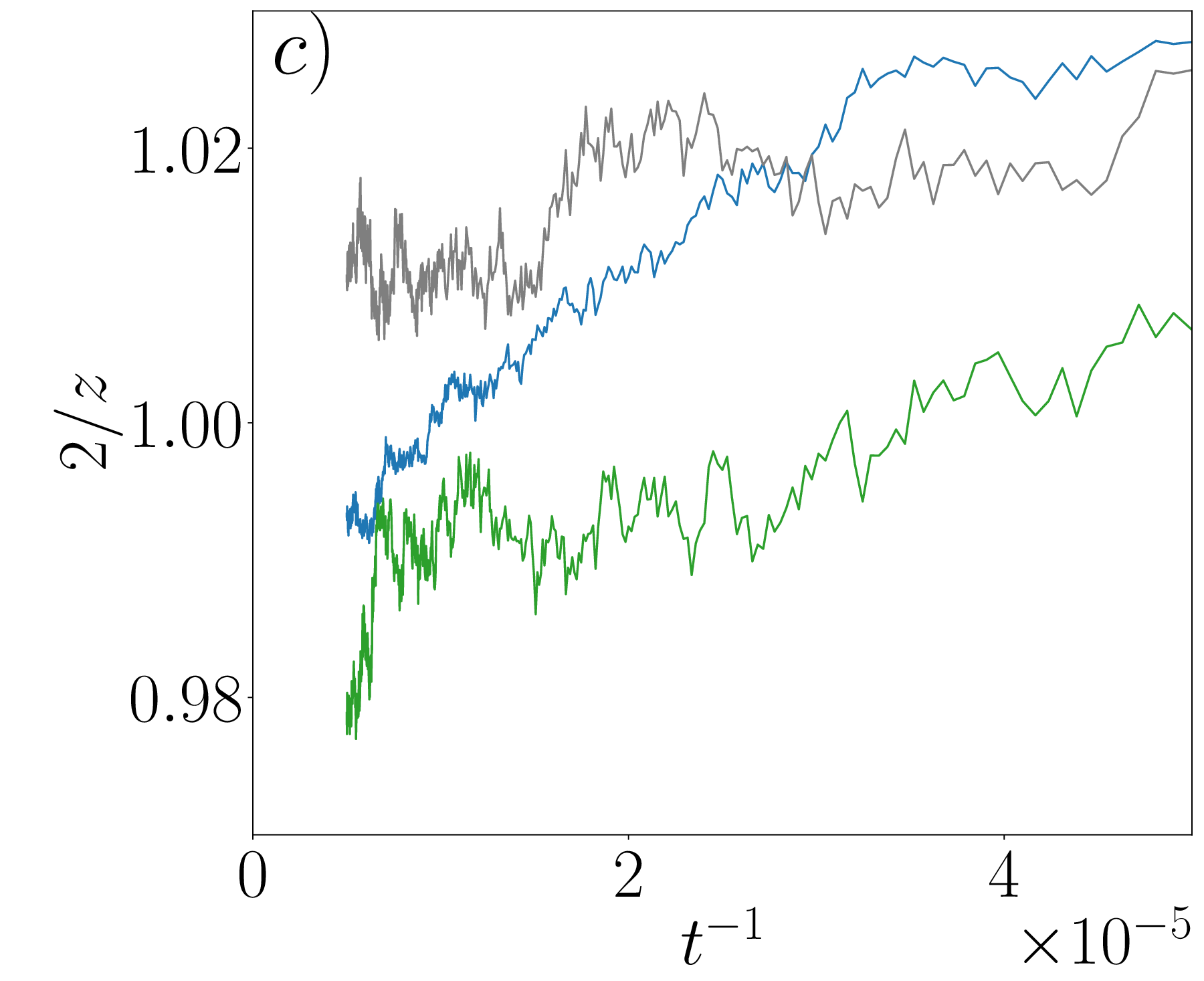}
    \includegraphics[width=0.34\textwidth]{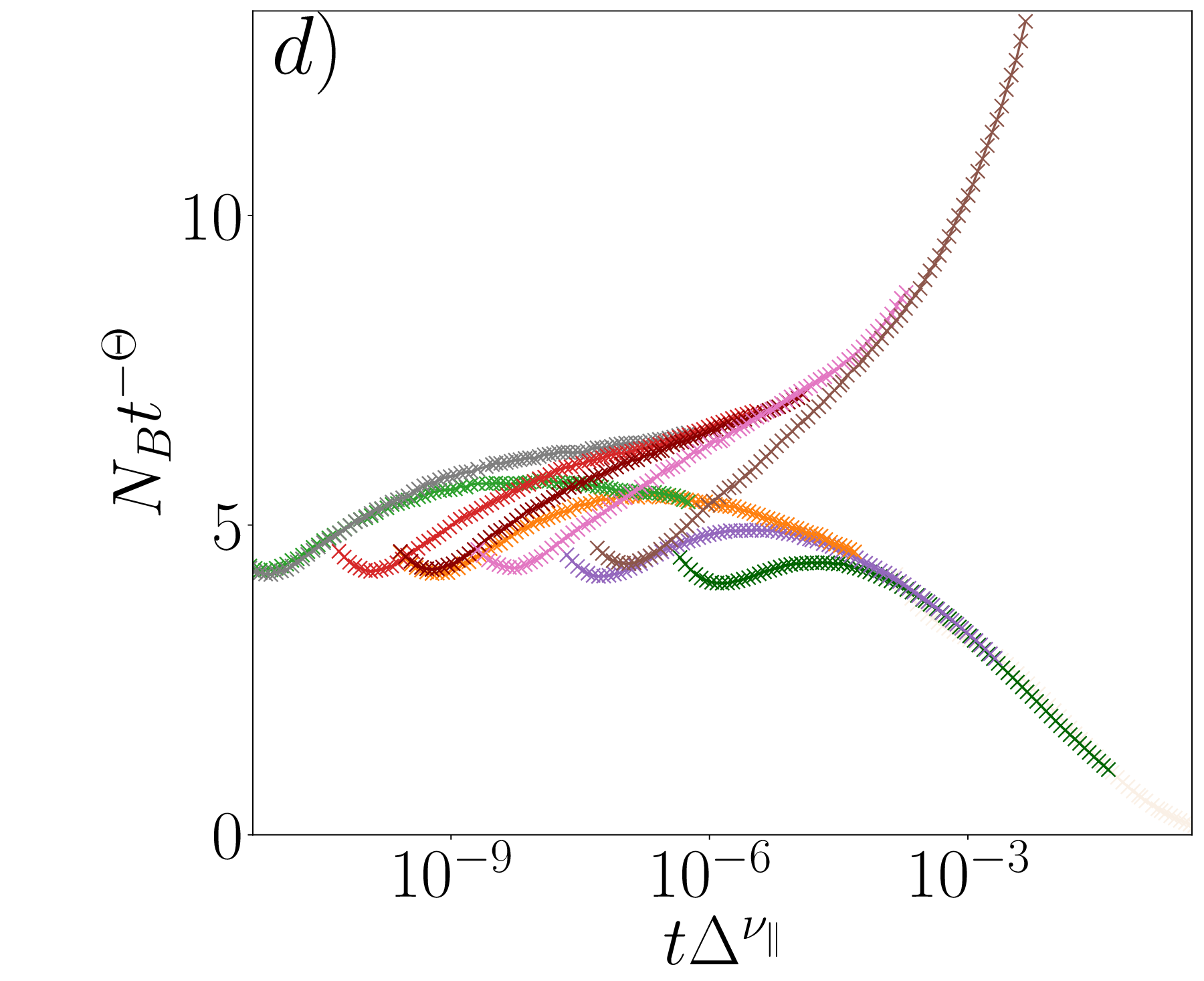}
    \caption{Critical properties of the pure DEP with \(D_A = D_B=1\), starting from an initial state where the number of particles per site follows a Poisson distribution with mean \(\rho=5\). 
    (a-c) Effective critical exponents yielding the estimates \(\lambda_c = 0.3495(5)\), \(\Theta = 0.32(1)\), \(\delta = 0.090(5)\), and \(2/z = 0.992(10)\). A total of \(4 \times 10^4\) samples were used for \(\lambda = 0.349\) and \(0.35\), and \(2 \times 10^5\) samples for \(\lambda = 0.3495\).  
    (d) Scaling analysis using \(\Theta = 0.32\) and \(\lambda_c = 0.3495\) to estimate \(\nu_{\parallel} = 4.1(2)\). Contamination rates vary between \(\lambda = 0.31\) and \(0.355\).}
    \label{fig:PureDa=Db}
\end{figure}

Following the methodology described in Sec.~\ref{subsection:estimateCriticalProperties}, we estimate \( \lambda_c = 0.3495 \), \( \Theta = 0.32(1) \), \( \delta = 0.090(5) \), and \( \nu_{\parallel} = 4.1(2) \) from Fig.~\ref{fig:PureDa=Db}, using data up to \( T_{\max} = 2 \times 10^5 \)~s near the critical point. Applying the hyperscaling relation \( \Theta = d/z - \alpha - \delta \), we find \( \alpha = 0.086(12) \). The exponents \( \alpha \) and \( \delta \) are consistent within error bars, as expected from time-reversal symmetry. These results are summarized in Table~\ref{tab:exponents1D} and are in agreement with the most recent numerical studies.

\begin{figure}[h]
    \centering
    \includegraphics[width=0.34\textwidth]{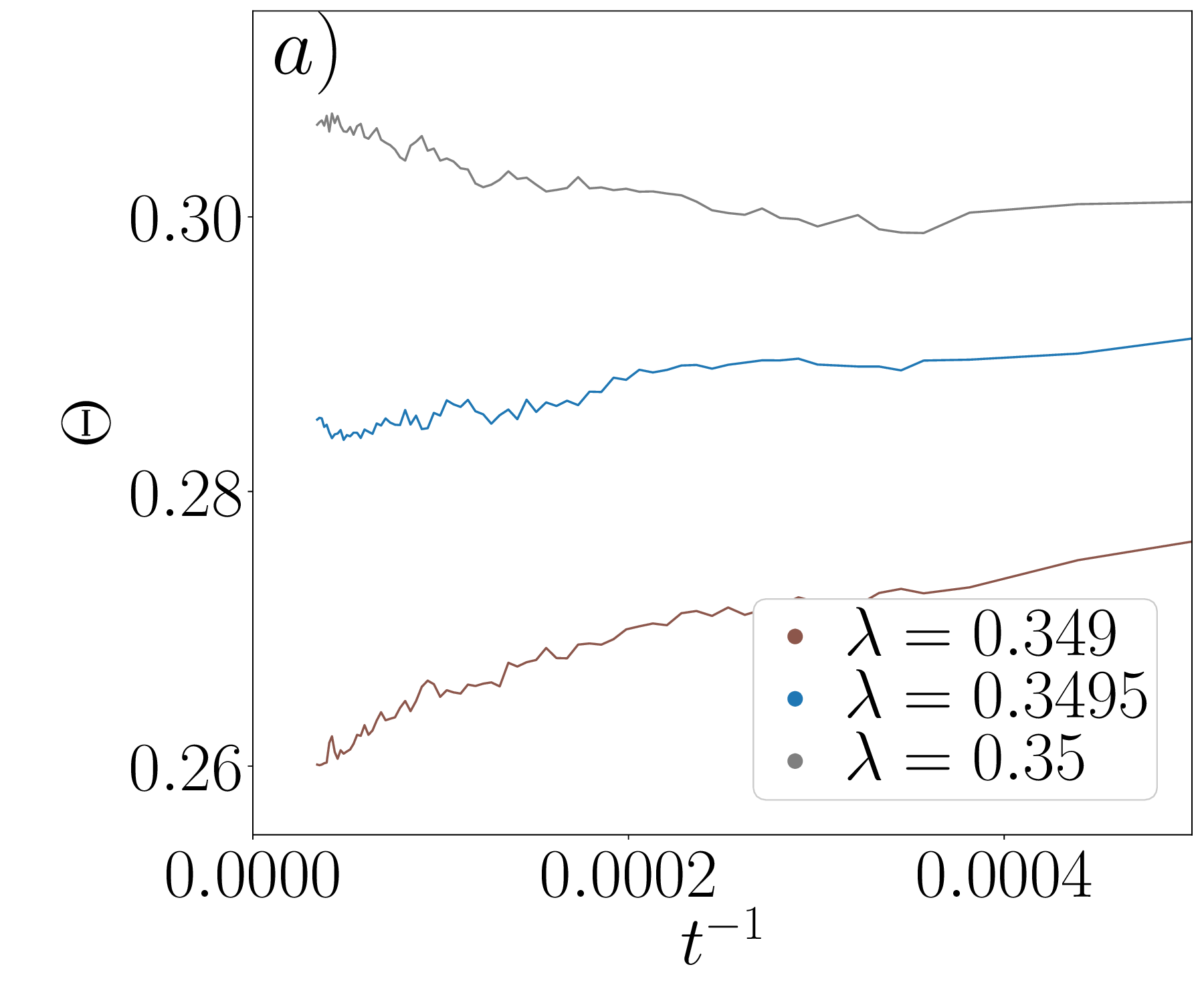}
    \includegraphics[width=0.34\textwidth]{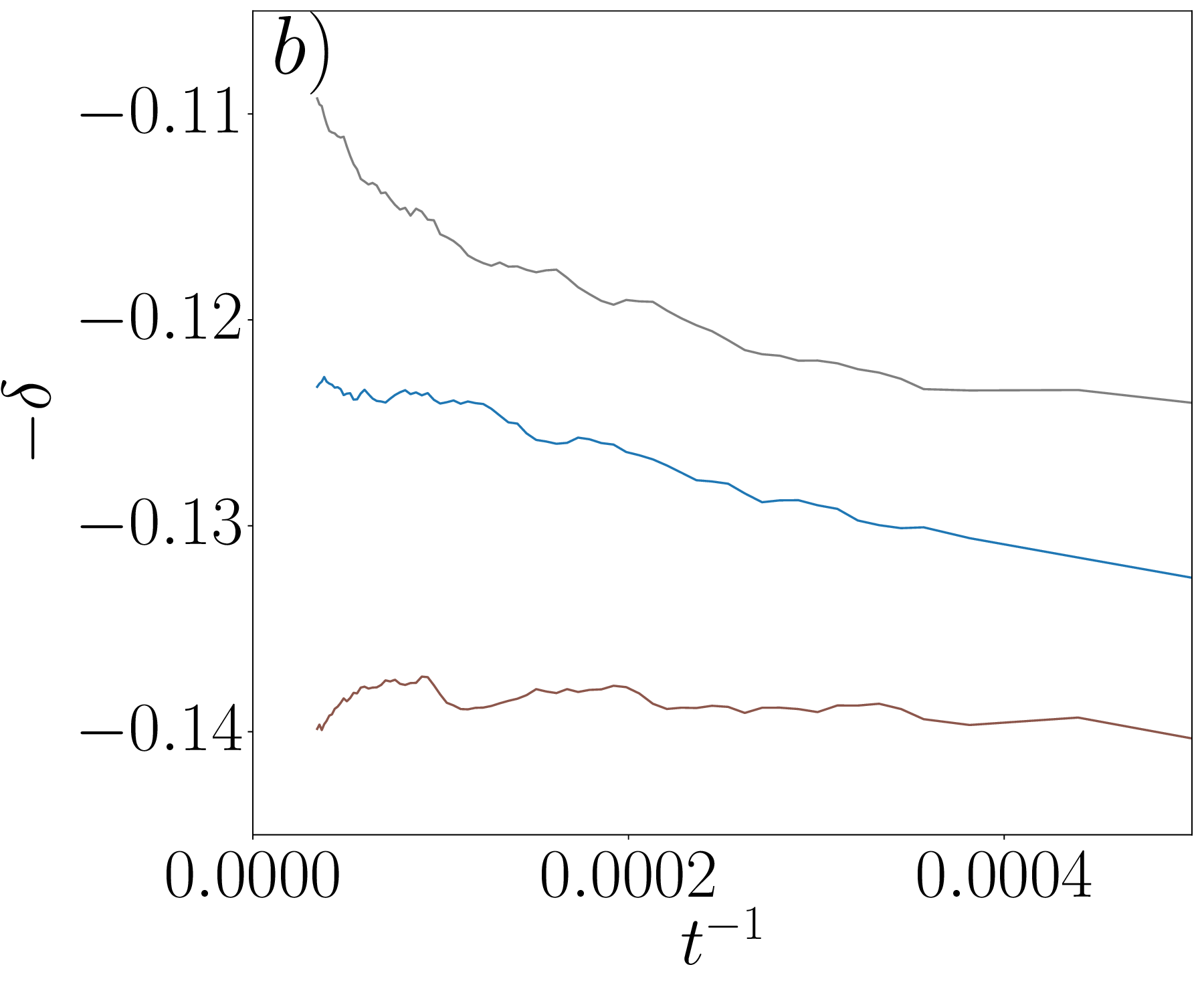}
    \includegraphics[width=0.34\textwidth]{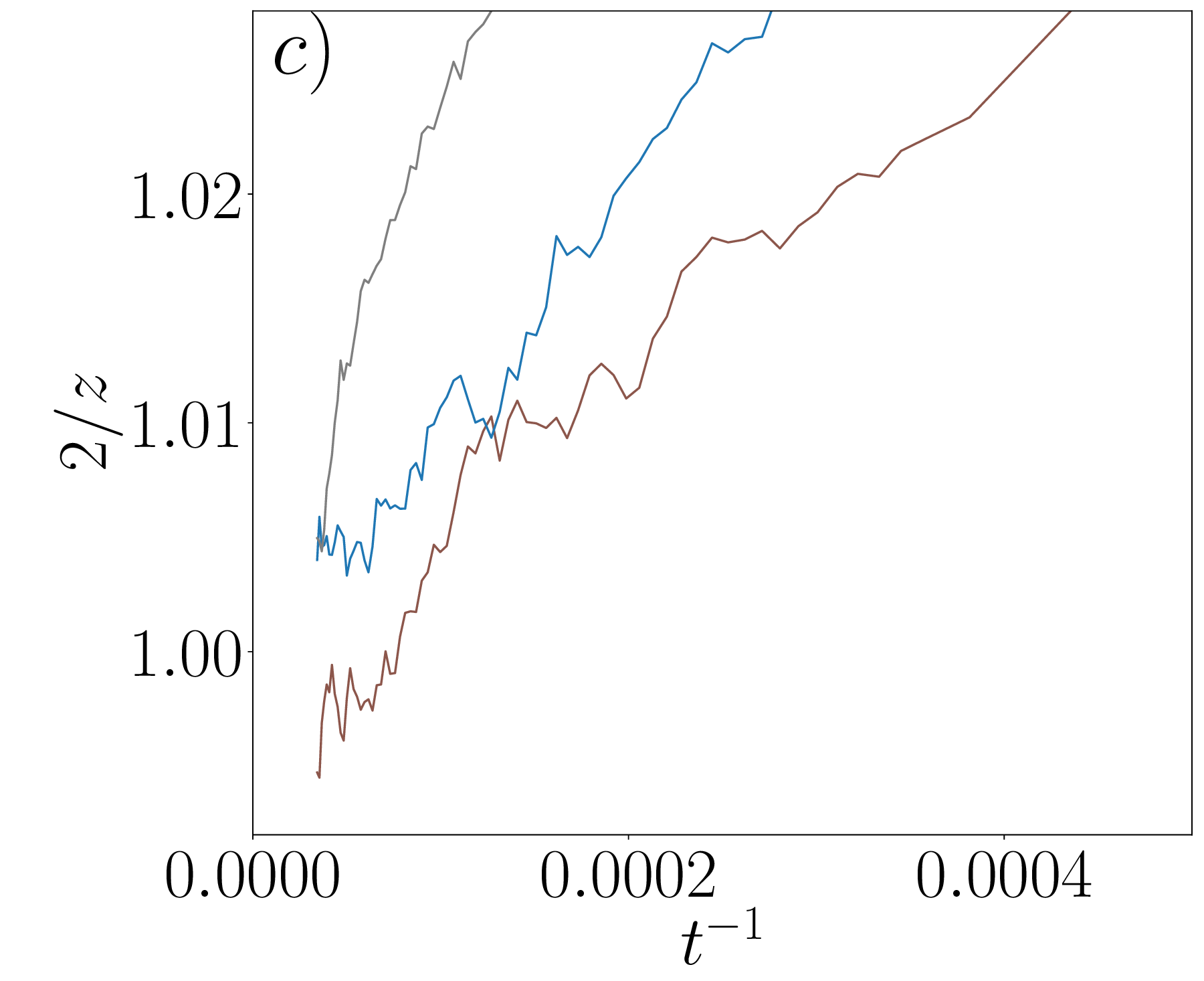}
    \includegraphics[width=0.34\textwidth]{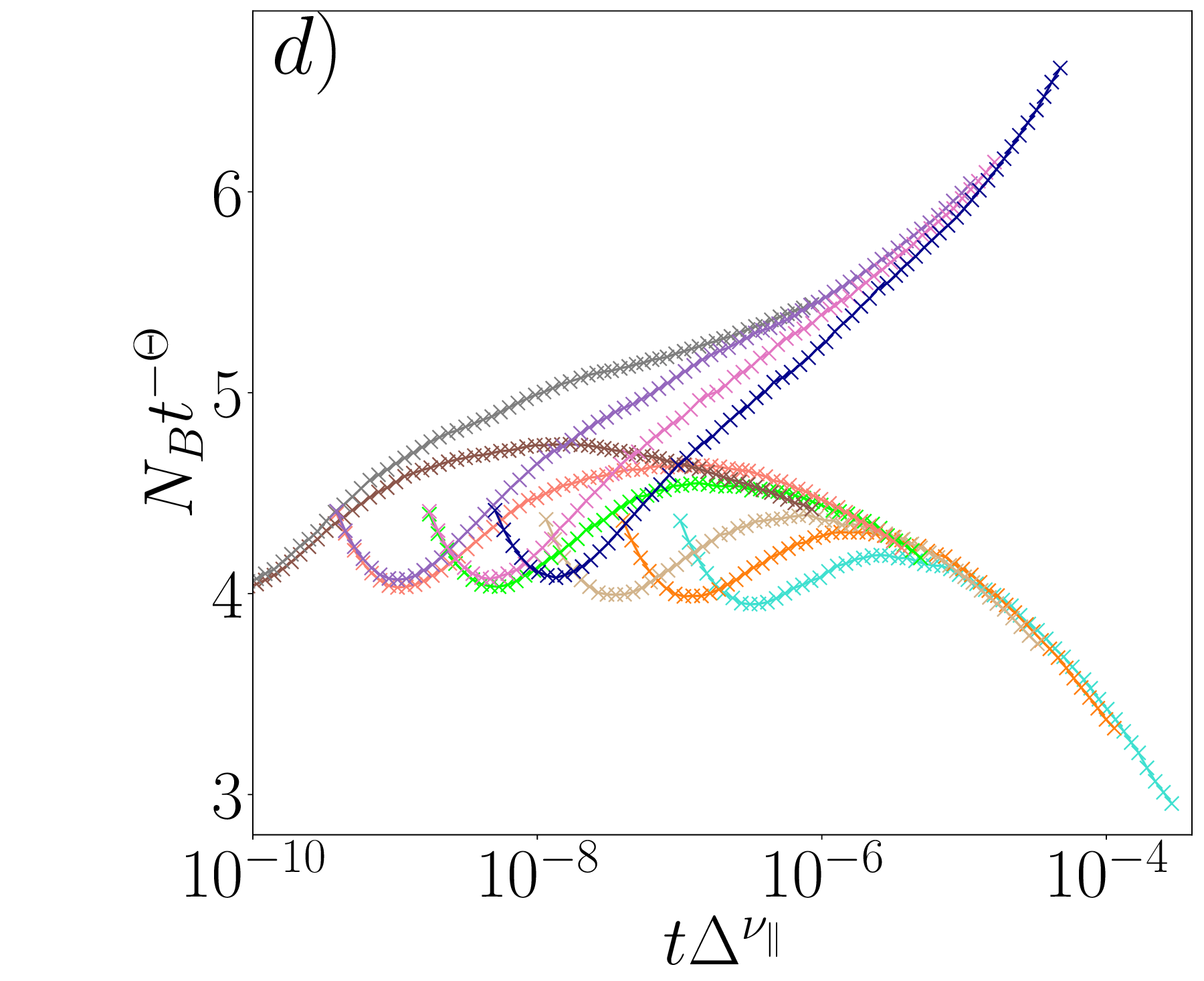}
    \caption{Critical properties of the pure DEP with \(D_A = D_B=1\), starting from an initial state with exactly $\rho=5$ particles per site. (a-c) Effective critical exponents yielding the estimates $\lambda_c=0.3495(5), \Theta=0.285(10), \delta= 0.122(10), 2/z=1.005(10)$. $5.10^5$ samples were used for $\lambda=0.349, 0.3495, 0.35$. (d) Scaling analysis using $\Theta=0.285$ and $\lambda_c=0.3495$ to estimate $\nu_{\parallel} = 3.7(4)$. Contamination rates vary between $\lambda=0.34$ to $0.352$.}
    \label{fig:PureDa=Db_homogeneousInitial}
\end{figure}

We also investigate the influence of initial conditions by starting with a lattice uniformly filled with exactly \( \rho = 5 \) particles per site, as shown in Fig.~\ref{fig:PureDa=Db_homogeneousInitial}. Simulations up to \( T_{\max} = 6 \times 10^4 \)~s yield \( \Theta_h = 0.285(10) \), \( \delta_h = 0.122(10) \), and \( 2/z_h = 1.005(10) \), leading to \( \alpha_h = 0.095(15) \). While the ratios \( \alpha/\alpha_h \) and \( z/z_h \) are consistent within error bars, \( \delta/\delta_h \) and \( \Theta/\Theta_h \) are not. This aligns with theoretical expectations, as time-reversal symmetry holds only for Poissonian initial distributions \cite{van_wijland_wilson_1998}. Moreover, a constant initial distribution is expected to yield a lower value of \( \Theta \) compared to a Poissonian distribution, which is consistent with our findings. Since this effect is relatively small, finite-size effects may explain why time-reversal symmetry was still observed in simulations starting from a homogeneously filled lattice in \cite{polovnikov_subdiffusive_2022}.

\subsection{$D_A<D_B$}

\begin{figure}[h]
    \centering
        \includegraphics[width=0.34\textwidth]{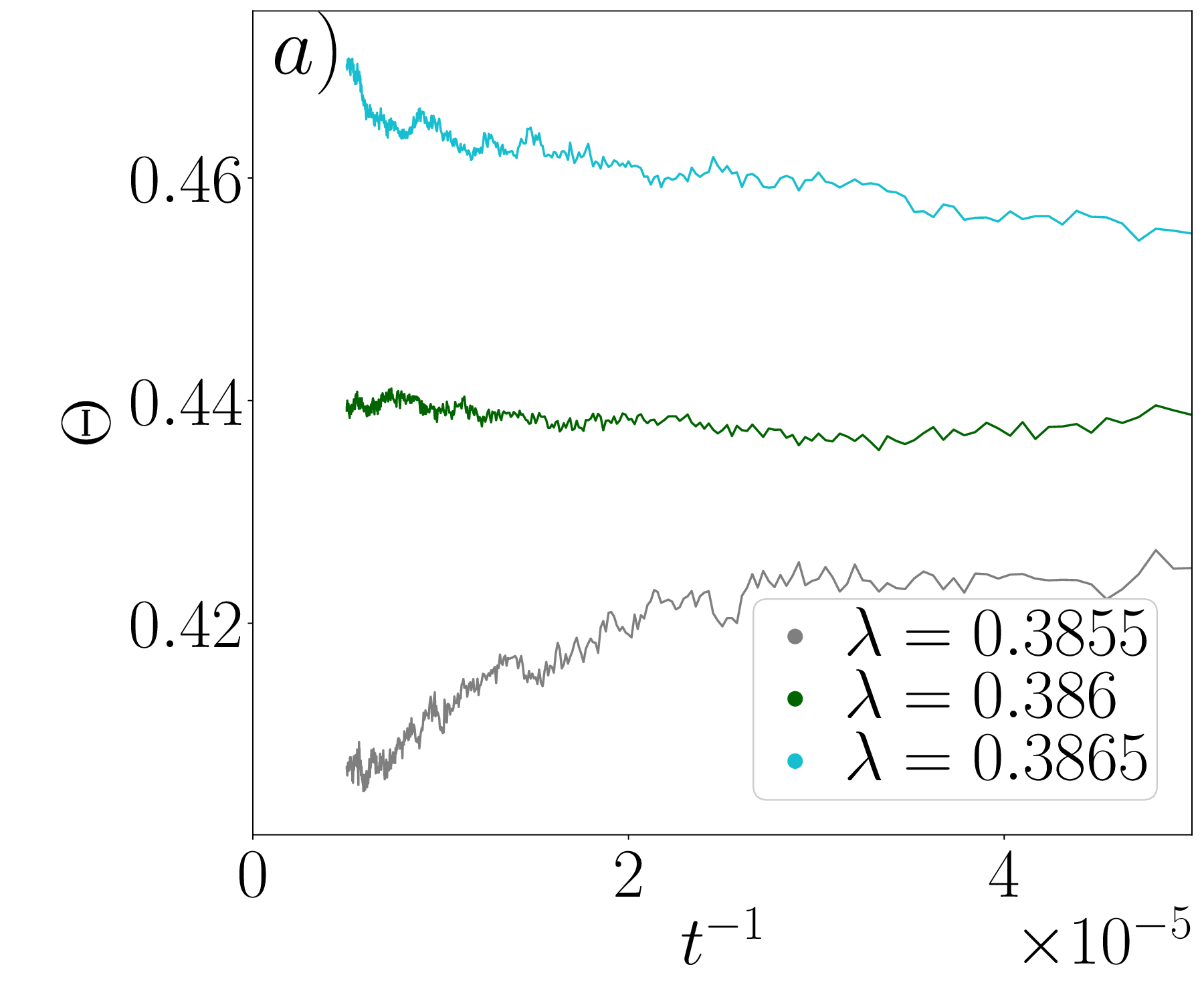}
        \includegraphics[width=0.34\textwidth]{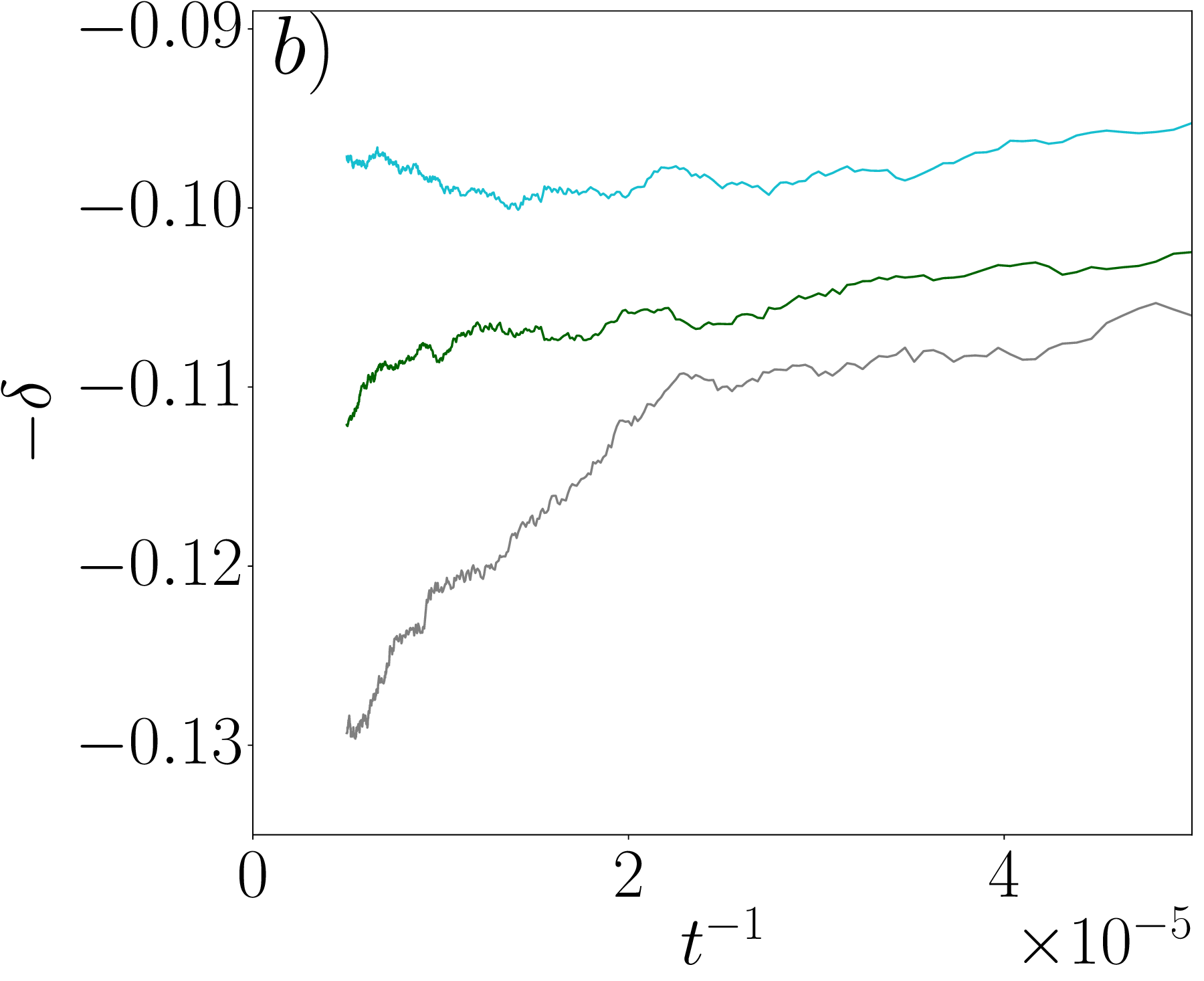}
        \includegraphics[width=0.34\textwidth]{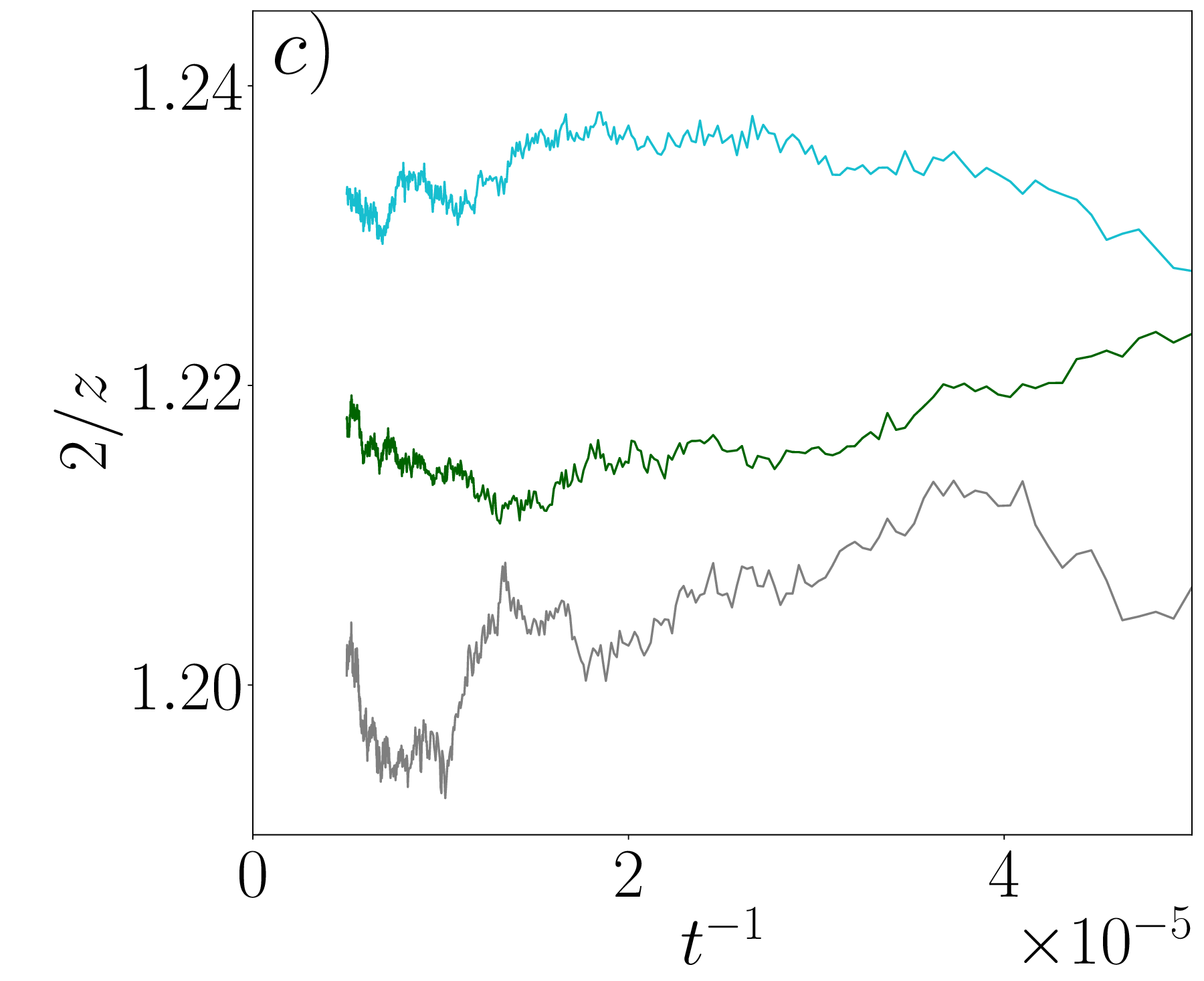}
        \includegraphics[width=0.34\textwidth]{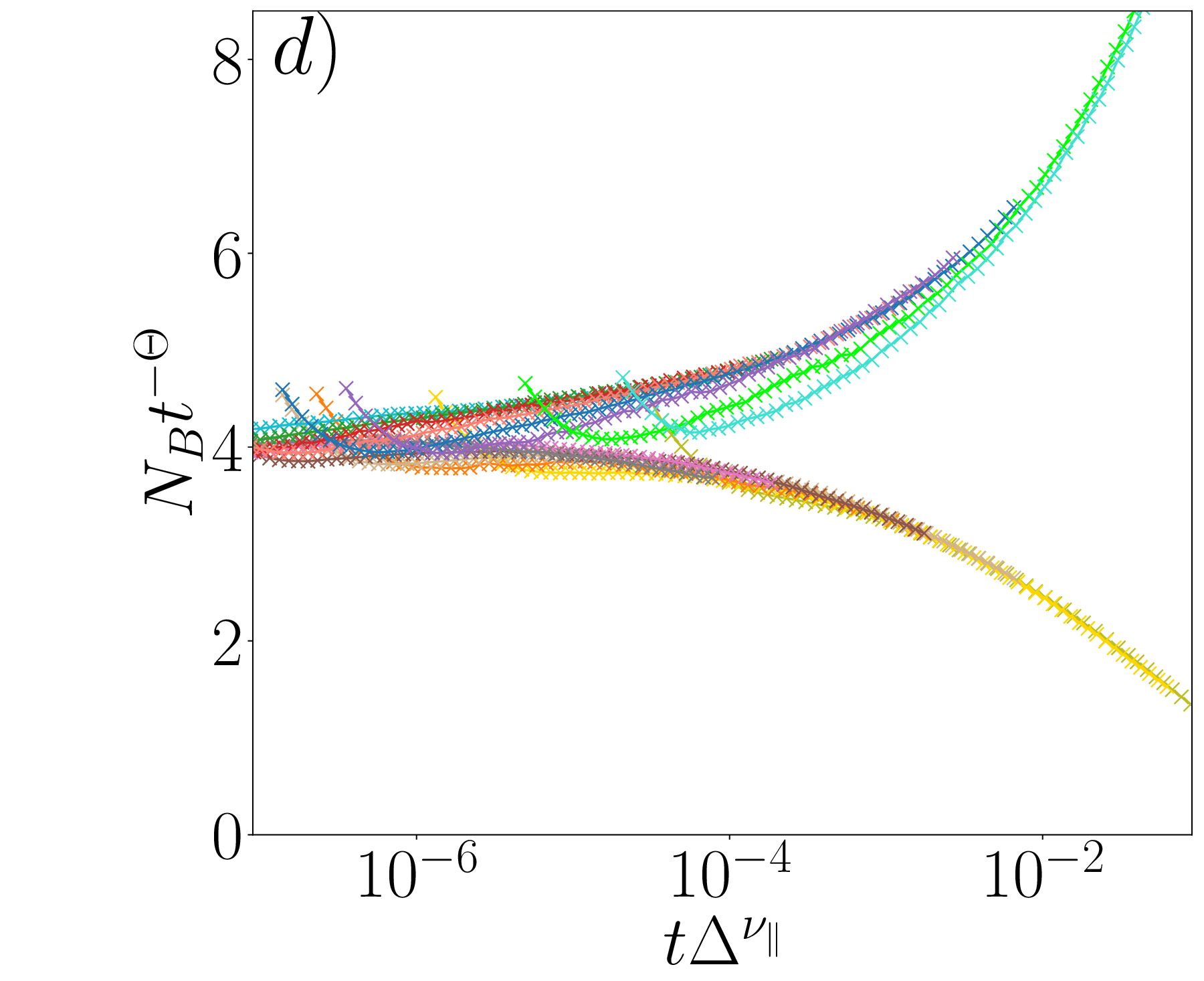}
    \caption{Critical properties of the pure DEP with \(0.375=D_A < D_B=1\). 
    (a-c) Effective critical exponents yielding the estimates \(\lambda_c = 0.3860(5)\), \(\Theta = 0.44(1)\), \(\delta = 0.11(1)\), and \(2/z = 1.22(2)\). A total of \(\sim 10^5\) samples were used for \(\lambda = 0.3855, 0.386, 0.3865\).  
    (d) Scaling analysis using \(\Theta = 0.44\) and \(\lambda_c = 0.386\) to estimate \(\nu_{\parallel} = 3.25(20)\). Contamination rates vary between \(\lambda = 0.37\) and \(0.40\).}
    \label{fig:PureDalDb}
\end{figure}

We apply the same procedure to determine the critical exponents as detailed in the previous section. Estimates of the critical exponents in Table~\ref{tab:exponents1D} are obtained from Fig.~\ref{fig:PureDalDb}. Specifically, our estimate of \(\nu_{\perp}\) agrees with \cite{filho_critical_2010} but not with \cite{maia_diffusive_2007}, while the opposite is true for \(z\). This discrepancy is significant, as spatial quenched disorder is considered relevant if \(\nu_{\perp} < 2\) at \(d = 1\). Our results suggest that such disorder is marginal, in line with \cite{filho_critical_2010}, rather than relevant as suggested by \cite{maia_diffusive_2007}. Evidence supporting the irrelevance of this disorder will be discussed in Section \ref{section:contaminationDisorder}.

\subsection{$D_A>D_B$}

\begin{figure}[h]
    \centering
        \includegraphics[width=0.34\textwidth]{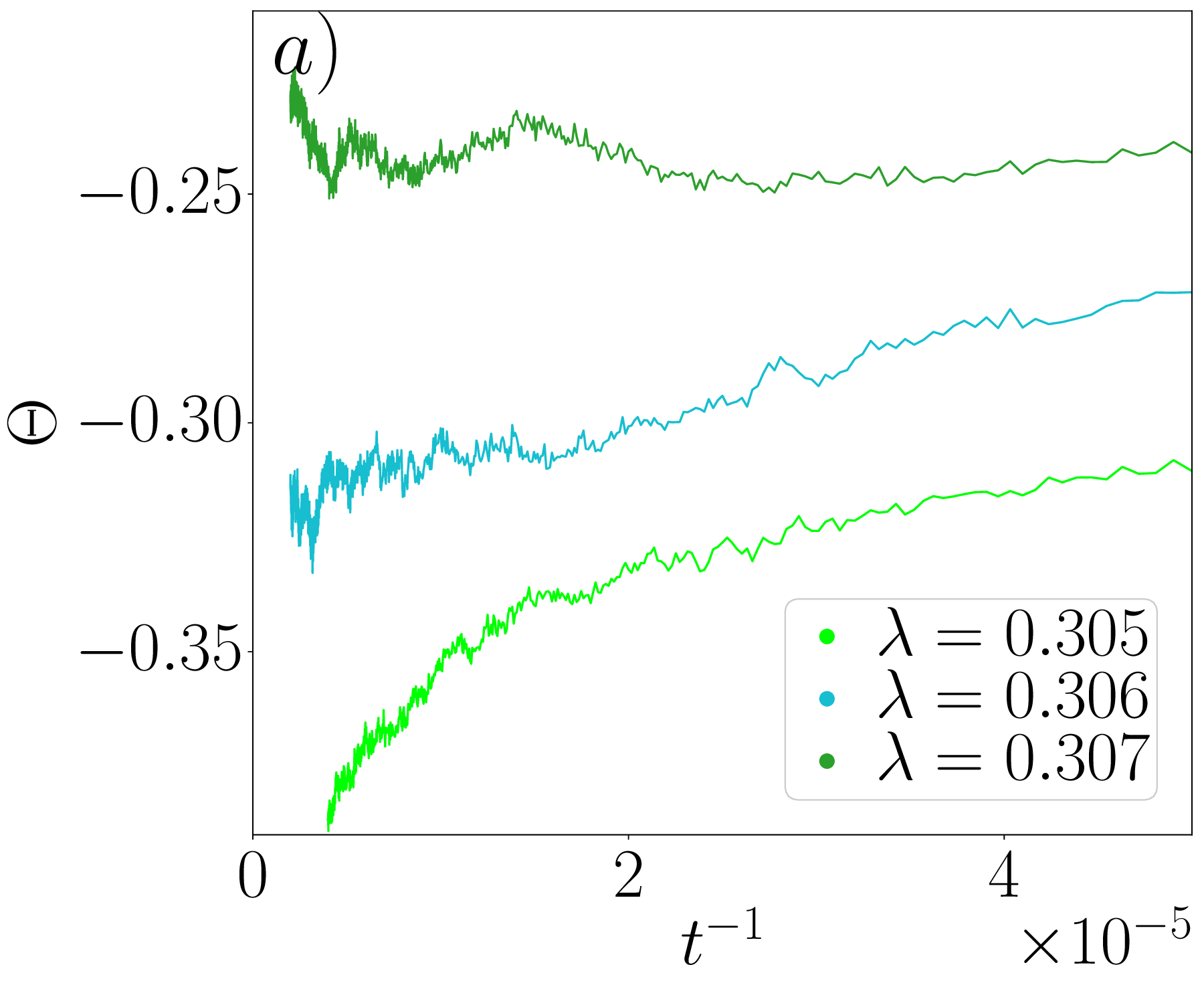}
        \centering
        \includegraphics[width=0.34\textwidth]{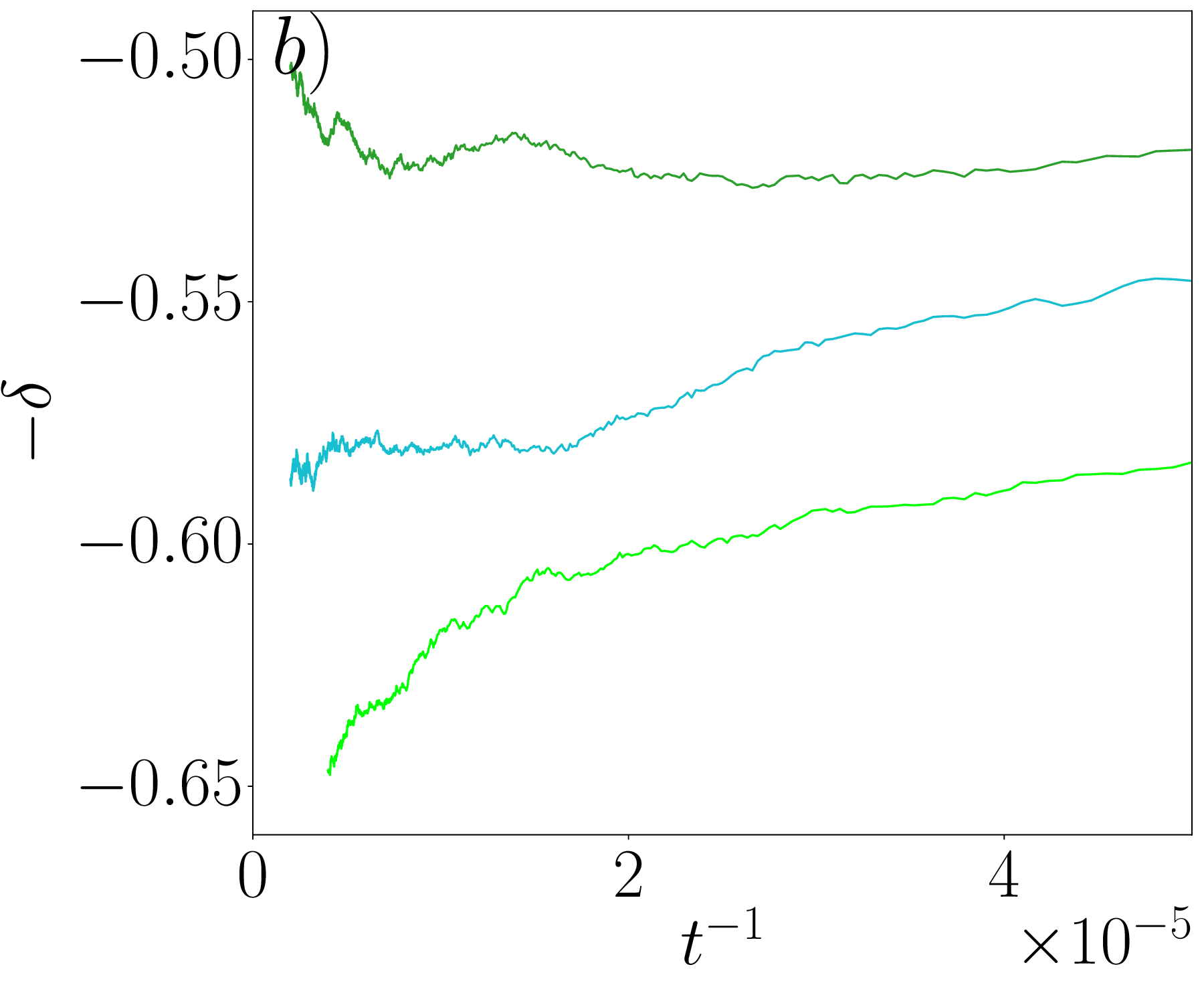}
        \centering
        \includegraphics[width=0.34\textwidth]{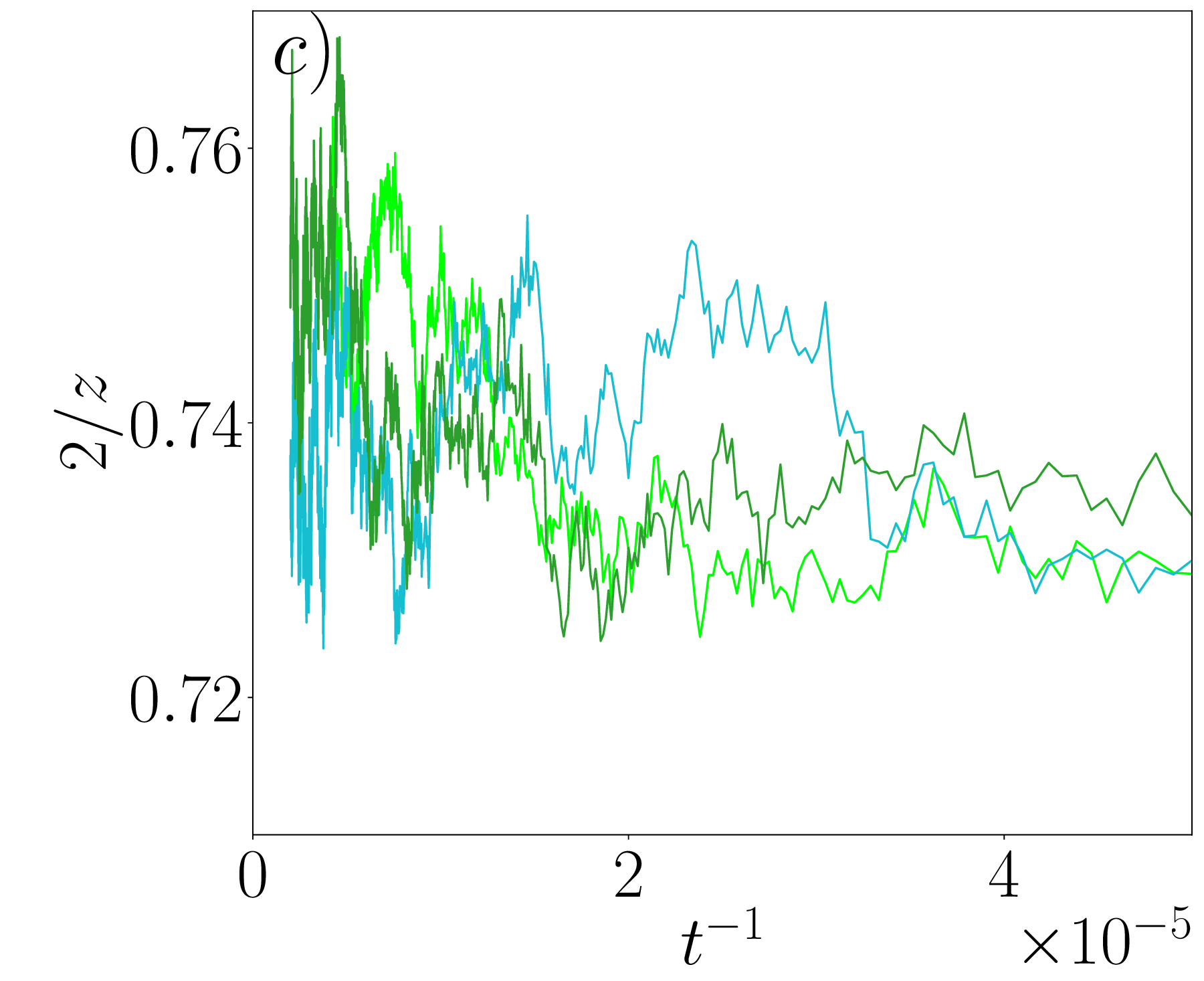}
        \centering
        \includegraphics[width=0.34\textwidth]{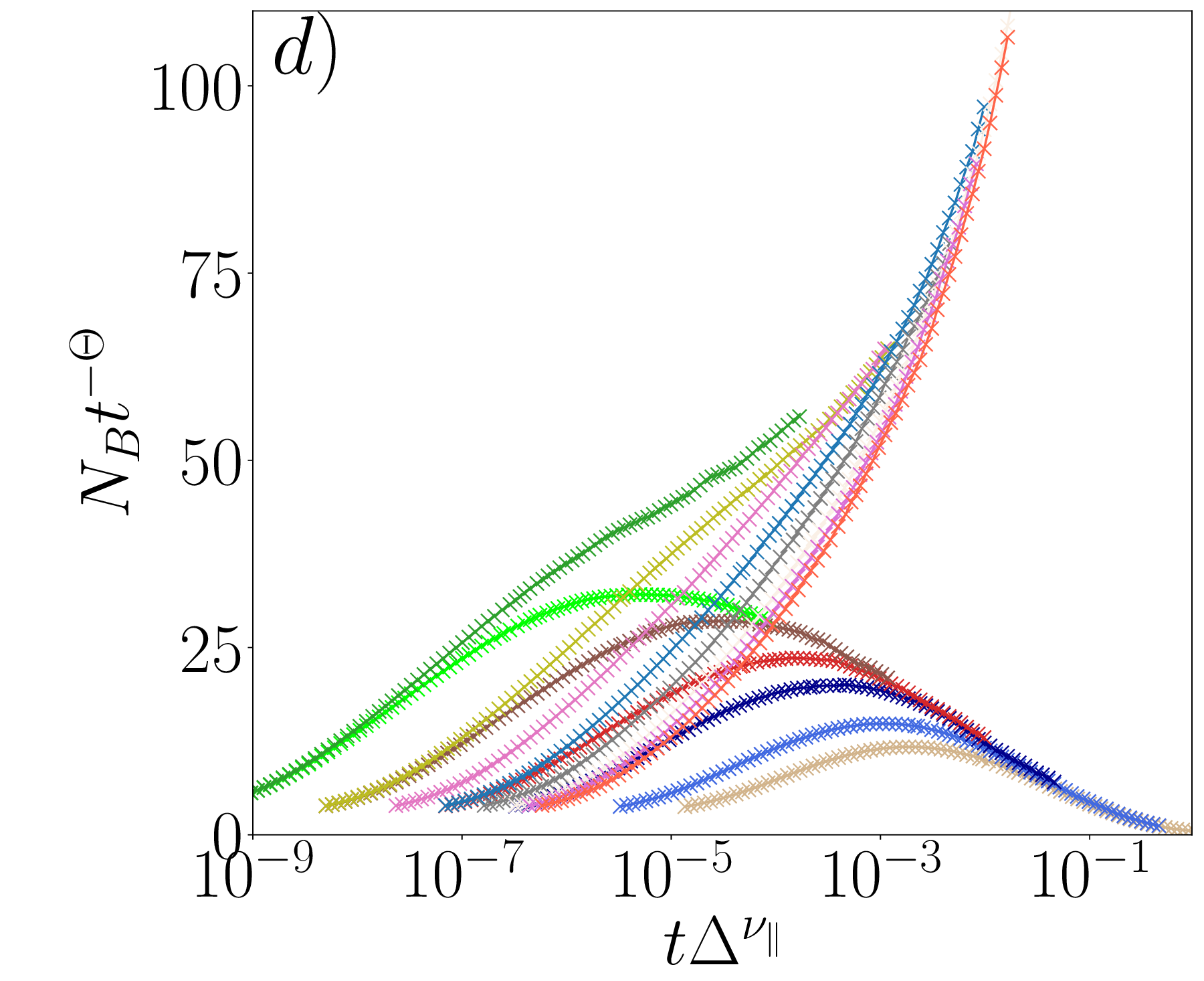}
    \caption{Critical properties of the pure DEP with \(1=D_A > D_B=0.375\). 
    (a-c) Effective critical exponents yielding the estimates \(\lambda_c = 0.3060(5)\), \(\Theta = -0.32(2)\), \(\delta = 0.58(2)\), and \(2/z = 0.74(2)\). A total of \(\sim 10^6\) samples were used for \(\lambda = 0.305, 0.306, 0.307\).
    (d) Scaling analysis using \(\Theta = -0.32\) and \(\lambda_c = 0.306\) to estimate \(\nu_{\parallel} = 3.8(2)\). Contamination rates vary between \(\lambda = 0.29\) and \(0.315\).}
    \label{fig:PureDblDa}
\end{figure}

In the case of \( D_A > D_B \), no evidence of a discontinuous transition is observed by contrast to the theoretical prediction \cite{van_wijland_wilson_1998}, but in agreement with other numerical studies (see Tab.~\ref{tab:exponents1D}). Exponents are extracted from Fig.~\ref{fig:PureDblDa}. We confirm the subdiffusive behavior reported in \cite{polovnikov_subdiffusive_2022}, while refining the estimation of the dynamical exponent. Our estimation is incompatible with theirs, which can be attributed to finite-size effects, absent in our new algorithm. Importantly, we also find that the correlation length exponent is less than 2.

\begin{figure}
    \centering
    \includegraphics[width=0.31\linewidth]{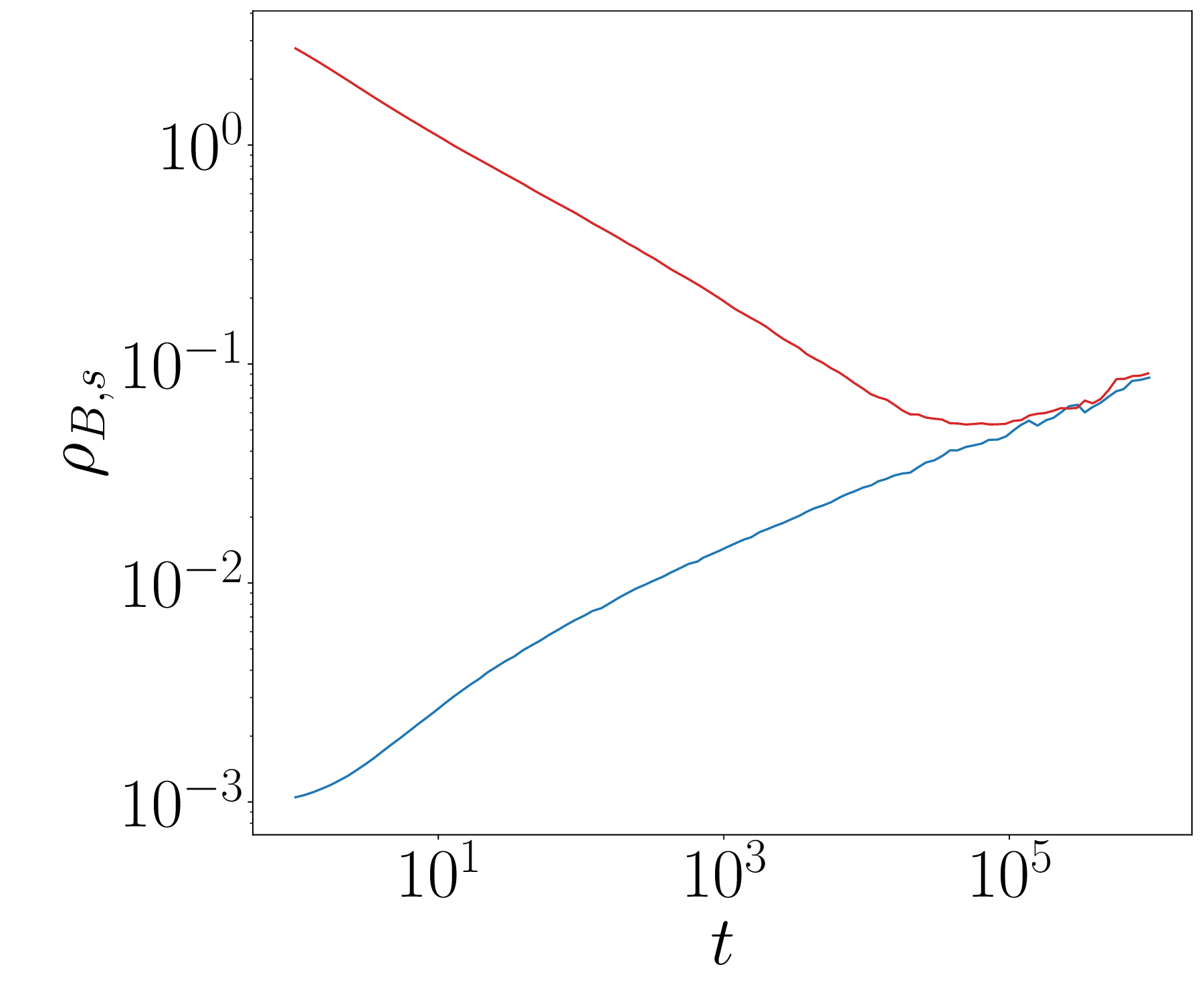}
    \caption{Evolution of the density of \(B\) particles, averaged over the surviving samples (\(\rho_{B,s}\)), starting from a fully infected lattice (red) or a single seed (blue) at the estimated critical point with \(L = 4096\), \(D_A = 1\), and \(D_B = 0.375\). Both curves converge to the same values at late times, indicating the absence of bistability. Note that the maximum simulation time is insufficient to reach the stationary state.}
    \label{fig:DalDb_NoFirstOrder}
\end{figure}

We further verify the absence of bistability at the critical point, as expected for a continuous phase transition, in Fig.~\ref{fig:DalDb_NoFirstOrder}. For a fixed system size of \( L = 4096 \), the evolution of the density of \( B \) particles, whether starting with only infected particles or a seed of infected particles, becomes similar after some time. Therefore, the stationary value of the density of \( B \) particles is unaffected by the initial conditions. This confirms the findings of \cite{dickman_nature_2008} for higher densities and larger system sizes.

\section{Contamination rate disorder}
\label{section:contaminationDisorder}

We present here the critical properties of DEP with a spatially dependent contamination rate, following the distribution described in Eq.~(2) of the main text, which we reproduce here for completeness:  
\begin{equation}
    \mathcal{P}(\lambda_i) = (1 - p_{\lambda}) \delta(\lambda_i - \lambda) + p_{\lambda} \delta(\lambda_i - c\lambda).
    \label{eq:DisorderDistributionLambda}
\end{equation}
For the numerical simulations, we set \( p_{\lambda} = 0.3 \) and \( c = 0.2 \), unless stated otherwise. To distinguish the dirty exponents from the pure exponents, we use the superscript \( \lambda \) on the critical exponents.  

\subsection{$D_A = D_B$}

\begin{figure}
    \centering
        \includegraphics[width=0.34\textwidth]{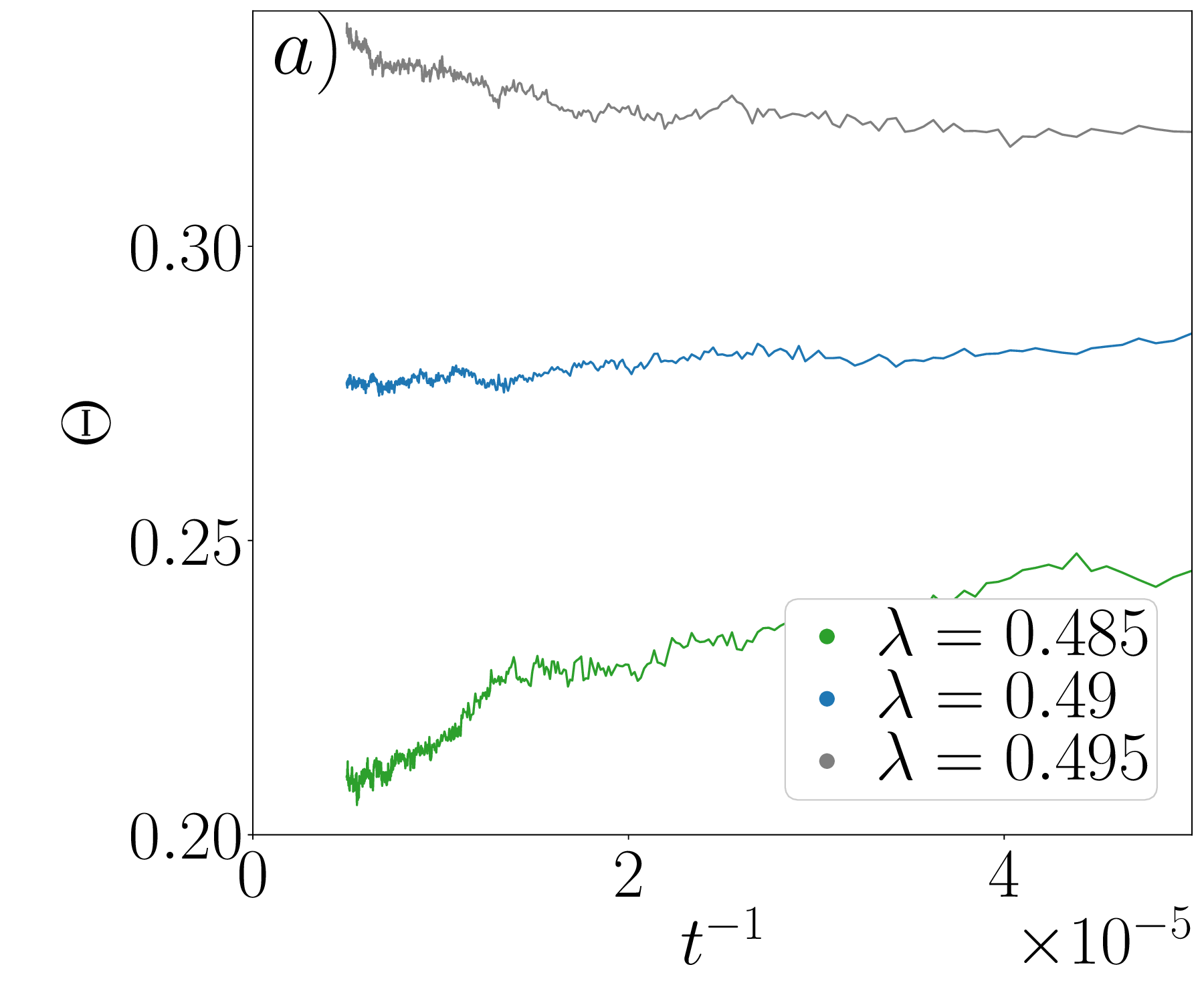}
        \centering
        \includegraphics[width=0.34\textwidth]{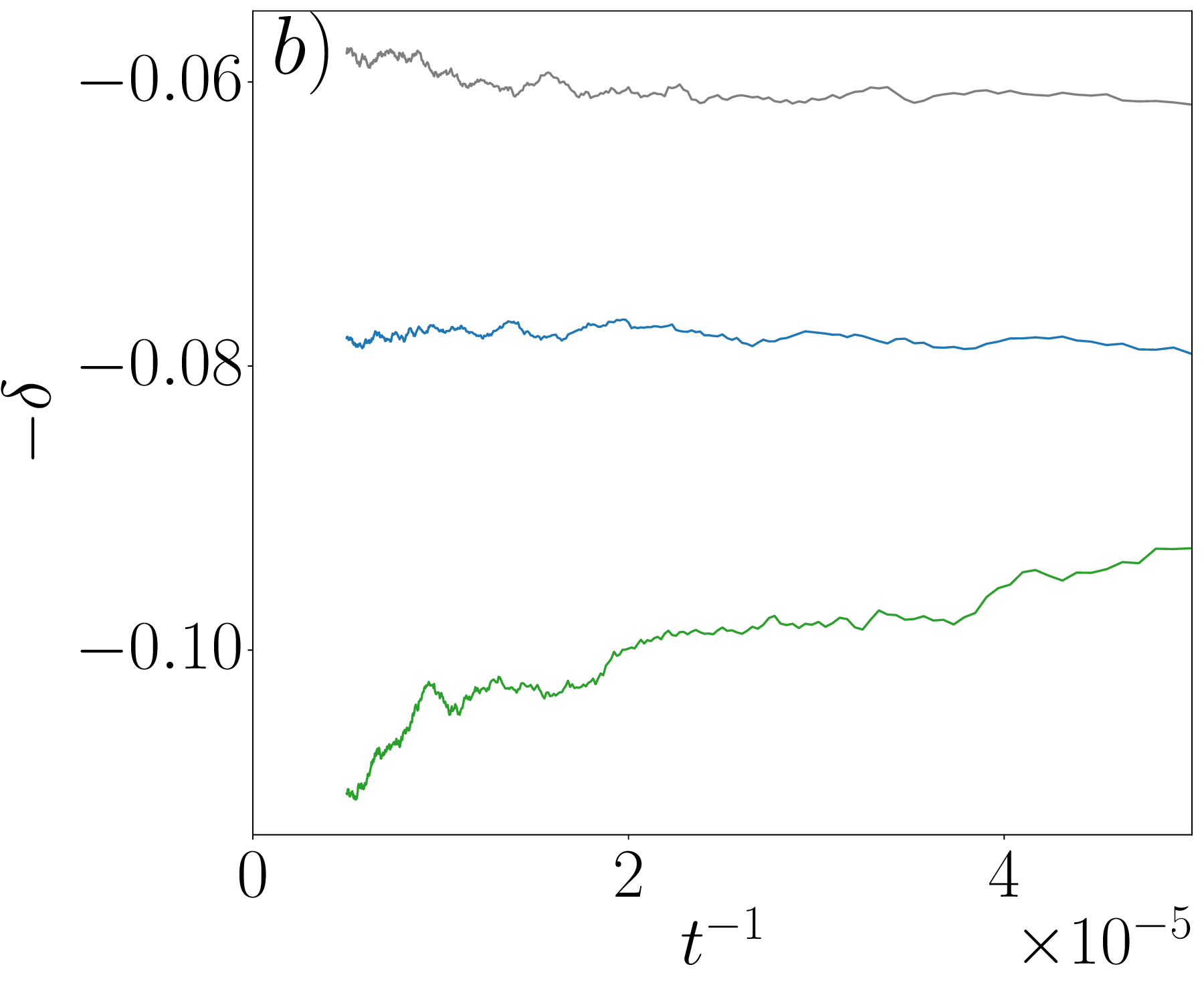}
        \centering
        \includegraphics[width=0.34\textwidth]{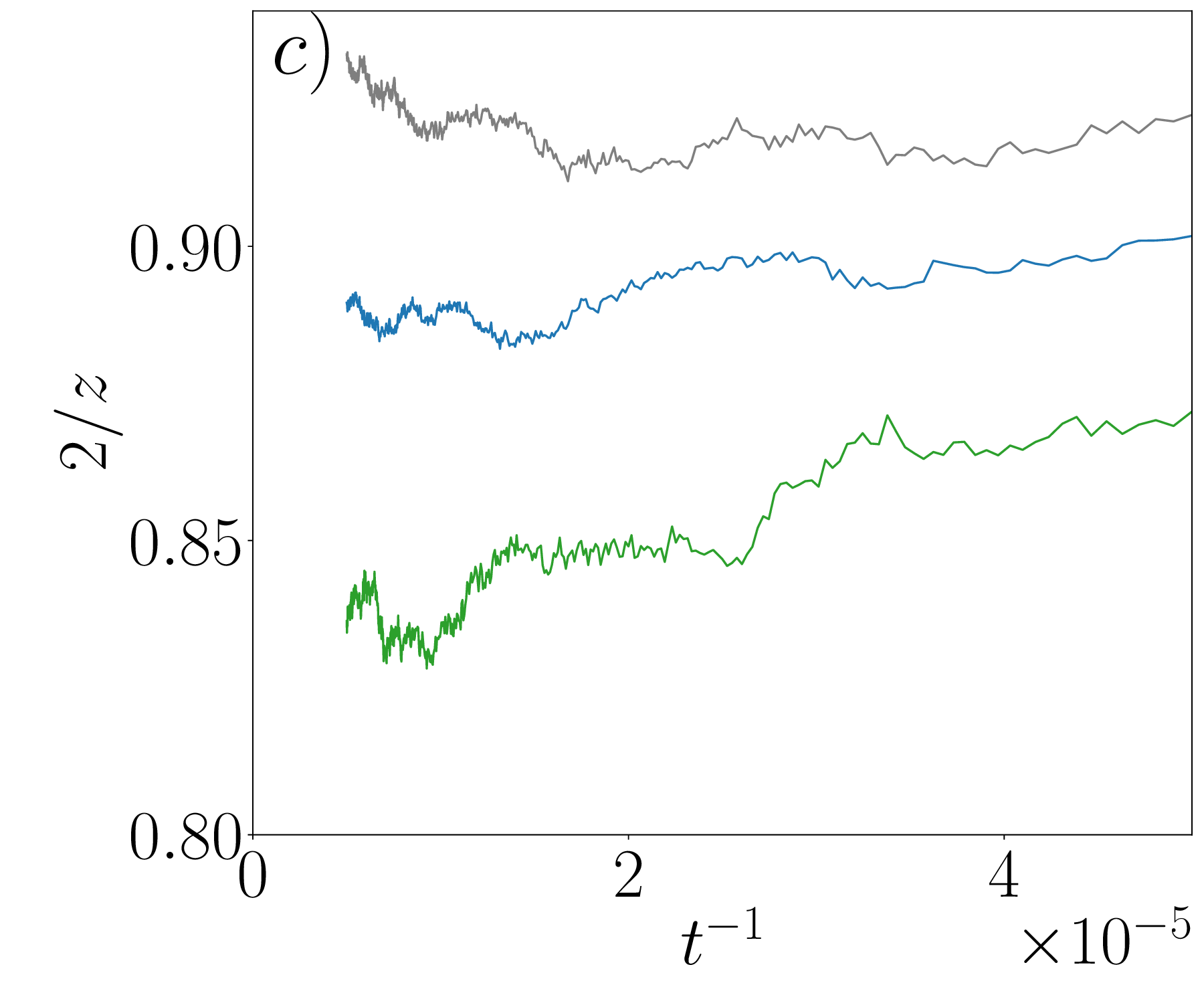}
        \centering
        \includegraphics[width=0.34\textwidth]{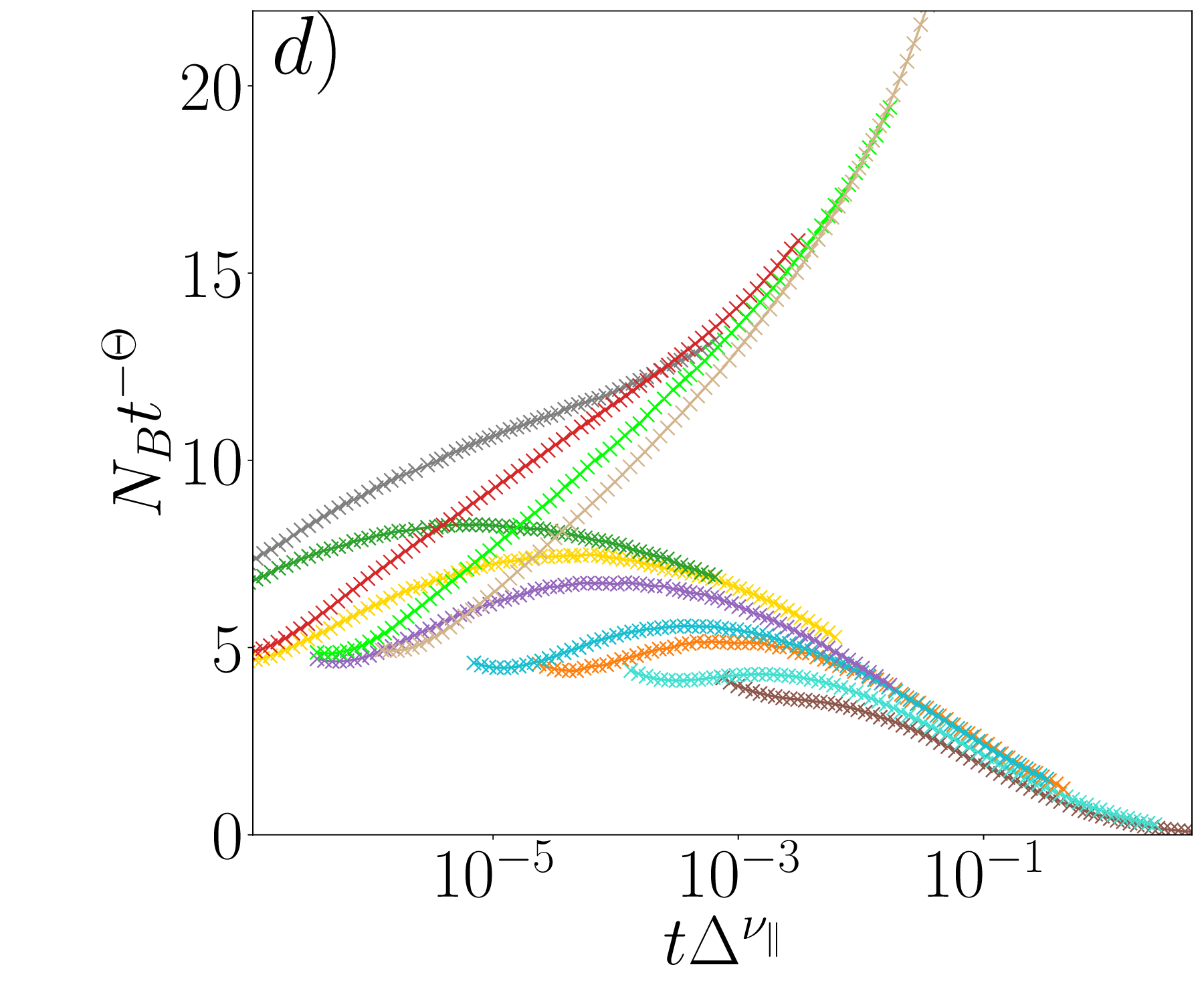}
    \caption{Critical properties of \(\lambda\)-disordered DEP with \(D_A = D_B = 1\). (a-c) Effective critical exponents yielding the estimates \(\lambda^{\lambda}_c = 0.490(5)\), \(\Theta^{\lambda} = 0.275(10)\), \(\delta^{\lambda} = 0.079(5)\), and \(2/z^{\lambda} = 0.89(2)\). A total of \(2 \times 10^3\) disorder realization were used for \(\lambda = 0.485\) and \(0.495\), and \(4 \times 10^3\) samples for \(\lambda = 0.49\). (d) Scaling analysis using \(\Theta^{\lambda} = 0.275\) and \(\lambda^{\lambda}_c = 0.49\) to estimate \(\nu^{\lambda}_{\parallel} = 4.25(40)\). Contamination rates range from \(\lambda = 0.4\) to \(0.51\).}
    \label{fig:LambdaDisorderDa=Db}
\end{figure}

Applying the procedure from Sec.~\ref{subsection:estimateCriticalProperties}, we estimate: \( \Theta^{\lambda} = 0.275(10) \), \( \delta^{\lambda} = 0.079(5) \), \( 2/z^{\lambda} = 0.89(2) \), \( \nu_{\parallel}^{\lambda} = 4.25(40) \), and \( \nu_{\perp}^{\lambda} = 1.96(18) \), as shown in Fig.~\ref{fig:LambdaDisorderDa=Db}. Using the hyperscaling relation, we find \( \alpha^{\lambda} = 0.091(15) \) and \( \beta^{\lambda}/\nu_{\perp}^{\lambda} = 0.20(3) \). In Fig.~\ref{fig:LambdaDisorderDa=Db}(d), tuning only \( \nu^{\lambda}_{\parallel} \) does not yield a satisfactory data collapse. This discrepancy may be explained by stronger corrections compared to the pure case. While longer simulations would be beneficial, we are limited by our algorithm, which effectively accounts for systems of size \( L > 5 \times 10^4 \).

The exponents \( \beta^{\lambda} \) and \( \nu_{\perp}^{\lambda} \) are consistent with the pure case within the error bars. Assuming time-reversal symmetry, we estimate \( \tilde{\Theta}^{\lambda} = \frac{d}{z} - 2\delta = 0.263(14) \) using the hyperscaling relation. Since \( \tilde{\Theta}^{\lambda} \) and \( \Theta^{\lambda} \) are in agreement within the error bars, time-reversal symmetry appears to be conserved. However, the dynamical exponent \( z^{\lambda} \) is clearly incompatible with the pure estimate \( 2/z = 0.992(10) \), and the large uncertainty in \( \nu_{\parallel}^{\lambda} \) makes it difficult to draw a definitive conclusion. Nevertheless, the increase in \( \nu_{\parallel}^{\lambda} \) compared to \( \nu_{\parallel} \) compensates for the increase in \( z^{\lambda} \) relative to \( z \), such that \( \nu_{\perp}^{\lambda} \approx \nu_{\perp} \), thereby satisfying the inequality \( d \nu_{\perp}^{\lambda} \geq 2 \), as expected for a disordered system \cite{chayes_finite-size_1986}.

\begin{figure}
    \centering
        \includegraphics[width=0.31\textwidth]{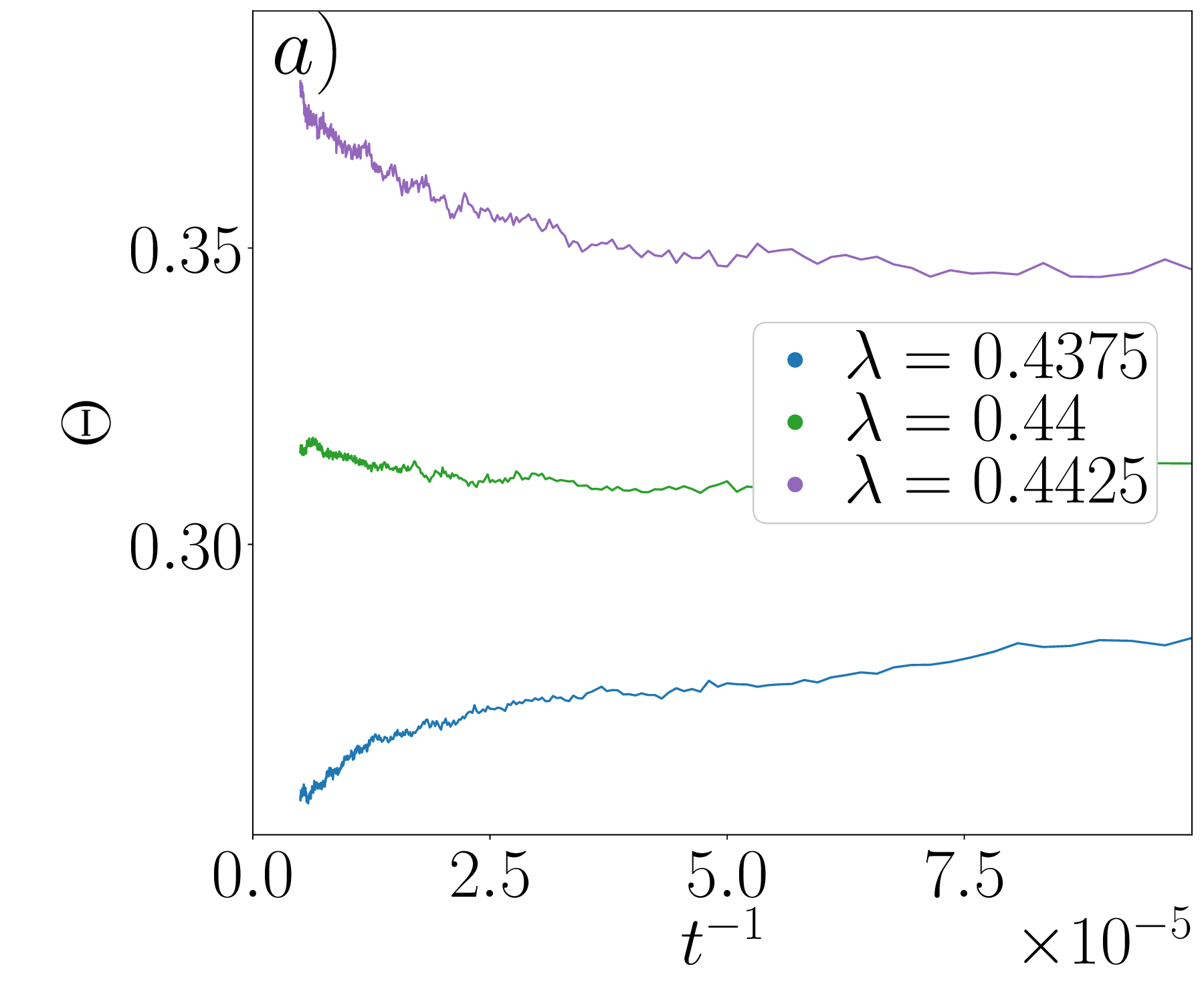}
        \centering
        \includegraphics[width=0.31\textwidth]{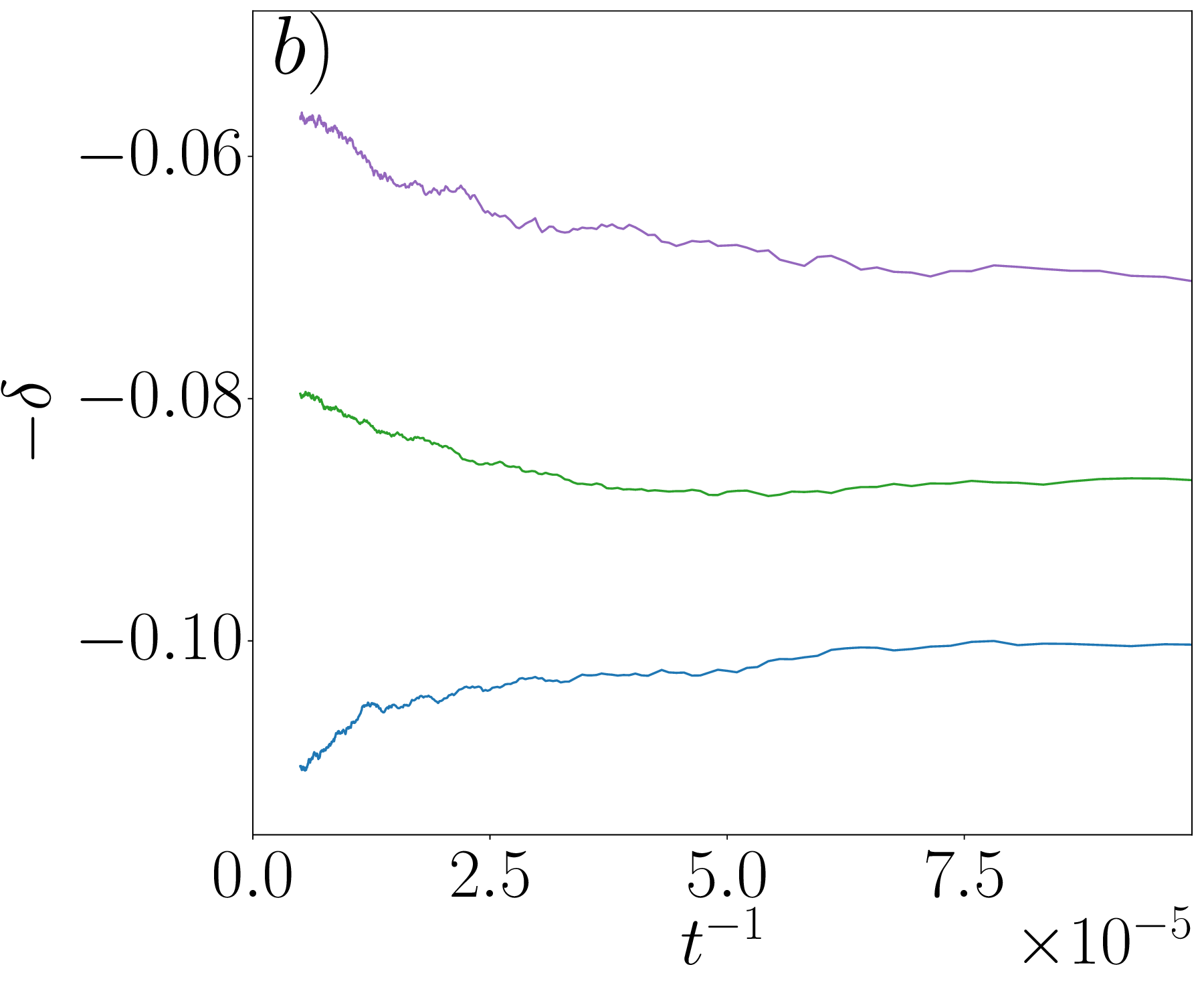}
        \centering
        \includegraphics[width=0.31\textwidth]{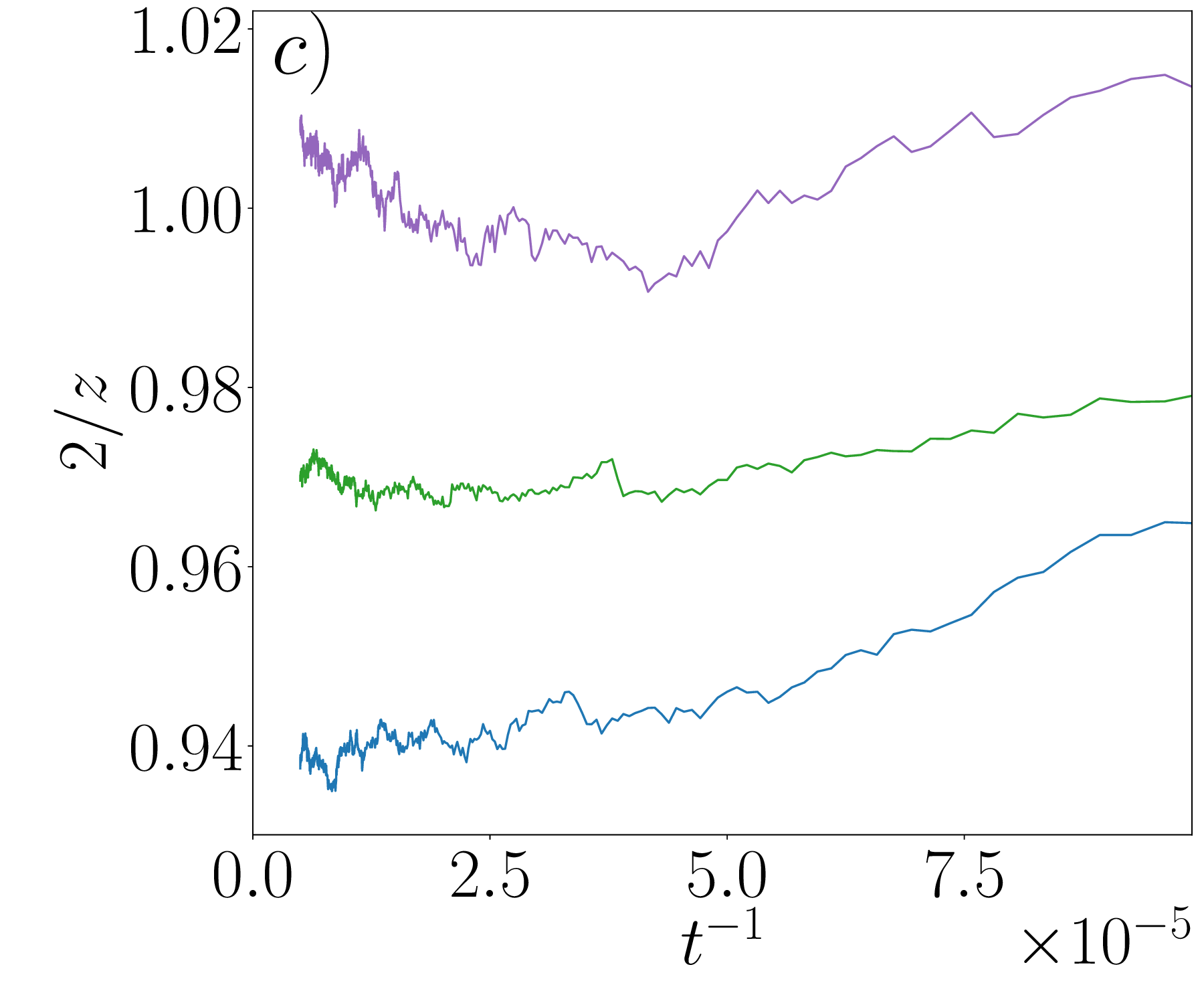}
    \caption{Same as in Figure \ref{fig:LambdaDisorderDa=Db} with a weaker disorder, $c=0.4$ and $p_{\lambda}=0.3$. (a-c) Effective critical exponents yielding the estimates \(\lambda_c = 0.440(5)\), $\Theta^{\lambda} = 0.32(2), \delta^{\lambda}=0.078(10)$ and to $2/z^{\lambda}=0.97(2)$. A total of \(5 \times 10^3\) disorder realization are used for $\lambda = 0.440$, and a few thousand otherwise. }
    \label{fig:LambdaWeakDisorderDa=Db}
\end{figure}

We investigated whether the critical properties depend on the disorder distribution by increasing the lowest contamination rate value to \( c = 0.4 \) instead of \( c = 0.2 \). Using only a few contamination rates, we estimate \( \lambda_c^{\lambda} = 0.44 \), \( \Theta^{\lambda} = 0.32(2) \), \( \delta^{\lambda} = 0.078(10) \), and \( 2/z^{\lambda} = 0.97(2) \) (see Fig.~\ref{fig:LambdaWeakDisorderDa=Db}). Applying the hyperscaling relation, we obtain \( \alpha^{\lambda} = 0.087(25) \) and \( \beta^{\lambda} / \nu_{\perp}^{\lambda} = 0.18(5) \). Due to insufficient data, we were unable to estimate \( \nu^{\lambda}_{\parallel} \).  
We find good agreement with the pure static exponent \( \beta/\nu_{\perp} \), while the dynamical exponent \( z^{\lambda} \) shows a slight deviation from the pure value but is still within the errors bars.  

We therefore conclude that the critical properties depend on the strength of the disorder, with the dynamical exponent \( z^{\lambda} \) being particularly sensitive, while the static exponents \( \beta^{\lambda} \) and \( \nu^{\lambda}_{\perp} \) remain consistent with the pure estimates for the disorder distribution studied. However, our data are not precise enough to rule out alternative scenarios, such as logarithmic corrections. Extrapolating to strong enough disorder distributions, the critical properties could be governed by an infinite disorder fixed point where $z$ is formally infinite. Such a dependence on the initial disorder strength has been observed in other disordered systems, such as the disordered quantum clock model \cite{senthil_critical_1996}.

\subsection{$D_A < D_B$}

\begin{figure}
    \centering
        \includegraphics[width=0.34\textwidth]{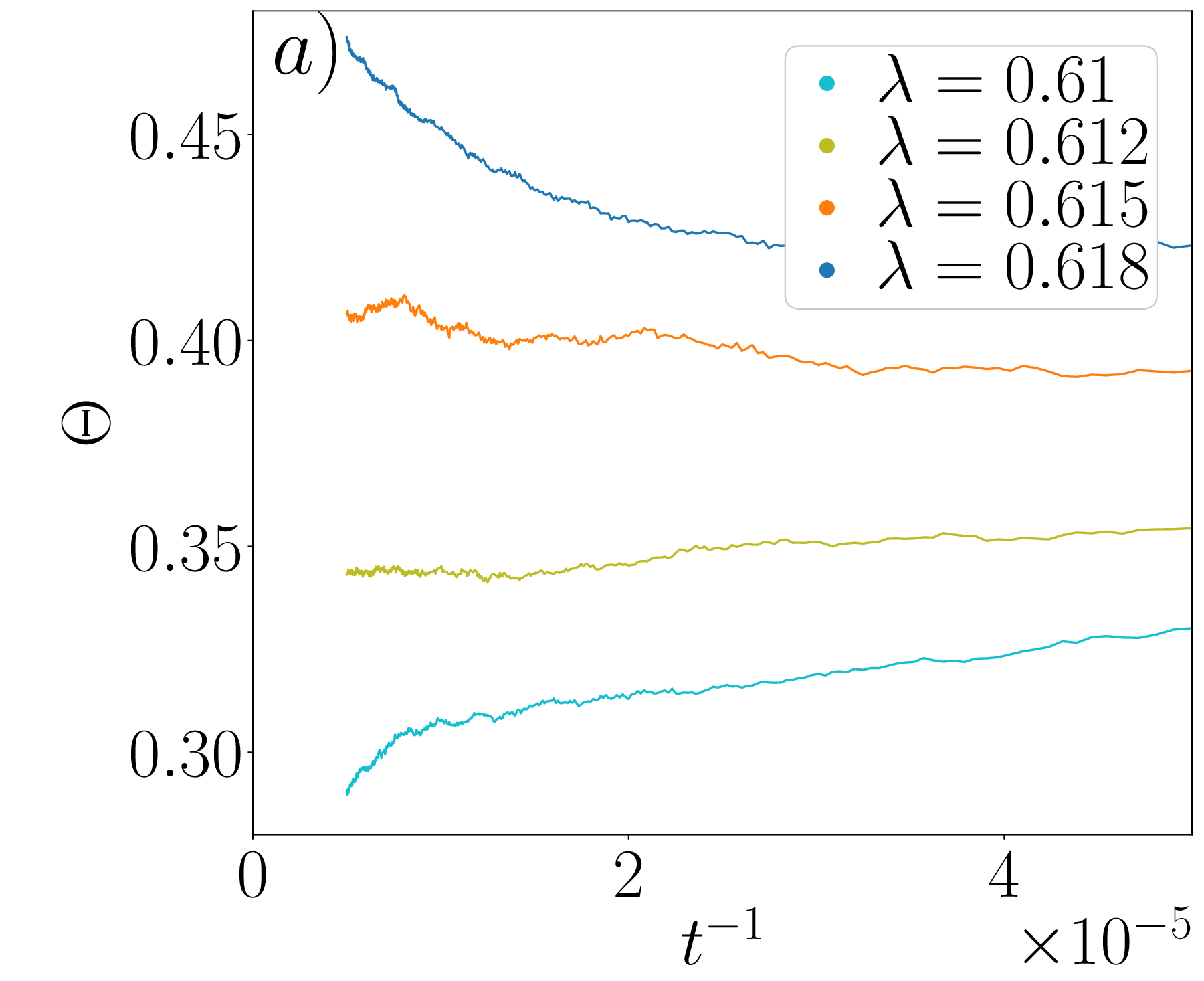}
        \centering
        \includegraphics[width=0.34\textwidth]{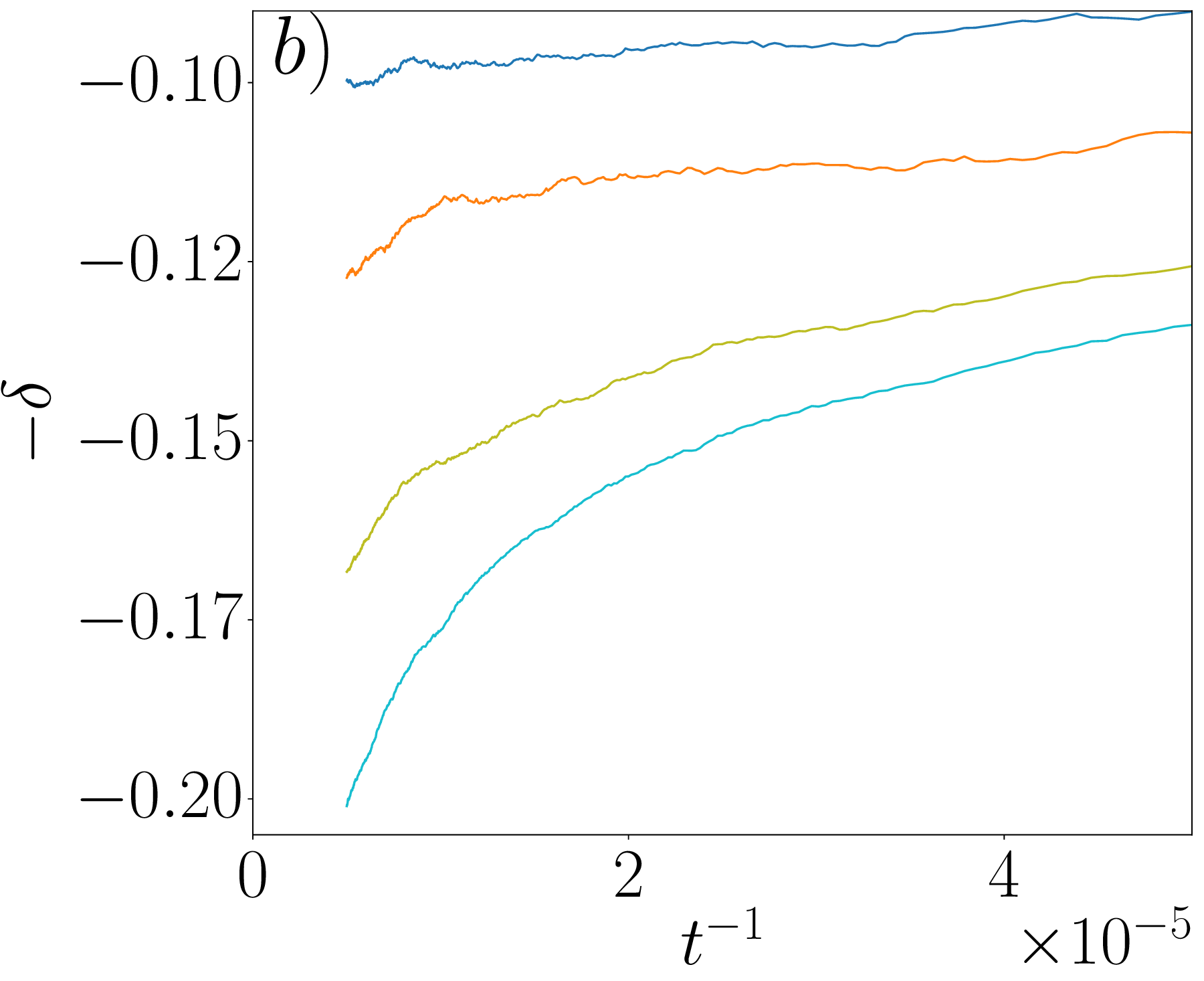}
        \centering
        \includegraphics[width=0.34\textwidth]{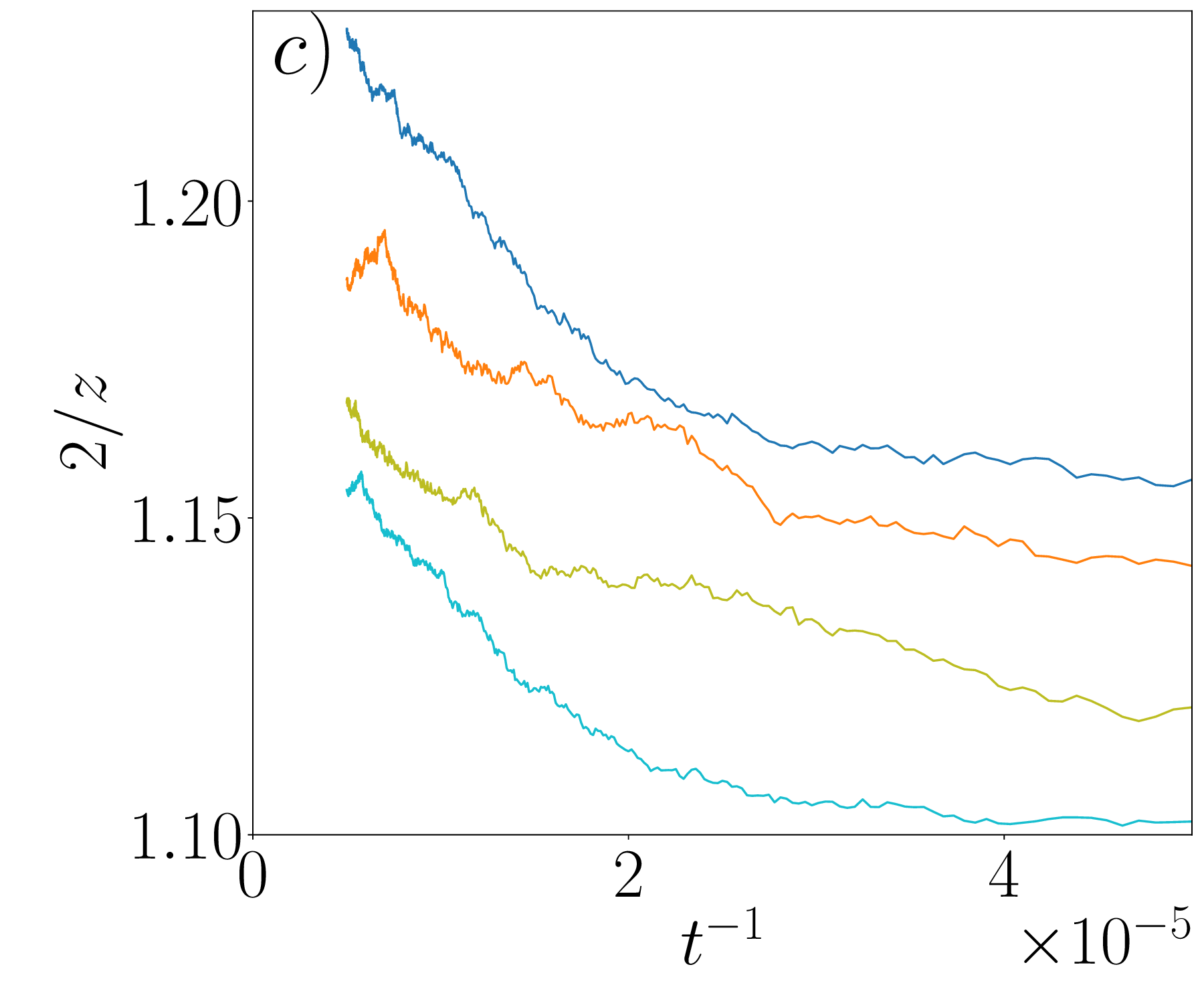}
        \centering
        \includegraphics[width=0.34\textwidth]{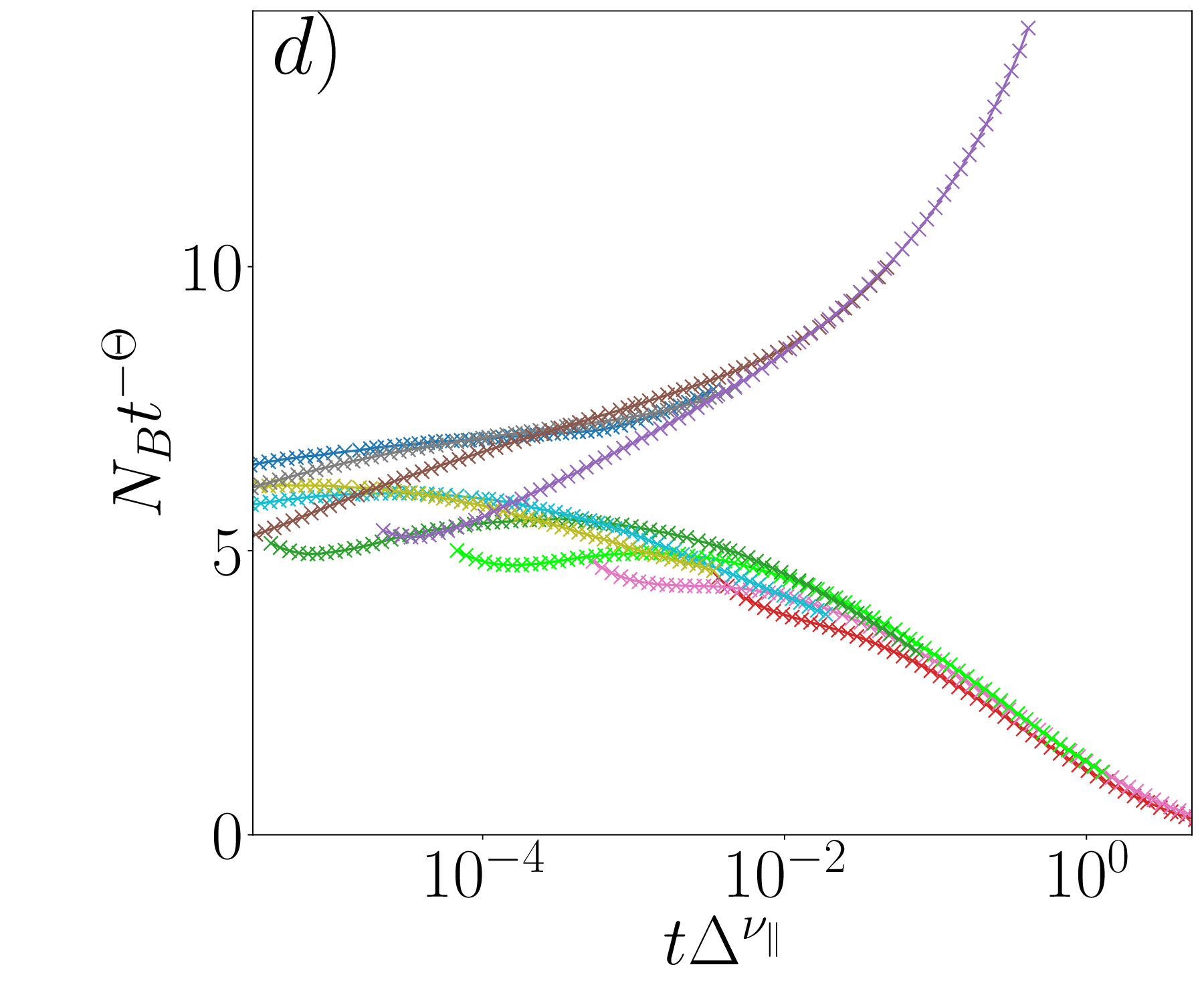}
    \caption{Critical properties of \(\lambda\)-disordered DEP with \(0.375 = D_A < D_B = 1\). (a-c) Effective critical exponents yielding the estimates \(\lambda_c = 0.615(3)\), \(\Theta = 0.41(2)\), \(\delta = 0.13(2)\), and \(2/z = 1.20(2)\). At least \(2 \times 10^3\) disorder realization were used for each contamination rate . (d) Scaling analysis using \(\Theta = 0.41\) and \(\lambda_c = 0.615\) to estimate \(\nu_{\parallel} = 3.35(20)\). Contamination rates range from \(\lambda = 0.5\) to \(0.64\).}
    \label{fig:LambdaDisorderDalDb}
\end{figure}

Determining the critical properties from Fig.~\ref{fig:LambdaDisorderDalDb} is more challenging, as both \( \Theta^{\lambda}(t) \) show no clear curvature up to \( T_{\max} = 2.10^5 \)~s at \( \lambda = 0.612 \) and \( 0.615 \), while \( \delta^{\lambda} \) exhibits downward curvature at these values. Since a better data collapse is observed at \( \lambda = 0.615 \), we estimate \( \lambda_c^{\lambda} = 0.615(3)\) and hence \(\Theta^{\lambda} = 0.41(2) \), \( \delta^{\lambda} = 0.13(2) \), \( 2/z^{\lambda} = 1.20(2) \), \( \nu^{\lambda}_{\parallel} = 3.35(20) \), and, using the hyperscaling relation, \( \alpha^{\lambda} = 0.06(3) \) and \( \beta^{\lambda}/\nu_{\perp}^{\lambda} = 0.10(5) \). The exponents \( z^{\lambda} = 1.67(3) \), \( \beta^{\lambda}/\nu_{\perp}^{\lambda} = 0.10(5) \), and \( \nu_{\perp}^{\lambda} = 2.01(12) \) are in good agreement with their corresponding pure values.  

We therefore conclude that disorder is not relevant. However, as before, we cannot rule out more complex scenarios, such as a slight dependence of the critical exponents on disorder or the necessity of logarithmic corrections, which could lead to different exponent values. The irrelevance of disorder is consistent with the expectation from the Harris criterion, based on our estimate of the pure correlation length exponent \( \nu_{\perp} = 2.01(12) \). Although this contrasts with the value \( \nu_{\perp} = 1.77(3) \) found in \cite{maia_diffusive_2007}, but their estimate of the dynamical exponent \( z = 1.6(2) \) agrees well with ours.  

\subsection{$D_A>D_B$}

\begin{figure}
    \centering
        \includegraphics[width=0.31\textwidth]{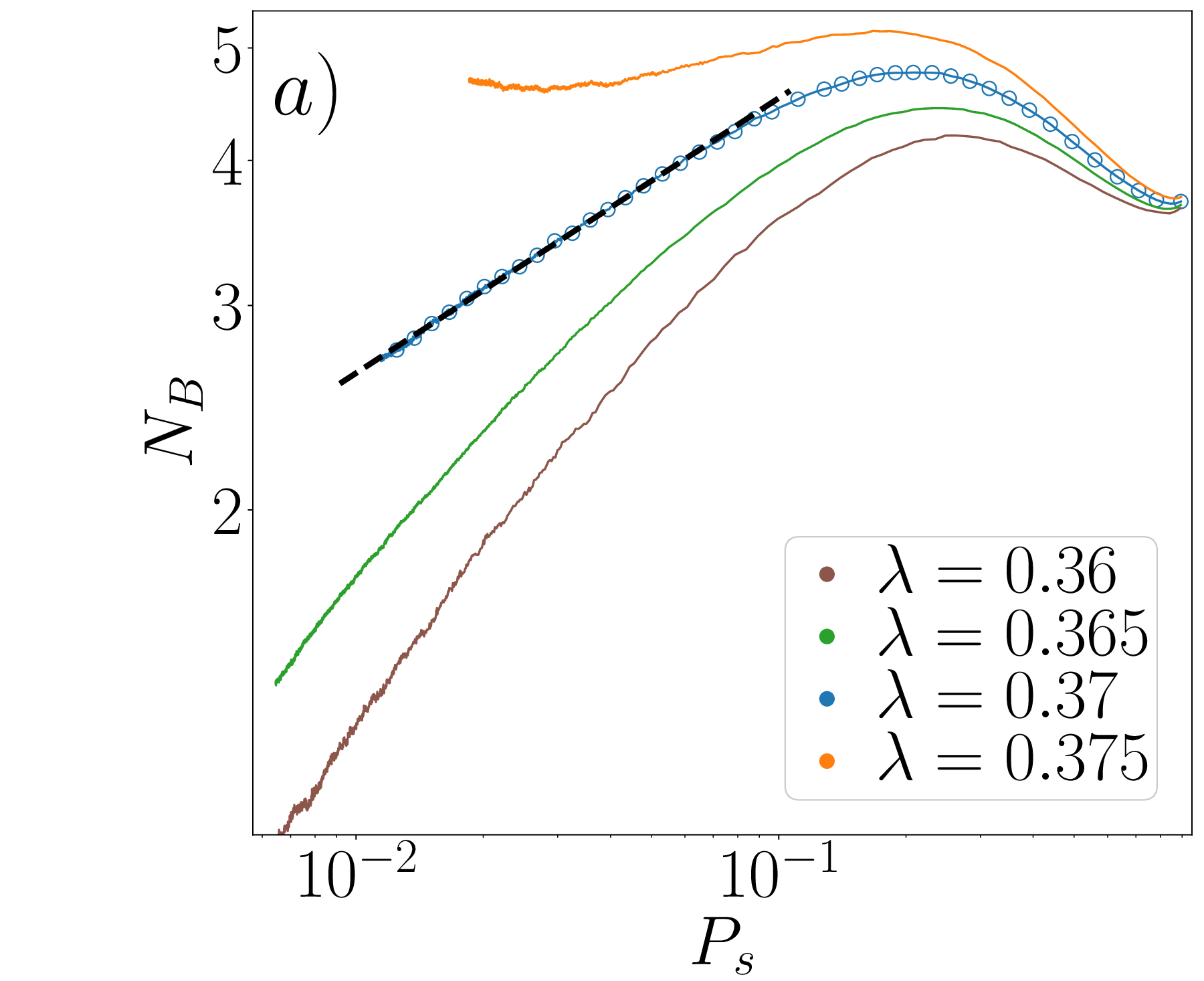}
        \centering
        \includegraphics[width=0.31\textwidth]{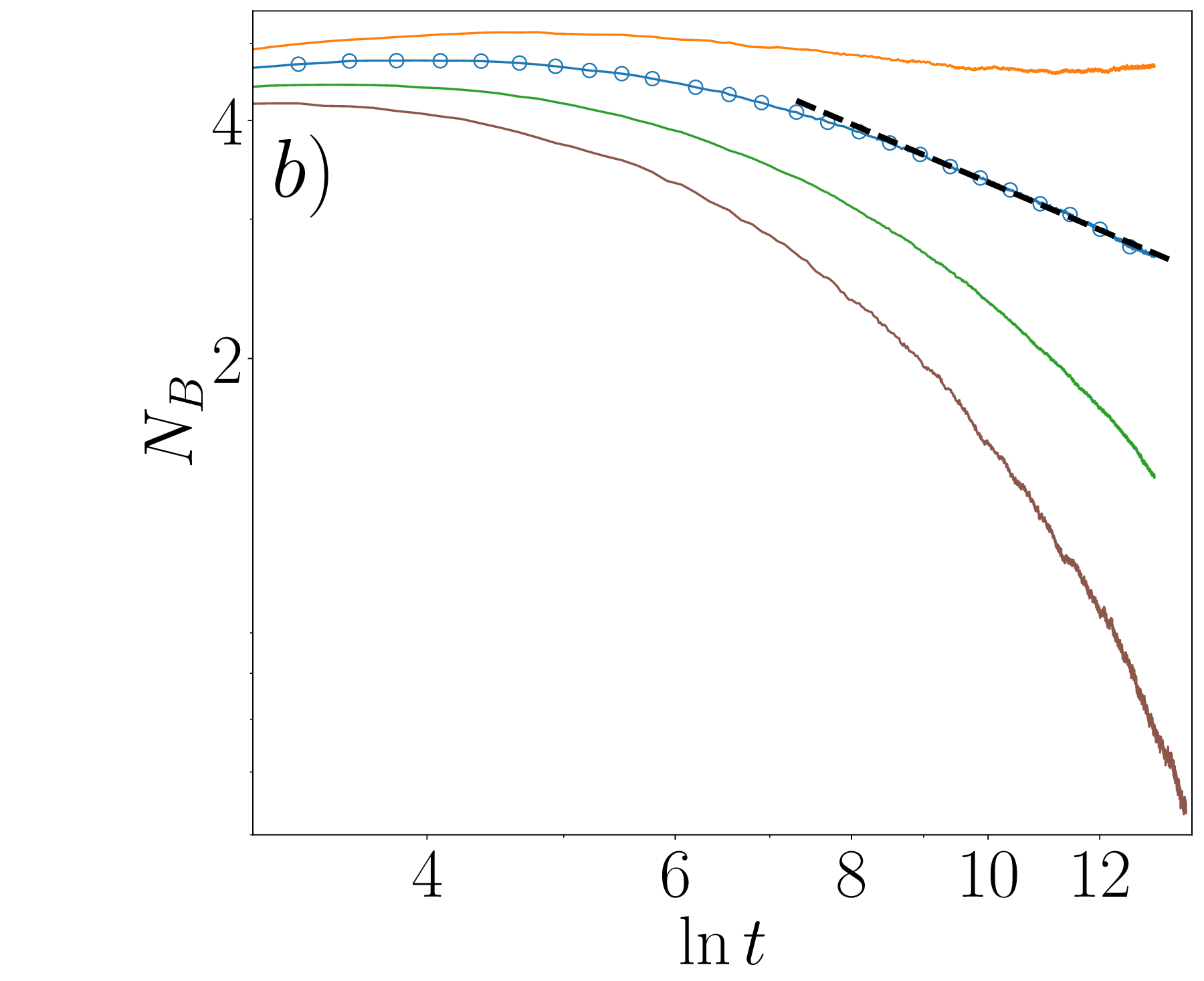}
        \centering
        \includegraphics[width=0.31\textwidth]{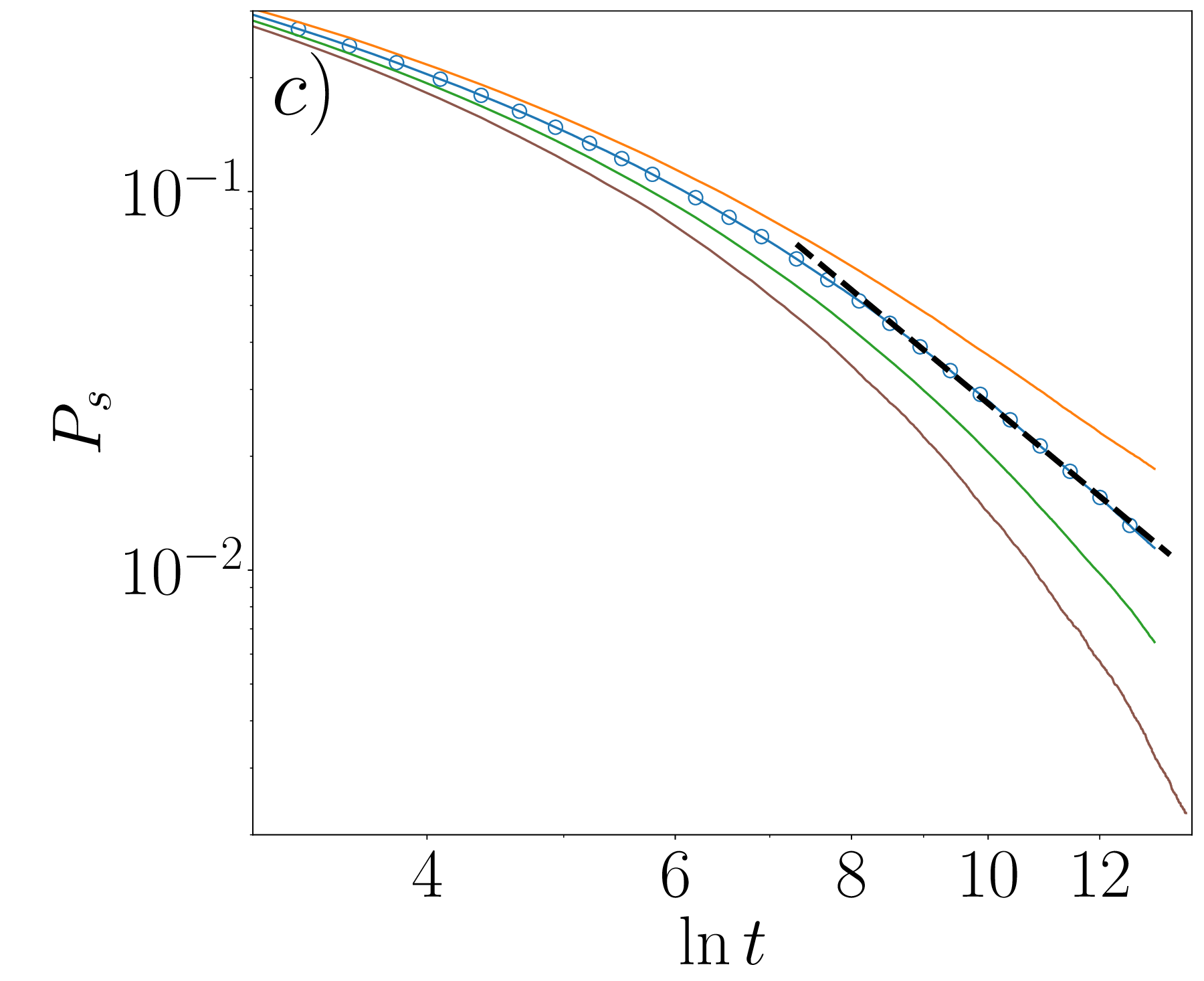}
    \caption{Critical properties of \(\lambda\)-disordered DEP with \(1 = D_A > D_B = 0.375\). (a) Evolution of the number of infected particles $N_B$ with $P_s$ to estimate the position of the critical point yielding to $\lambda_c = 0.370(5)$ (circle). (b-c) Variation of $N_B/P_s$ with $\ln t$ in a log-log scale to check if a straight line is obtained as expected at an IDFP. Data obtained at $\lambda=0.365, 0.37, 0.375$ have been averaged over more than 20000 samples with $T_{\max}=5.10^5$s and 2500 samples at $\lambda=0.36$ with $T_{\max}=10^6$s.}
    \label{fig:LambdaDisorderDblDa}
\end{figure}

We provide additional data to Fig.~2 in the main text. The critical point $\lambda_c = 0.370(5)$ (see Fig.~\ref{fig:LambdaDisorderDblDa}(a)) is estimated from the relation \( N_B \sim P_s^{-\overline{\Theta}/\overline{\delta}} \), which holds also at an infinite disorder fixed point (IDFP), independently of the initial time \( t_0 \) \cite{vojta_infinite-randomness_2009}. We further investigate signatures of activated dynamical scaling at the critical point, as described in Eq.~(3) of the main text. In both Fig.~\ref{fig:LambdaDisorderDblDa}(b) and (c), straight lines at $\lambda_c$ are barely discernible. To resolve this, it would be necessary to increase the maximum simulation time by several orders of magnitude to surpass the microscopic time scale \( t_0 \) and observe the logarithmic growth. Therefore, estimating the critical exponents is beyond the scope of this study. Nevertheless, our data suggest that the critical exponent \( \overline{\Theta} \) is negative. In contrast, this exponent is positive at the IDFP of the disordered contact process \cite{hooyberghs_strong_2003, vojta_critical_2005}, indicating that the IDFP observed in our study appears to be genuine.

\section{One-specie diffusion rate disorder}

We present only some of the most relevant cases in Table~\ref{tab:One-speicieDisorder} and provide details on unexpected cases. For the numerical details of the pure-like cases, we followed the same procedure as before, generally using fewer computational resources (lower \( T_{\max} \) and/or fewer disorder realizations). Our goal was not to obtain precise estimates of the critical exponents but rather to determine whether they align with one of the three pure universality classes or deviate significantly. That being said, we cannot rule out the possibility that the disordered exponents may vary slowly with disorder strength.

For completeness, we recall the expression for the effective macroscopic diffusion rate in the case of a binary disorder distribution (see Eq.~\ref{eq:DisorderDistributionDiffusion}):  
\begin{equation}
    \frac{1}{D_X^{\text{eff}}} = \frac{p_X}{D_X^0}+\frac{1-p_X}{D_X^1},
    \label{eq:effectiveDiffusionRate}
\end{equation}
where $X=A,B$.

\begin{table}[h]
\centering
\setlength\extrarowheight{-3pt}
\begin{tabular}{|c|c|c|c|c|c|c|}
\hline
 $D_{A,0}$ & $D_{A,1}$ & $D_{B,0}$ & $D_{B,1}$ & $p_{A/B}$ & Type & CP \\
\hline
0.08 & 0.46 & 1 & - & 0.05 & I & Pure \\
0.25 & 6 & 1 & - & 0.65 & I & Pure \\
0.18 & 1000 & 1 & - & 0.25 & I & $/$ \\
0.18 & 1000 & 1 & - & 0.85 & I & Pure \\
1 & - & 0.25 & 6 & 0.05 & I & Pure \\
1 & - & 1.4 & 2000 & 0.5 & I & Pure \\
0.25 & 6 & 1 & - & 0.05 & II & Slow \\
1.4 & 2.96 & 1 & - & 0.05 & II & Slow \\
0.85 & 2000 & 1 & - & 0.65 & II & Slow \\
1.4 & 2000 & 1 & - & 0.5 & II & Slow \\
1.4 & 2000 & 1 & - & 0.8 & II & Slow \\
1 & - & 0.25 & 1 & 0.55 & II & Pure \\
1 & - & 0.25 & 6 & 0.65 & II & Pure \\
1 & - & 0.18 & 1000 & 0.5 & II & Pure \\
1 & - & 0.2 & 1 & 0.2 & II & Slow$^*$ \\
\hline
\end{tabular}
\caption{Summary of the critical properties (CP) for various disorder distributions, see Eq.~\ref{eq:DisorderDistributionDiffusion}. The system is classified as Type I if \(D_A^{\text{eff}} < D_B\) or \(D_A < D_B^{\text{eff}}\), and as Type II if \(D_A^{\text{eff}} > D_B\) or \(D_A > D_B^{\text{eff}}\). The term \textit{Slow} refers to systems where the effective dynamical exponent \(2/z(t)\) consistently decreases over time at the critical point, which is compatible with a logarithmic growth of \(R^2\). The symbol "$/$" denotes unconventional behavior, which is attributed to the absence of a phase transition (no active state). $^*$ refers to block disorder. }
\label{tab:One-speicieDisorder}
\end{table}

\subsection{$D_A$ disorder}

\begin{figure}
    \centering
    \includegraphics[width=0.32\columnwidth]{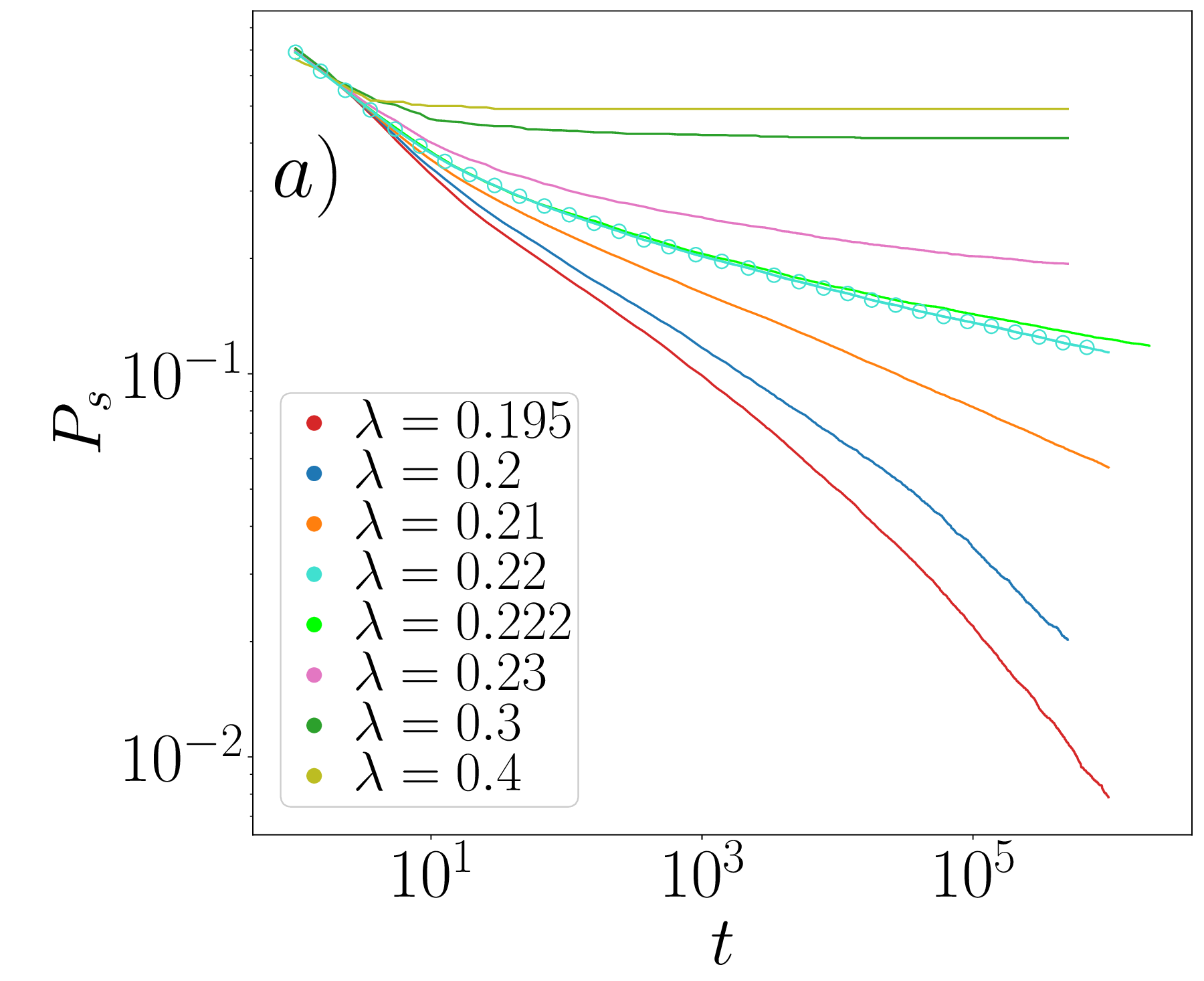}
    \includegraphics[width=0.32\columnwidth]{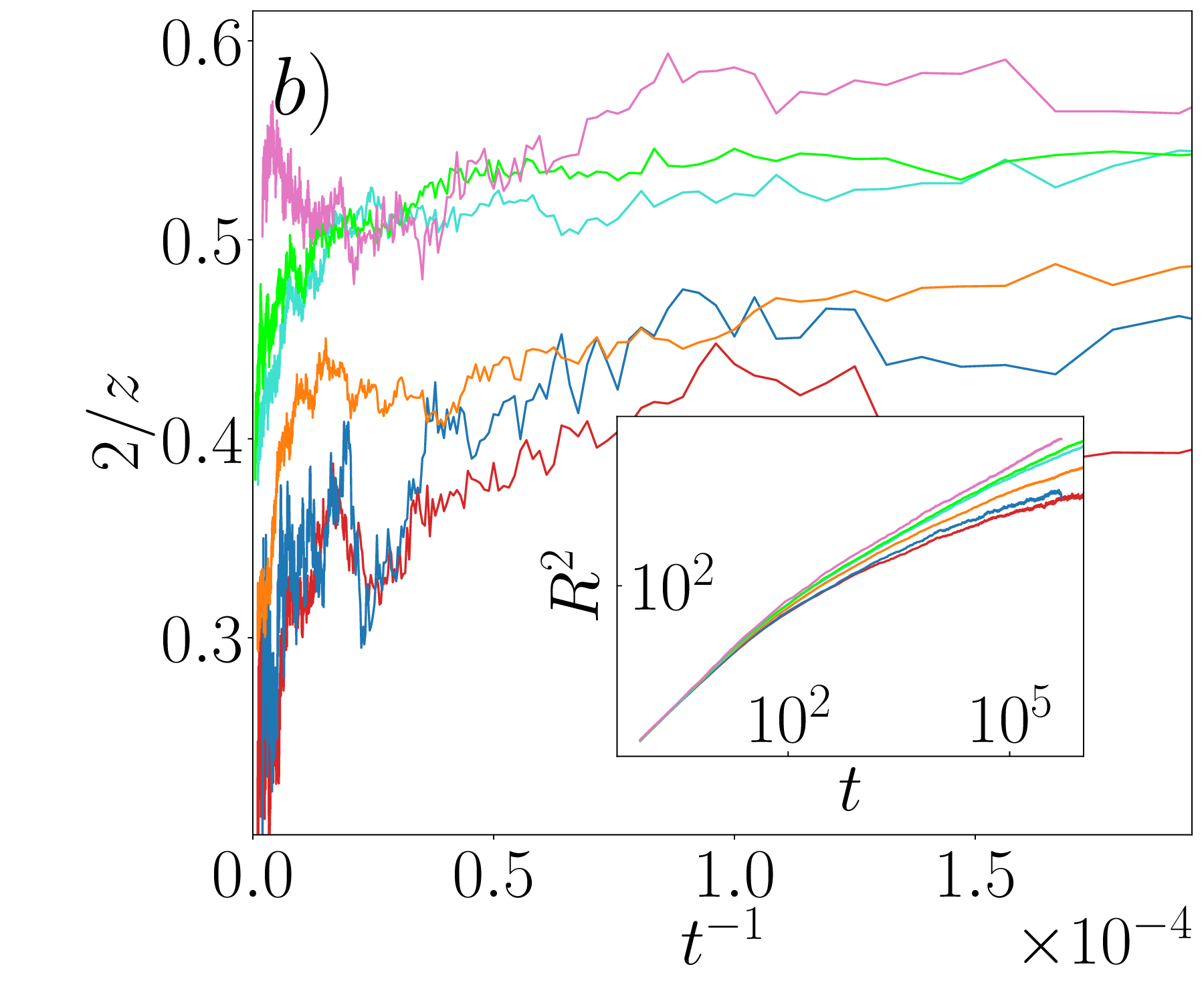}
    \includegraphics[width=0.32\columnwidth]{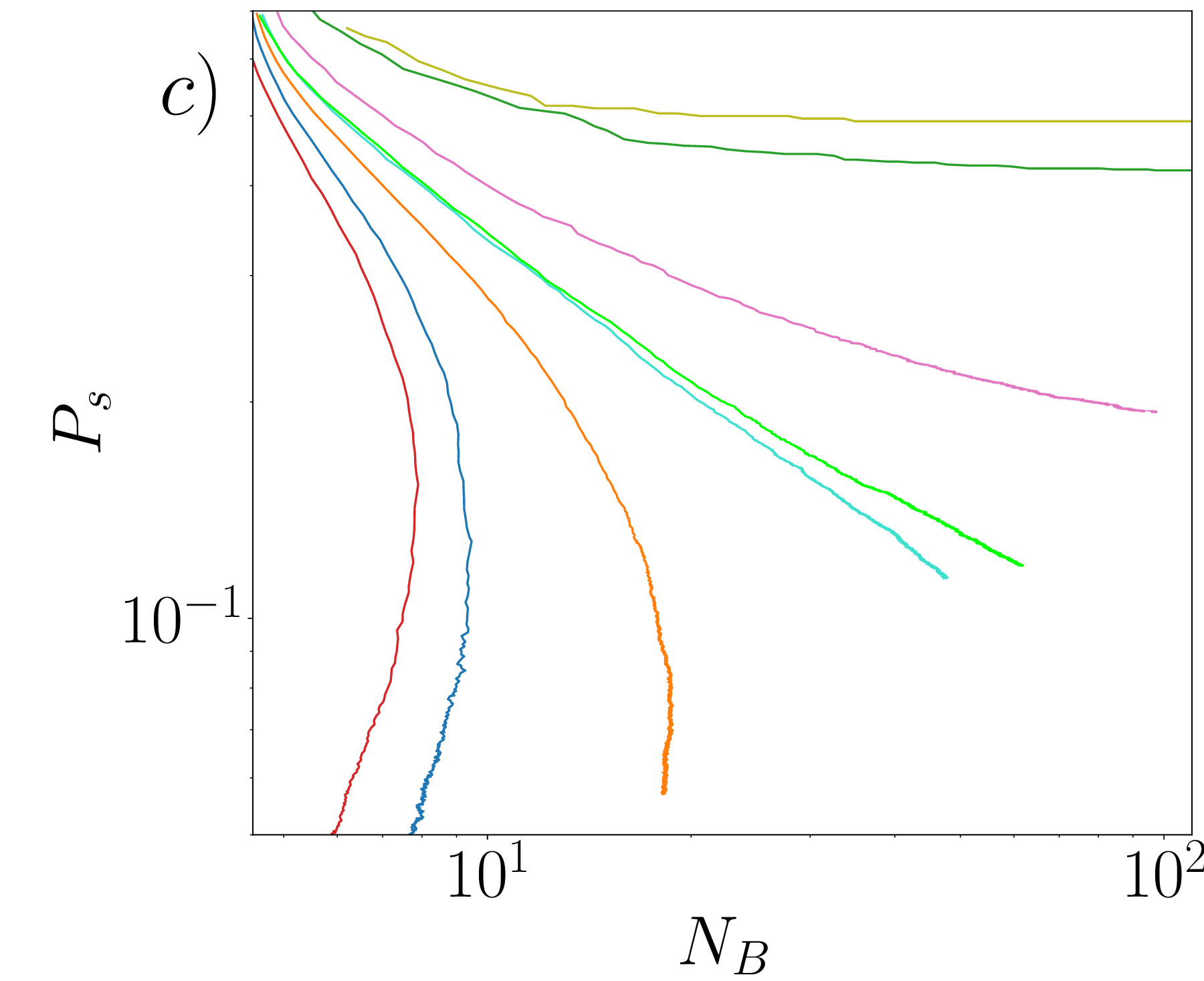}
    \includegraphics[width=0.32\columnwidth]{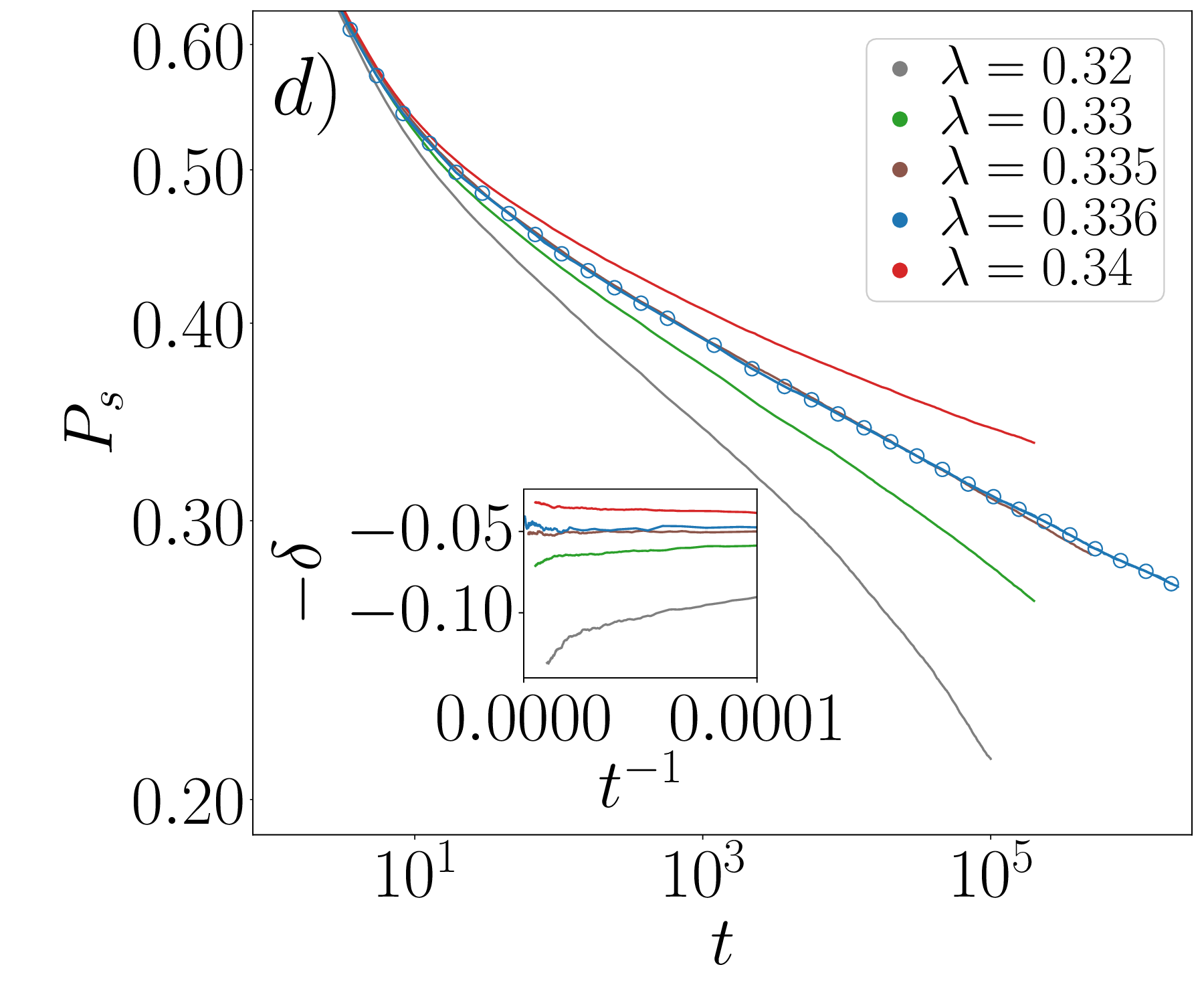}
    \includegraphics[width=0.32\columnwidth]{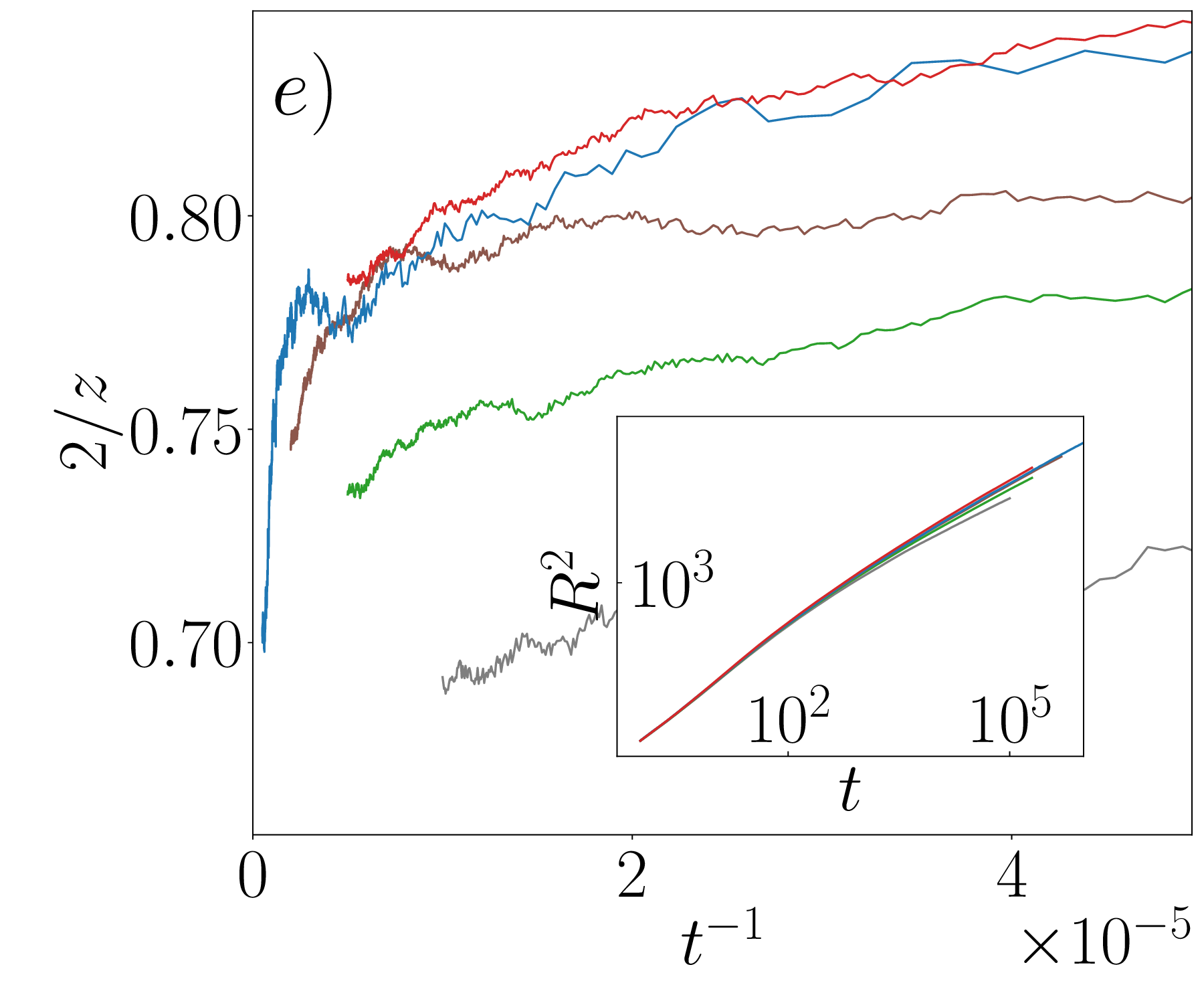}
    \includegraphics[width=0.32\columnwidth]{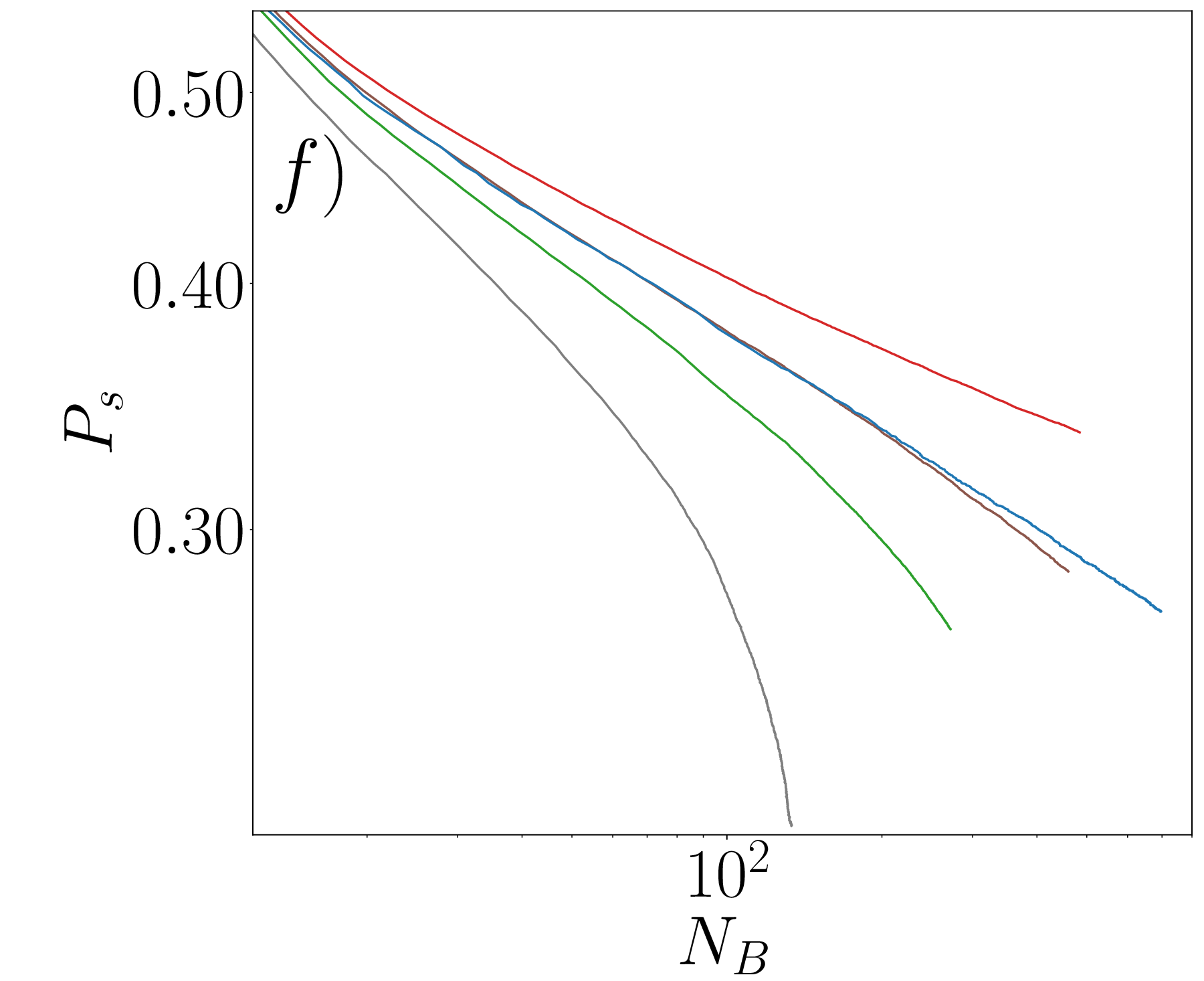}
    \caption{    \textbf{Up panel:} \(D_A\)-disorder with \(D_A^0=1.4\), \(D_A^1=2000\), \(p_A=0.5\), \(D_B=1\).  
    (a) No clear power-law decay of survival probability \(P_s\) at \(\lambda_c \approx 0.22\) (circle), while a satisfying decay is observed at \(\lambda = 0.21\) in the absorbing phase. This is the first evidence of an Infinite Disorder Fixed Point surrounded by a Griffith phase.  
    (b) Effective dynamical exponent \(2/z(t)\) decays to abnormally low values near and below the critical point, implying no power-law growth of \(R^2\) (inset).  
    (c) \(P_s\) vs \(N_b\) in log-log scale is used to estimate \(\lambda_c \approx 0.22\). 2000 disorder configurations were used for \(\lambda = 0.21, 0.22, 0.222\), with 10 runs per configuration.  
    \textbf{Down panel (d,e,f):} Same as (a,b,c) with \(D_A^0=0.85\), \(D_A^1=2000\), \(p_A=0.65\), \(D_B=1\).  
    (d) Variation of critical exponent \(\delta(t)\) (inset) shows strong deviation from the pure value (\(\delta \approx 0.58\)).  
    (e) Effective exponent \(2/z(t)\) at \(\lambda_c \approx 0.336\) remains relatively large (\(\sim 0.7\)) at late times, close to the pure value (\(\approx 0.74\)). But its sharp decrease contrasts with the pure case (see Fig.~\ref{fig:PureDblDa}(c)), further supporting a \textit{slow} scenario.  
    (f) \(P_s\) vs \(N_b\) to estimate \(\lambda_c \approx 0.336\).  \(10^3\) disorder realizations were used for \(\lambda = 0.336\).}
    \label{fig:DaInfDisorderDblDa}
\end{figure}

\begin{figure}
    \centering
    \includegraphics[width=0.32\columnwidth]{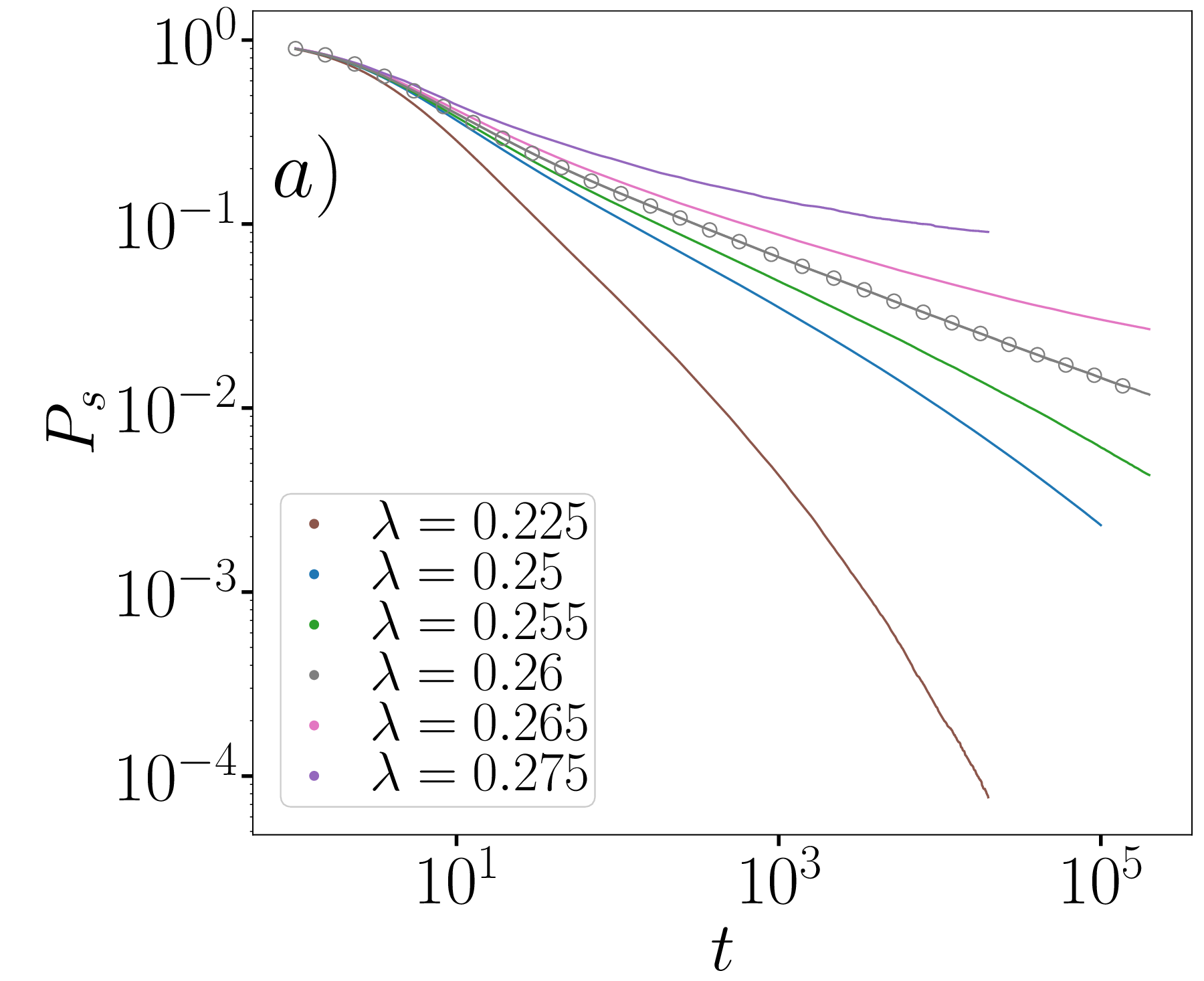}
    \includegraphics[width=0.32\columnwidth]{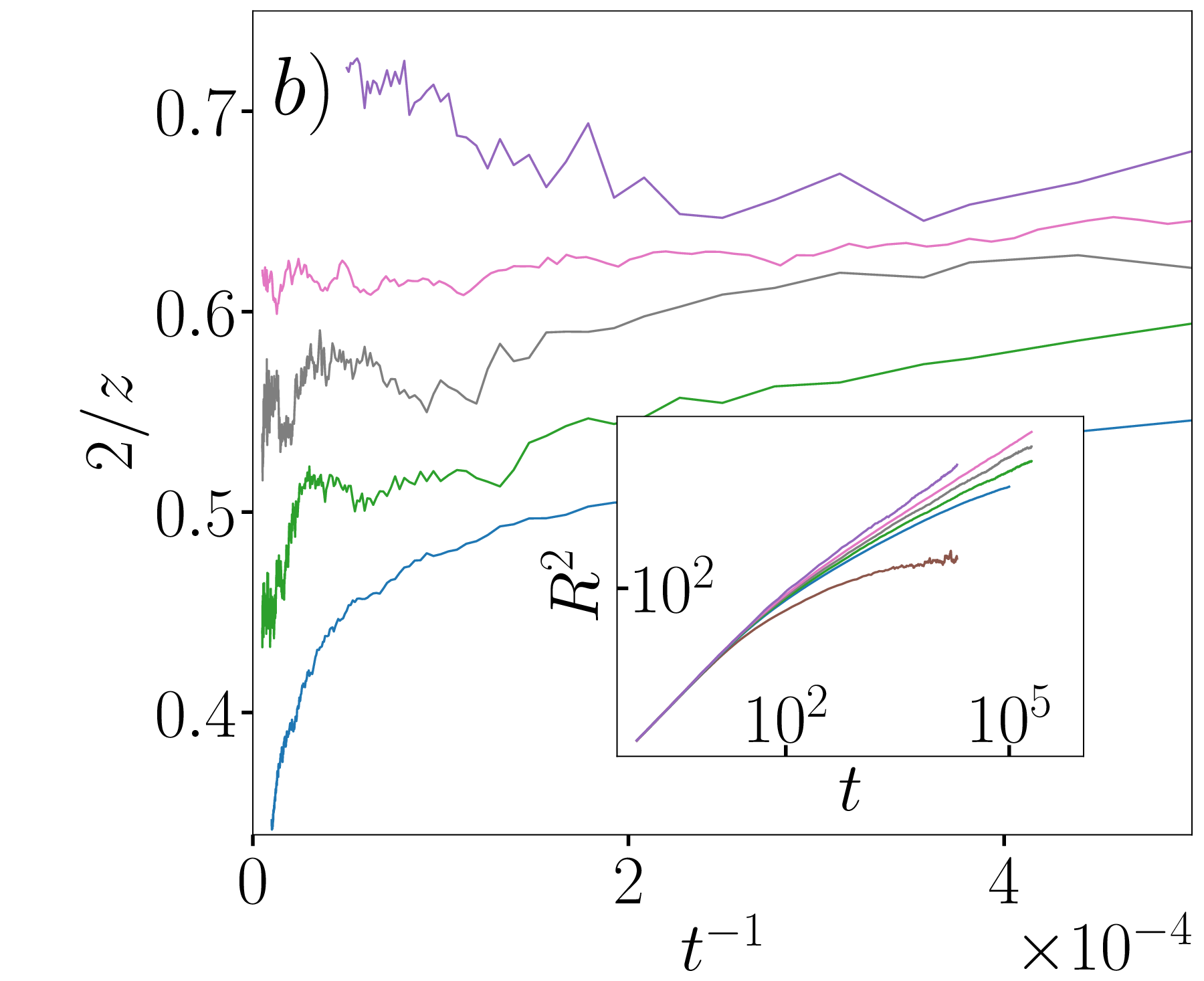}
    \includegraphics[width=0.32\columnwidth]{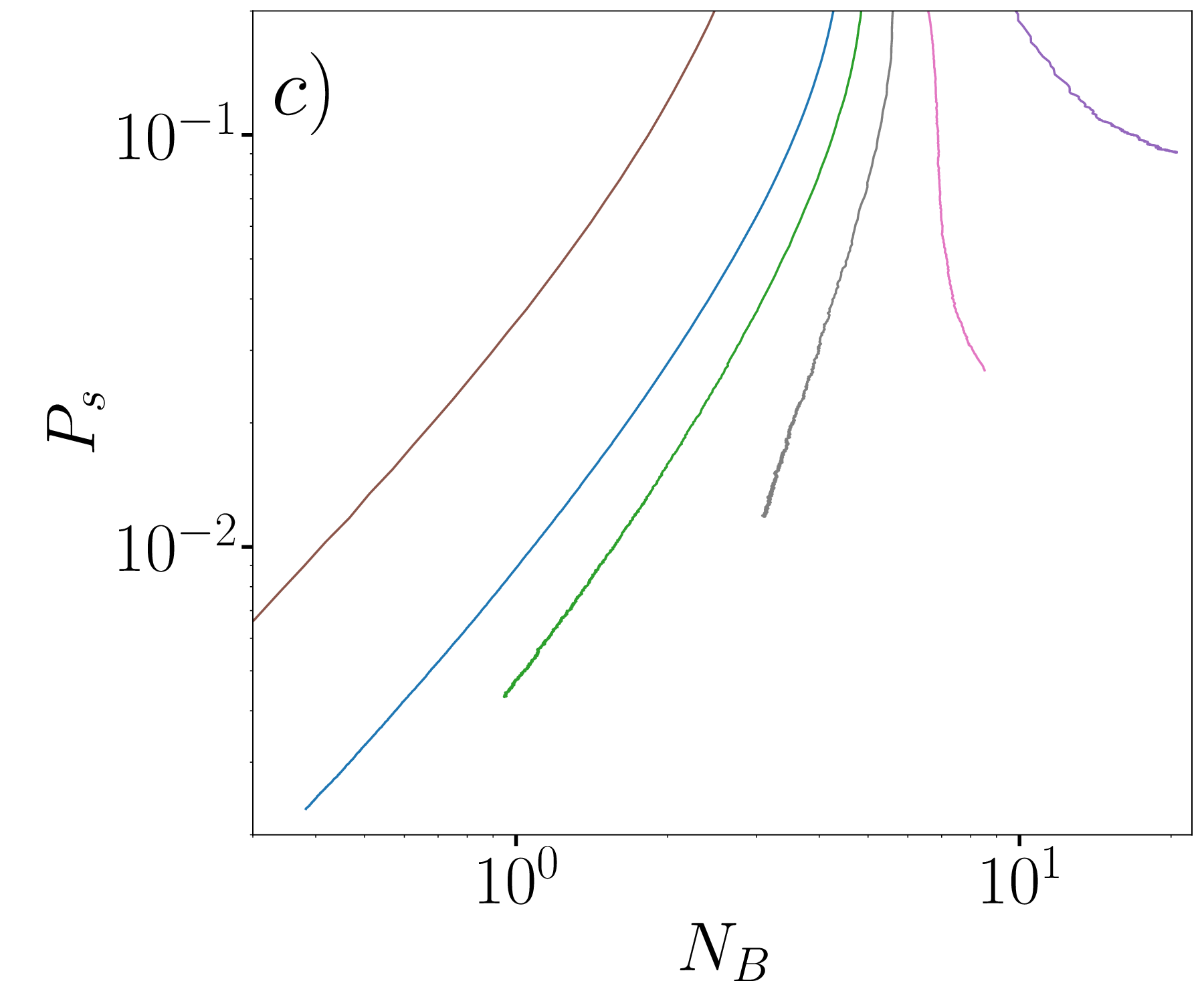}
    \caption{ \(D_A\)-disorder with \(D_A^0=1.4\), \(D_A^1=2.96\), \(p_A=0.05\), \(D_B=1\).  
    Similar legend as for the Up panel Fig.~\ref{fig:DaInfDisorderDblDa}. At least 4000 disorder configurations were used for \(\lambda = 0.255, 0.26, 0.265\), with 64 runs per configuration.}
    \label{fig:DaDisorderDblDa}
\end{figure}

\begin{figure}
    \centering
    \includegraphics[width=0.32\columnwidth]{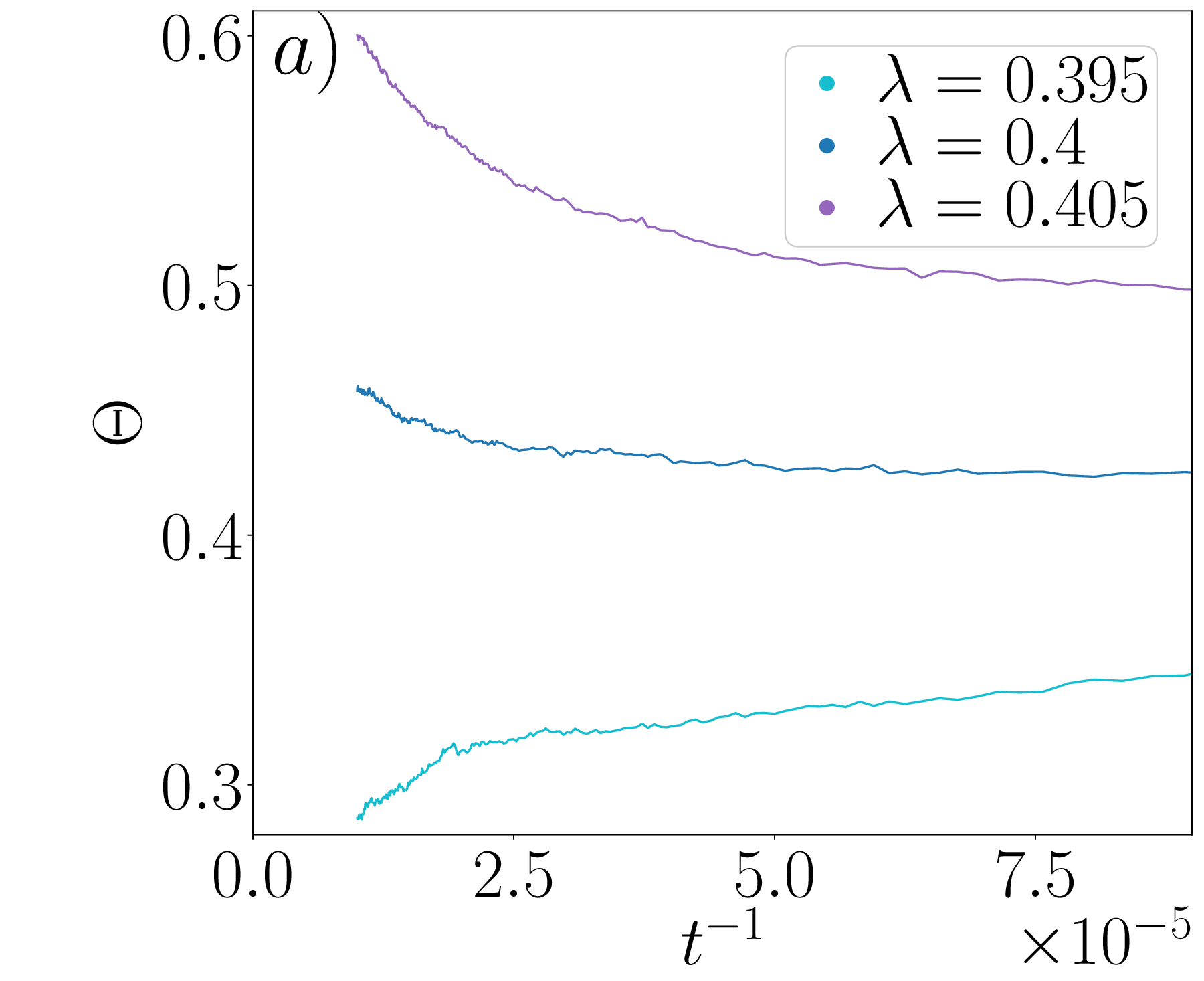}
    \includegraphics[width=0.32\columnwidth]{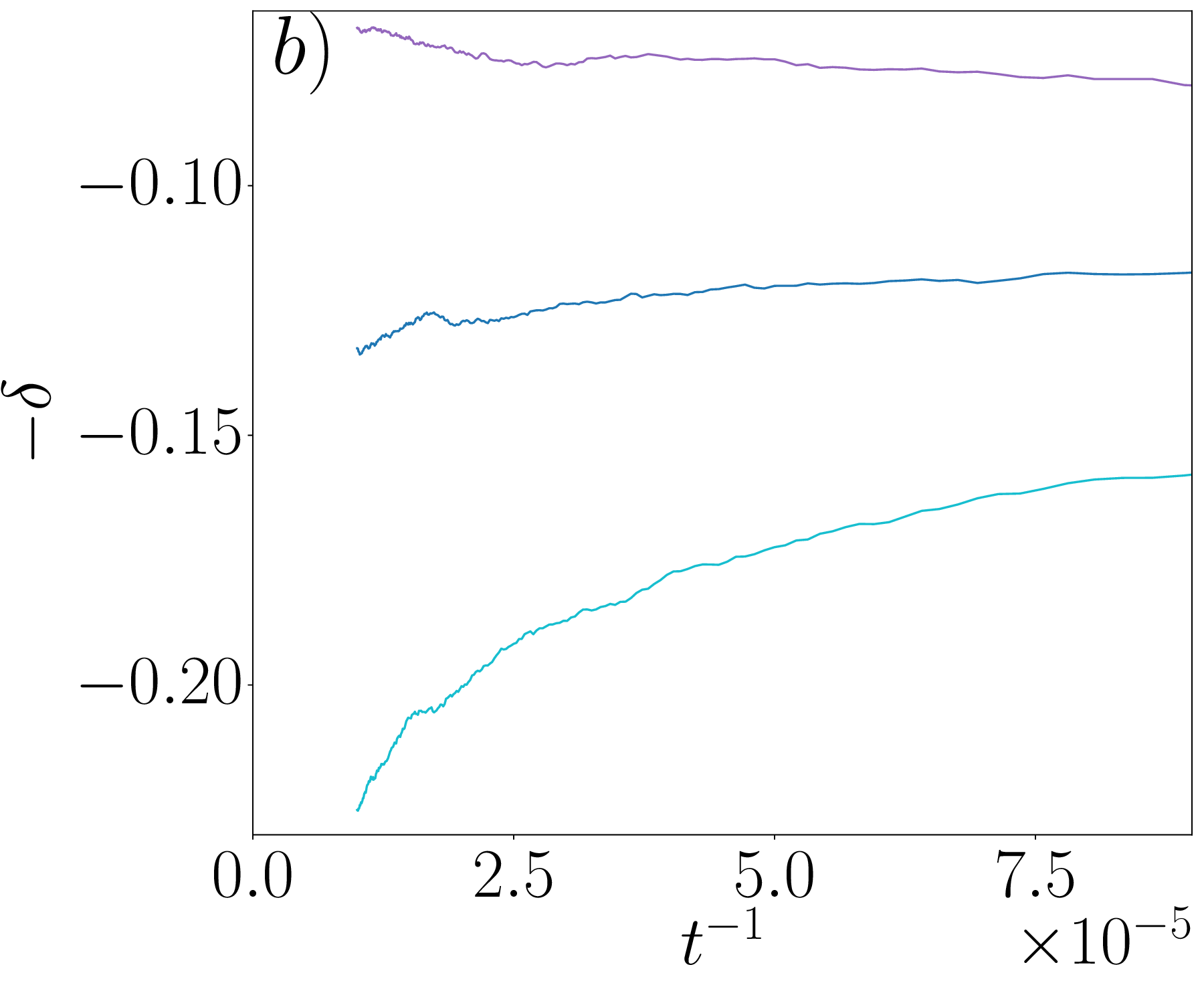}
    \includegraphics[width=0.32\columnwidth]{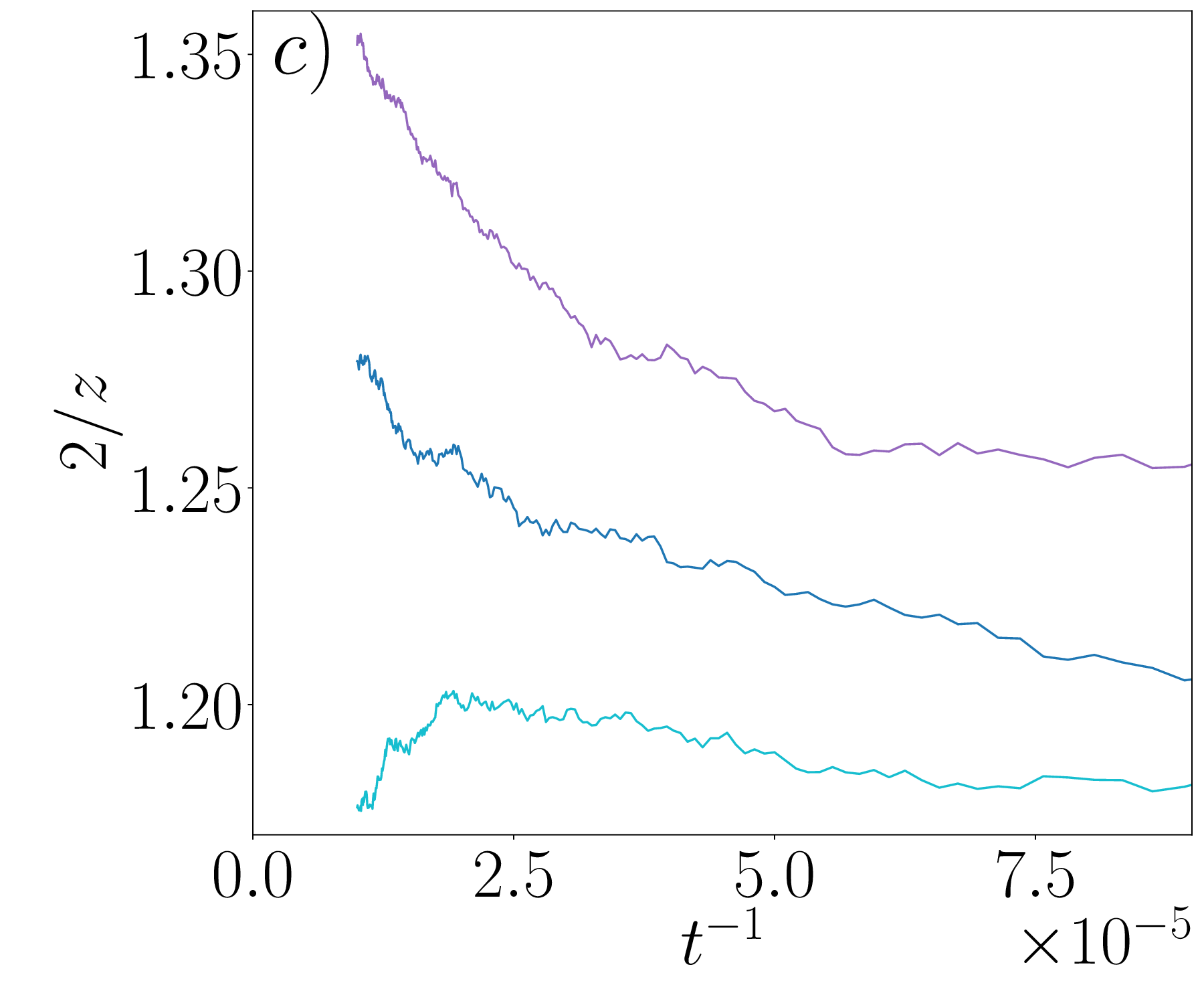}
    \caption{Critical properties of the \(D_A\)-disordered DEP with $D_A^0=0.18, D_A^1=1000, p_A=0.85, D_B=1$. 
    (a-c) Effective critical exponents yielding the estimates \(\lambda_c = 0.400(5)\), \(\Theta = 0.48(4)\), \(\delta = 0.14(2)\), and \(2/z = 1.29(5)\). These estimates are in relatively good agreement with the pure exponents. The observed differences can be attributed to the uncertainty in determining the exact position of the critical point. Additionally, the upward curvature of $\theta(t)$ at late times for $\lambda = 0.4$ suggests that the critical point occurs at a lower contamination rate. $\sim 10^3$ disorder realizations were performed at $\lambda= 0.395 ,0.4 ,0.405$.}
    \label{fig:DaInfDisorderDalDb}
\end{figure}

\begin{figure}
    \centering
    \includegraphics[width=0.32\columnwidth]{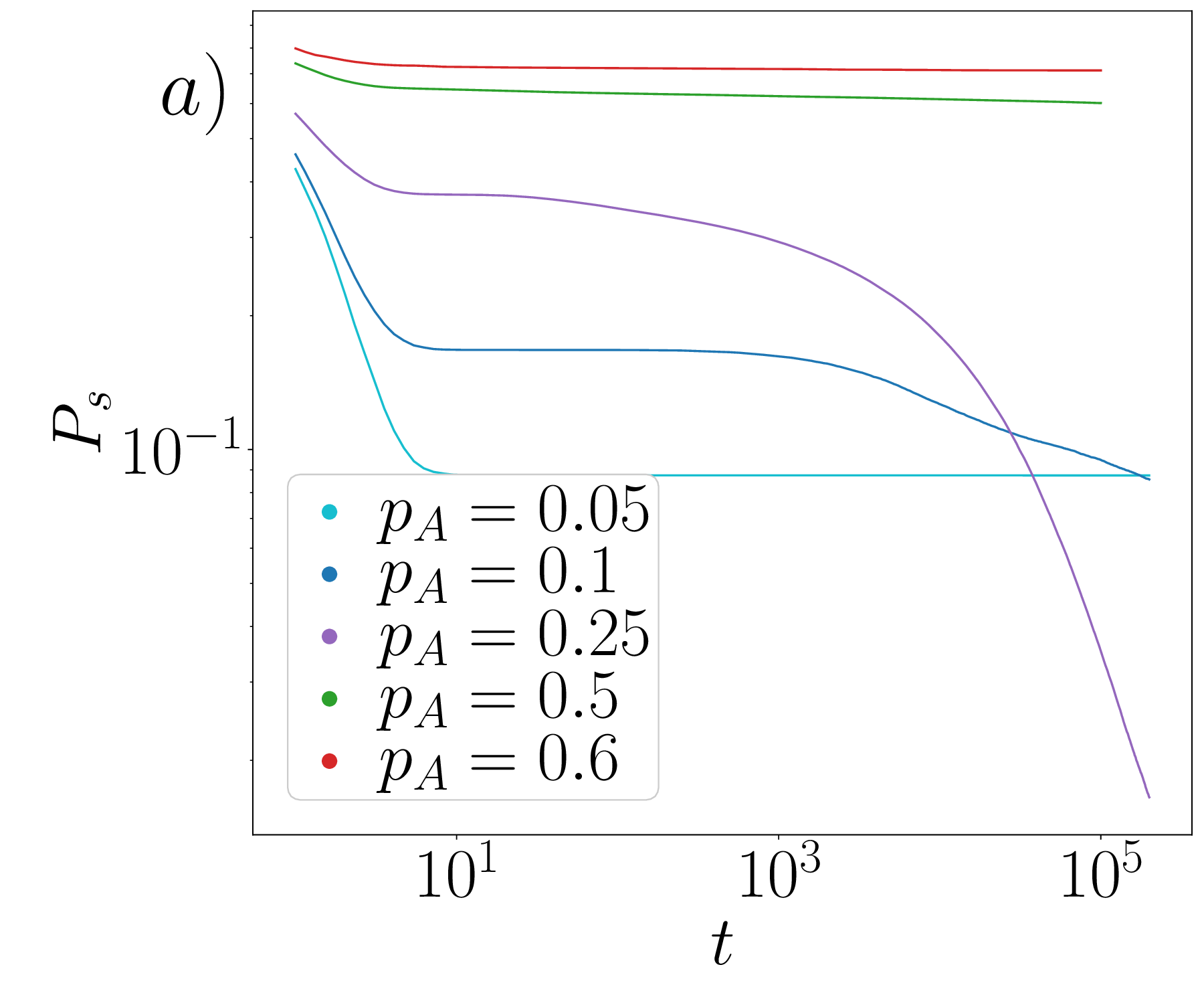}
    \includegraphics[width=0.32\columnwidth]{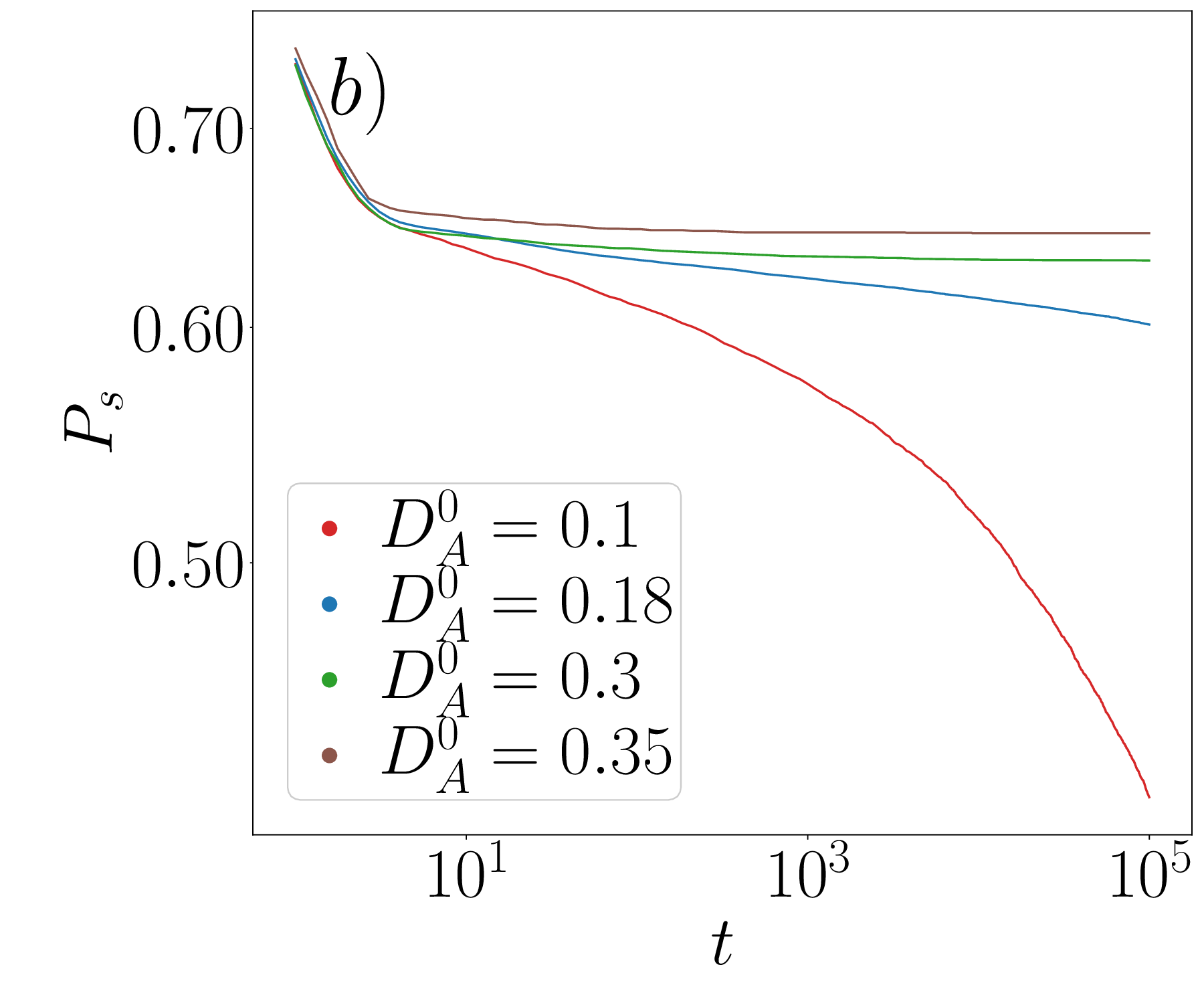}
    \includegraphics[width=0.32\columnwidth]{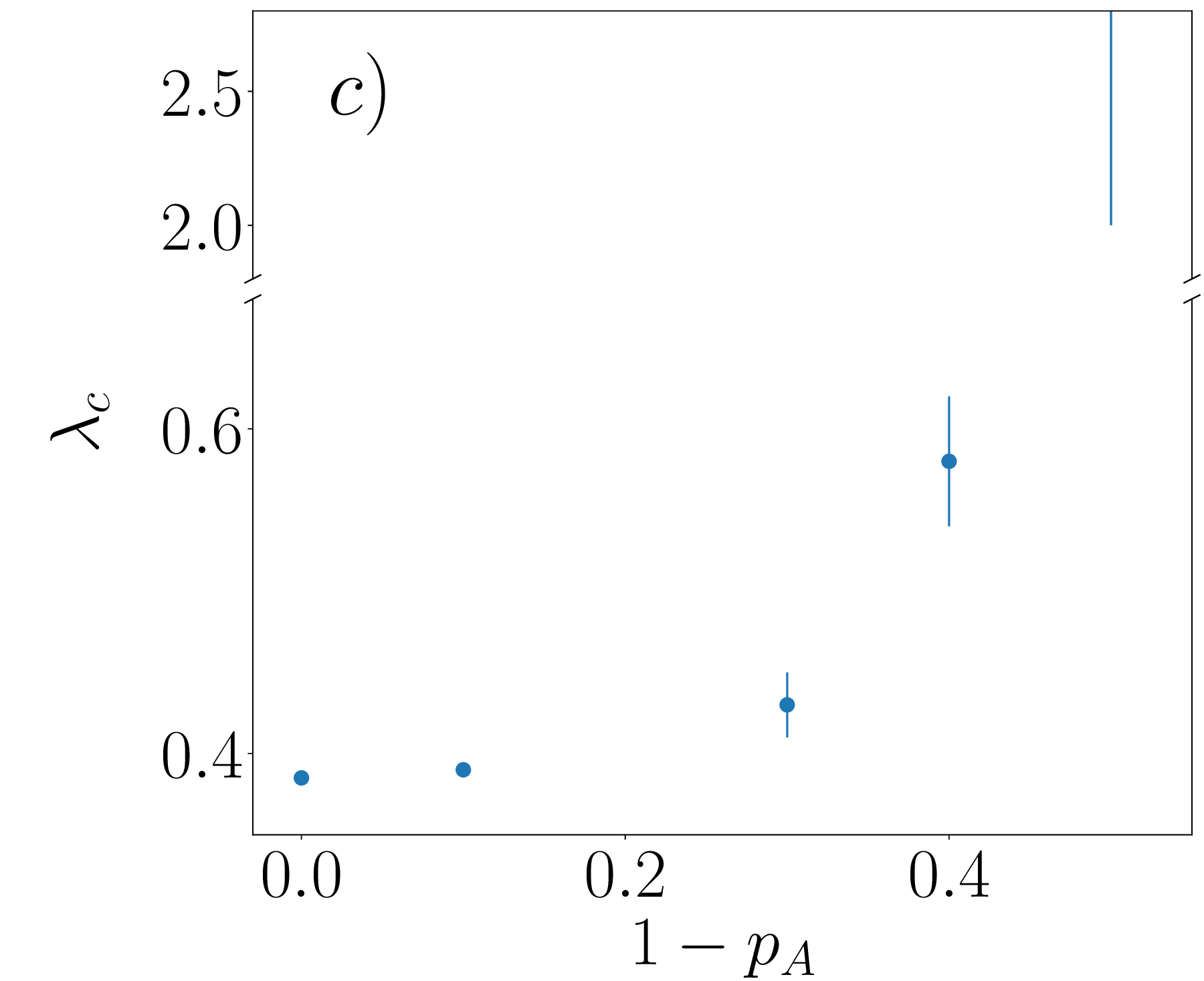}
    \caption{(a-b) Time evolution of the survival probability \(P_s\) for the \(D_A\)-disordered DEP with \(D_A^1 = 5000\), \(D_B = 1\), and \(\lambda = 50\). (a) \(D_A^0\) is fixed at 0.18. For large \(p_A\), \(P_s\) converges to a non-zero value. For intermediate \(p_A\), \(P_s\) continuously decreases up to the maximum time considered. Given the large \(\lambda\), we interpret this decay as a lack of an active phase. For low \(p_A\), the dynamics are too slow due to the high particle density per site, preventing us from observing whether \(P_s\) will ultimately decay or not within the maximum time considered.
    (b) Same as (a), with \(p_A = 0.5\) fixed. When \(D_A^0\) is too small, \(P_s\) continuously decays at late times, which we also interpret as a lack of an active phase.
    (c) Evolution of the critical point \( \lambda_c \) as a function of \( 1 - p_A \), at fixed \( D_A^{\text{eff}} \approx 0.36 \), \( D_A^1 = 1000 \), and \( D_B = 1 \). The value of \( D_A^0(p_A) \) is adjusted to maintain constant \( D_A^{\text{eff}} \) (see Eq.~\ref{eq:effectiveDiffusionRate}).  The critical point \( \lambda_c \) increases sharply with \( 1 - p_A \) around $0.4$, further suggesting the absence of a phase transition at larger values.}
    \label{fig:DaInfDisorderDalDbNoActive}
\end{figure}

Cases with \( D_A^1 \gg 1 \) were used to construct the phase diagram in Fig.~(3) of the main text. Specifically, for \( D_A^{\text{eff}} > D_B \), and especially when \( D_A^0 < D_B < D_A^1 \), the critical properties align with an IDFP scenario [Fig.~\ref{fig:DaInfDisorderDblDa}]. Signature of an IDFP is also found using a smaller disorder variance ($D_A^1/D_A^1 \simeq 1$) [Fig.~\ref{fig:DaDisorderDblDa}].
For \( D_A^{\text{eff}} < D_B \), disorder is not relevant if disorder remains weak (see Fig.~\ref{fig:DaInfDisorderDalDb}). 
When the strength of spatial heterogeneities increases, either by lowering \(p_A\) or \(D_A^0\), the critical point shifts to higher contamination rates, ultimately suppressing the active phase. This is evidenced by the continuous decay of \(P_s\) at high \(\lambda\) (\(\lambda = 50\)) when \(p_A\) or \(D_A^0\) are sufficiently low, as shown in Fig.~\ref{fig:DaInfDisorderDalDbNoActive}.

For lower \( D_A^1 \) (comparable to the other rates), our numerical results remain consistent with previous observations: disorder is relevant for \( D_A^{\text{eff}} > D_B \) and irrelevant for \( D_A^{\text{eff}} < D_B \), see Tab.~\ref{tab:One-speicieDisorder}. But we have not systematically searched for the absence of a transition, which might occur at very low $D_A^0$ or small $p_A$ to create a highly heterogeneous environment. 

\subsection{Upper bound of $R^2$ for $D_A$ disorder}

\begin{figure}
    \centering
        \includegraphics[width=0.31\textwidth]{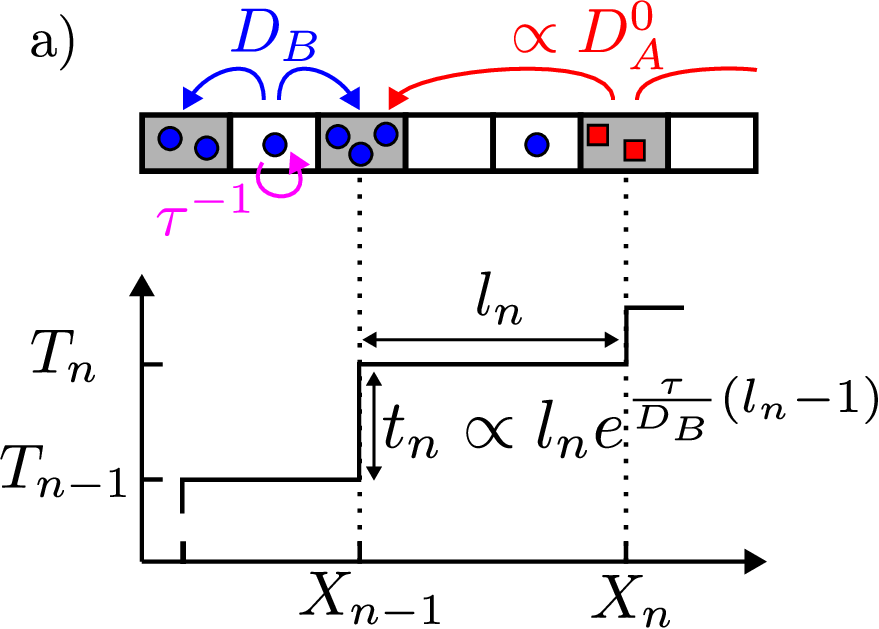}
        \includegraphics[width=0.31\textwidth]{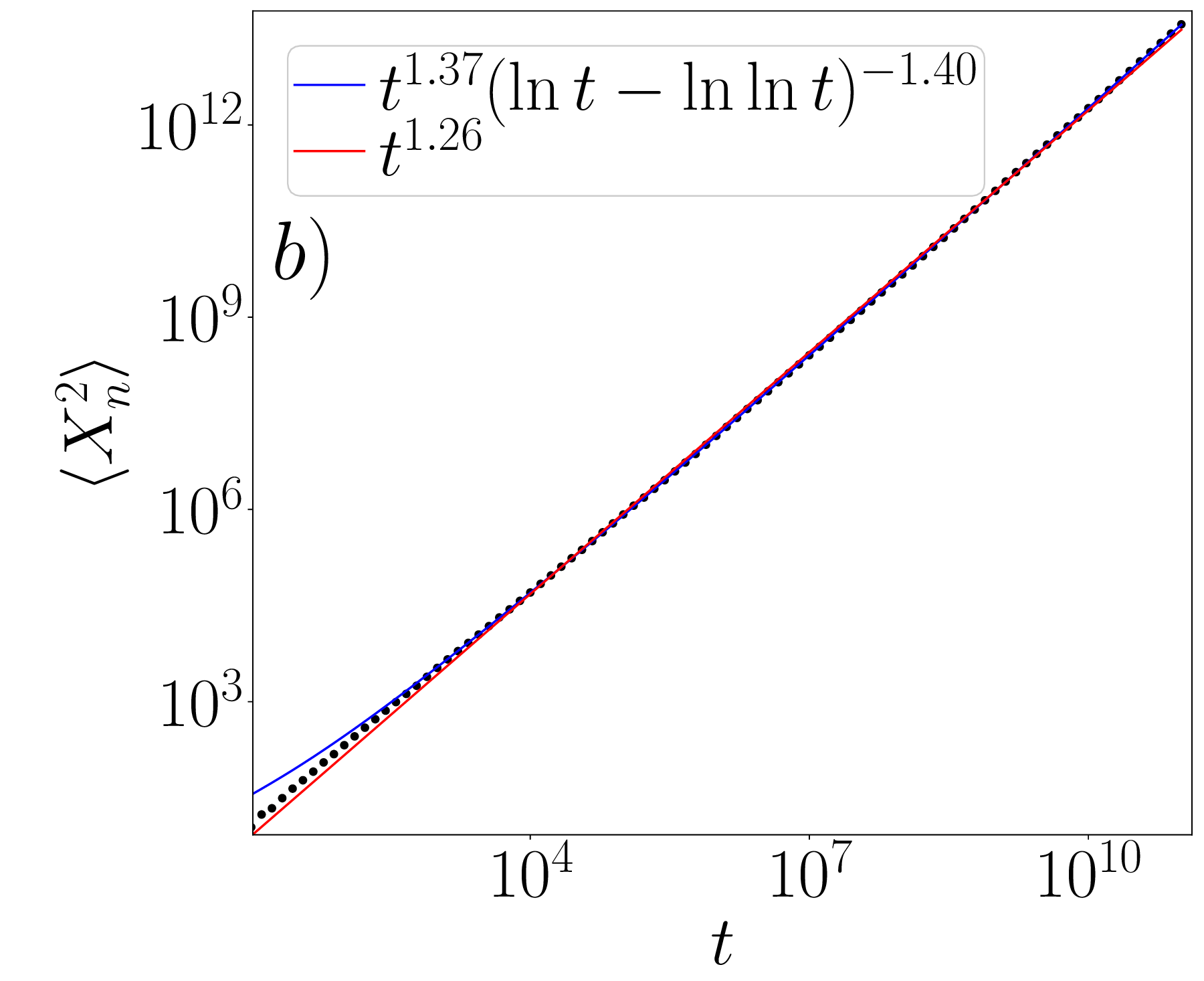}
        \includegraphics[width=0.31\textwidth]{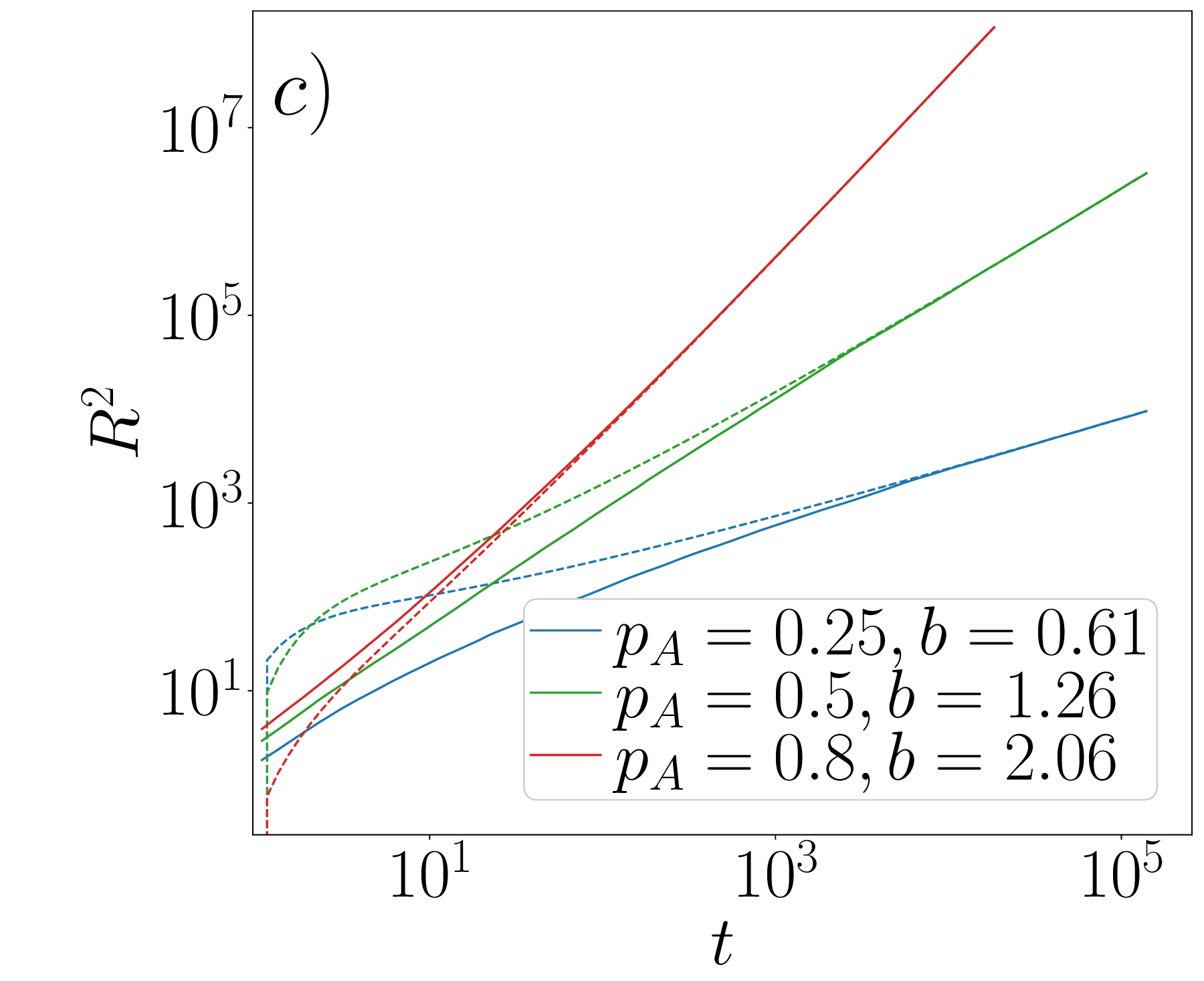}
    \caption{(a) Up: Schematic of the \(D_A\)-disorder DEP dynamics with \(\lambda \gg 1\) and \(D_A^1 \to \infty\), where sites are either unoccupied, occupied by only \(A\) particles (grey regions), or by only \(B\) particles. Down: Mapping to an effective one-particle model, where \(X_{n-1}\) represents the position of the last infected grey site at time \(t\). The distance between grey regions corresponds to the jump distance \(l_n\) that the effective particle must make, with waiting time \(t_n\) analogous to the time before a \(B\) particle reaches the next grey region.  (b) Average over 5000 independent trajectories of \(\langle X_n^2(t) \rangle\), where \(X_n\) and \(t=T_n\) are sum of the random variables $l_n$ drawn from a geometric distribution with parameter $p=1/2$ and $t_n$ from a Geometric distribution with parameter $p_{\text{hops}}(l_n)$, see Eq.~\ref{eq:p_hops_sinai}. Two fits are shown starting from \(t_{\min} = 800\): a simple power-law (red), which poorly fits the data, and a power-law with logarithmic correction (blue), which matches the data better. The exponents obtained from the latter are close to the predicted value of \(-2\ln(1-p) \approx 1.39\).  
    (c) Solid lines represent data from DEP simulations with \(D_A^0 = 1.4\), \(D_A^1 = 2000\), \(D_B = 1\), \(\lambda = 10\), and \(p_A\) as indicated in the legend. Dashed lines show power-law fits with logarithmic correction (Eq.~\ref{eq:powerLawLog}). The exponent \(b\) from the fits agrees reasonably well with the predicted upper bound \(\min(-2\ln(1-p_A), 2)\) corresponding to $2, 1.39, 0.57$ resp. for $p_A=0.8, 0.5, 0.25$. Main difference are attributed to the limited maximum time used \(T_{\max}\). }
    \label{fig:SinaiLike_Xsquare_t}
\end{figure}

The goal of this subsection is to explain the slow behavior observed in the numerical simulations. We consider a \(D_A\)-disorder, where \(D_A^1 = \infty\) and \(D_A^0\) and \(p_A\) are controllable parameters. Under this disorder, healthy particles are confined to slow sites. We assume \(\lambda = \infty\) with \(D_A^1 / \lambda \gg 1\), ensuring that each site at any given time contains either only healthy particles, only infected particles, or no particles. Crucially, we assume (to be discussed later) that once a slow site becomes infected, it remains infected indefinitely, implying the persistent presence of at least two infected particles in an infected slow site. A schematic of this simplified model is shown in Fig.~\ref{fig:SinaiLike_Xsquare_t}a).

The key quantity of interest in this framework is the position of the infection front, $x_f$. The time evolution of $x_f$ is directly related to the time evolution of $R^2$, the mean-square displacement. Specifically, $x_f^2(t)$ provides an upper bound for $R^2(t)$ as $R^2(t) \leq (x_f(t) - x_0)^2$.

To study the scaling behavior of \( x_f(t) \), we focus on the dynamics of a single particle in an inhomogeneous environment (see Fig.~\ref{fig:SinaiLike_Xsquare_t}a)). This particle hops to the next slow site, located at a distance \( l_n \), in \( t_n \) seconds. The distance \( l_n \), representing the separation between two consecutive slow sites, is independently drawn from a geometric distribution:

\[
P_1(X = u) = (1 - p)^{u - 1} p, \quad u \geq 1,
\]

where \( p \) denotes the probability of encountering a slow site.

At each step, the particle has a probability \( p_{\text{hops}}(l_n) \) of reaching the next slow site at a distance \( l_n \). The time \( t_n \), defined as the number of attempts before a success, required for this hop is independently drawn from another geometric distribution:

\[
P_2(X = u) = (1 - p_{\text{hops}})^{u - 1} p_{\text{hops}}, \quad u \geq 1,
\]

where \( p_{\text{hops}} \) represents the probability that an infected particle will reach the next slow site by simple diffusion before recovering.

To determine \( p_{\text{hops}} \), we use results from random walk  \cite{feller_introduction_1968}. In this model, a random walker starts at \( x = 1 \) between two absorbing sites located at \( x = 0 \) and \( x = l_n \). The expected number of steps required to reach an absorbing site is \( l_n - 1 \), and the probability of reaching \( x = l_n \) is \( 1 / l_n \). Each step takes \( 1 / D_B \) seconds on average, while the recovery rate is \( 1 / \tau \). Consequently, the probability that the particle remains infected after \( (l_n - 1) / D_B \) seconds is

\[
\exp\left(-\frac{\tau (l_n - 1)}{D_B}\right).
\]

Thus, the probability that an infected particle reaches the next slow site is:

\begin{equation}
    p_{\text{hops}}(l_n) = \frac{1}{l_n} e^{-k (l_n - 1)},
    \label{eq:p_hops_sinai}
\end{equation}
where \( k = \tau / D_B \).

In the context of the actual dynamics in the Diffusive Epidemic Process (DEP), we assumed that the effective particle attempts a hop at every time step. In reality, this hopping frequency would depend on the density of infected particles on the last infected slow site. However, since we assumed the density to be finite, constant (at least on average) and always greater than two, this approximation does not affect the scaling behavior. Furthermore, we neglected the time required for an infected particle to reach the next slow site. This simplification is justified by the exponential decay of the hopping probability with \( l_n \), which dominates the quadratic growth of the average number of steps with \( l_n \).

The position of the front \( x_f=X_n \) and the actual time $T_N$ are given by the sum of random variables:
\begin{equation}
    X_N = \sum_{n=1}^N l_n, \quad T_N = \sum_{n=1}^N t_n,
    \label{eq:EffectiveTrajectory}
\end{equation}

The average jump length \( \langle l_n \rangle = \langle l_0 \rangle = 1/p \) is always finite. However, for the average time duration \( \langle t_n \rangle =  \langle t_0 \rangle \), we calculate:

\begin{align*}
    \sum_{t=1}^{\infty} t P_2(X=t) &= \sum_{t=1}^{\infty} \sum_{n=1}^{\infty} t(1-p)^{n-1}p \left(1-\frac{1}{n}e^{-k(n-1)}\right)^{t-1} \frac{1}{n} e^{-k(n-1)} \\
    &= \sum_{n=1}^{\infty} np e^{(n-1)[k + \ln(1-p)]} \\
    &= \frac{p}{\left(1 - e^{k + \ln(1-p)}\right)^2}.
\end{align*}

The interchange of the two sums to obtain the second line is valid only if the series converges, i.e. if \( k + \ln(1-p) < 0 \); otherwise, it diverges. 

When \( \langle t_0 \rangle \) is finite, then \(\langle T_n \rangle = n \langle t_0 \rangle\) and \(\langle X_n^2 \rangle \sim n^2 \langle l_0 \rangle^2\) at large $n$, so that:
\[
    \langle X_n^2 \rangle \sim \langle T_n \rangle^2.
\]

If the mean is infinite, we expect the value \( T_n = \sum_{i=1}^n t_i \) to be governed by its largest term. This largest value is likely correlated with the largest jump \( l_i \) observed in \( X_n = \sum_{i=1}^n l_i \). To estimate this, we calculate the expected maximum value \( l_c(n) = \mathbb{E}[\max(l_1, \dots, l_n)] \), which is given by \cite{eisenberg_expectation_2008}:
\[
    l_c(n) = \sum_{i=1}^n \binom{n}{i}(-1)^{i+1} \frac{1}{1 - (1-p)^i}.
\]

For large \( n \), an approximation of \( l_c(n) \) is \cite{eisenberg_expectation_2008}:
\[
    l_c(n) \approx -\frac{1}{\ln(1-p)} H_n + \frac{1}{2},
\]
where \( H_n \) is the \( n \)-th harmonic number, \( H_n = \sum_{k=1}^n \frac{1}{k} \). Using the asymptotic expansion of \( H_n \), we obtain:
\[
    l_c(n) \approx -\frac{1}{\ln(1-p)}\left[\ln n + \gamma + \frac{1}{2n} - \frac{1}{12n^2} + \frac{1}{120n^4} - \cdots \right] + \frac{1}{2},
\]
where \( \gamma \) is the Euler-Mascheroni constant. Thus, the asymptotic behavior is:
\begin{equation}
    l_c(n) \sim -\frac{\ln n}{\ln(1-p)}.
    \label{eq:expectedSize}
\end{equation}

Next, we estimate \( T_n \) using the largest value \( l_c(n) \):
\begin{equation}
    T_n = \sum_{i=1}^n t_i \approx t_k(l_c(n)) = l_c(n) e^{k(l_c(n)-1)},
\end{equation}
where we assume that the expected value of \( t_k(l_c(n)) \) is approximately \( 1/\tilde{p} \), with \( \tilde{p} = \frac{1}{l_c(n)} e^{-k(l_c(n)-1)} \).

Lastly, to estimate the infinite average \( \langle T_n \rangle \), we assume that the typical value of the sum is dominated by its largest term and find
\begin{align}
    \langle X_n^2 \rangle &\sim \frac{n^2}{p^2}, \\
    \langle T_n \rangle &\sim -\frac{\ln n}{\ln(1-p)} e^k n^{-k/\ln(1-p)}.
    \label{eq:th_data_model}
\end{align}

To express \( n \) as a function of \( \langle T_n \rangle \equiv t \), we define \( \alpha = -k/\ln(1-p) \) and invert the relation \( t = n^\alpha \ln n \). This yields:
\begin{equation}
    n = \left(\frac{\alpha t}{W(\alpha t)}\right)^{1/\alpha},
\end{equation}
where \( W(z) \) is the Lambert W function. For large \( z \), \( W(z) \approx \ln z - \ln \ln z \). Substituting this asymptotic form, we approximate:
\begin{equation}
    n \approx \frac{t^{1/\alpha}}{(\ln t - \ln \ln t)^{1/\alpha}},
\end{equation}
which leads to:
\begin{equation}
    \langle X_n^2 \rangle \sim \left[\frac{t}{\ln t - \ln \ln t}\right]^{-2\ln(1-p)/k}.
    \label{eq:th_model_scaling}
\end{equation}

Finally, combining the results for \( k + \ln(1-p) < 0 \) and \( k + \ln(1-p) > 0 \), we arrive at:
\begin{equation}
    \langle x_{f}^2 \rangle \sim 
    \begin{cases} 
      t^2 & \text{if } k + \ln(1-p) < 0, \\ 
      \left[\frac{t}{\ln t - \ln \ln t}\right]^{-2\ln(1-p)/k} & \text{if } k + \ln(1-p) > 0.
   \end{cases}
   \label{eq:UpperBoundRsquare}
\end{equation}

In Fig.~\ref{fig:SinaiLike_Xsquare_t}b), we present data from 5000 trajectories simulated up to a maximum time of \( T_{\max} = 10^{11} \). Each trajectory consists of a sequence \((l_1, t_1), \dots, (l_n, t_n)\) such that \( T_n = \sum_{i=1}^n t_i > T_{\max} \). For each jump, \( l_i \) is drawn independently from a geometric distribution with parameter \( p = 0.5 \), and \( t_i \) is sampled from another geometric distribution with parameter \( p_{\text{escape}} = \frac{1}{l_i} e^{-k(l_i-1)} \) and $k=1$. 

Each data point represents the average squared position, \(\left(\sum_{i=0}^{n_c} l_i\right)^2\), over all trajectories at specific times \( t \), where \( n_c \) is the largest index satisfying \( \sum_{i=1}^{n_c} t_i \leq t < \sum_{i=1}^{n_c+1} t_i \).

We performed two fits to the data: one using a simple power law, \( f_1(t) = a t^{-b} \), and another incorporating a logarithmic correction:
\begin{equation}
    f_2(t) = a t^{-b} (\ln t - \ln \ln t)^c.
    \label{eq:powerLawLog}
\end{equation}
To ensure consistency with the analytical prediction \( b \approx -c \) for \( f_2(t) \), we tuned \( t_{\min} \), the minimum time used in the fits. Setting \( t_{\min} = 800 \), we obtained \( a \approx 1.37 \) (blue line), which aligns well with the predicted exponent \( -2\ln(1-p) \approx 1.39 \). In contrast, the fit using a simple power law (red line) does not adequately capture the data.

To verify if this upper bound aligns with DEP data, we use \( D_A^0 = 1.4 \), \( D_A^1 = 2000 \), \( D_B = 1 \), \( \lambda = 10 \), and vary \( p_A \). The mean-square displacement \( R^2 \) (solid lines) is shown in Fig.~\ref{fig:SinaiLike_Xsquare_t}c). The predicted upper bounds for \( p_A = 0.25 \), \( 0.5 \), and \( 0.8 \) are \( 0.57 \), \( 1.39 \), and \( 2 \), respectively. Fitting Eq.~\ref{eq:powerLawLog} yields \( b \approx 0.61 \), \( 1.26 \), and \( 2.06 \), in good agreement with the predictions. \( t_{\min} \) was tuned to minimize the discrepancy between \( b \) and \( -c \) in the fit. For all \( p_A \) values, \( t_{\min} \) was large, leaving only a few points for fitting.

As a result, the critical exponent \(z\) satisfies \(z > -\frac{\tau}{D_B \ln(1-p)}\) when \(\frac{\tau}{D_B} + \ln(1-p) > 0\). This bound depends explicitly on the disorder distribution (\(p\)) and diverges as \(p \to 0\). Hence, the critical properties must differ from those of the pure system when the disorder is sufficiently strong (corresponding here to small \(p\)). However, this analytical result does not provide information about possible deviations from the pure critical behavior at weaker disorder.

\subsection{Absence of active phase for $D_A$ disorder}
\label{appendix:NoTransition}

We revisit the assumption that an infected slow site remains indefinitely infected. For \( \lambda \gg 1 \), recovery occurs only when the particle count drops below two, governed by density fluctuations. Such an event is more likely when the average density is low. If the infected region remains stable (i.e., does not grow) for long enough, the average infected particle density should satisfy \( \rho_B D_B = \rho_A D_A^0 \). Thus, a larger ratio \( D_B/D_A^0 \) leads to a lower \( \rho_B \), increasing the likelihood of site recovery.  
From Eq.~\ref{eq:expectedSize}, the expected distance between the last infected slow site and its healthy neighbor strongly depends on \( p_A \), leading to an exponential dependence on the infection time, whereas \( D_A^0 \) has only a linear effect (through the number of infection attempts per unit time). However, \( D_A^0 \) significantly influences the survival time of infected clusters. To estimate this effect, consider \( N \) particles diffusing between two sites with rates \( D_1 \) and \( D_2 \). The extinction time for site 1, initially containing one particle, follows the stochastic logistic model \cite{norden_distribution_1982}:
\begin{equation}
    \tau_1 = \frac{1}{D_2} \frac{(D_2/D_1 + 1)^N - 1}{N}.
\end{equation}
Applying this result:
\begin{itemize}
    \item Inside an infected region (\( D_1 = D_2 = D_B \), \( N = N_B \)), the recovery time depends exponentially on \( N \), thus decreasing exponentially with \( D_A^0 \) and \( 1/p_A \).
    \item At the interface between infected and healthy regions (\( D_1 = D_B \), \( D_2 = D_A^0 \), \( N = N_{\text{tot}} \)), the recovery time decreases exponentially with \( D_A^0 \) when \( D_A^0 < D_B \). And \(N\) also decreases with \( D_A^0 \) (if $D_A^0<D_B$) which further reduces the recovery time. 
\end{itemize}
Fixing \( p_A \), this simplified model suggests an exponential dependence of recovery time on \( D_A^0 \), implying a critical value \( D_{A,c}^0(p_A) \) below which infected sites recover before the infection can spread further, thereby suppressing the active phase. If \( \tau/D_B + \ln(1-p_A) < 0 \), the equilibrium assumption breaks down, and this conclusion no longer hold.  

For fixed \( D_A^0 \), the dependence on \( p_A \) is more intricate, as \( p_A \) strongly affects both the survival time of infected sites and the time required to infect the next site. Further investigation is needed to determine whether lowering \( p_A \) further can restore the active phase (see Fig.~\ref{fig:DaInfDisorderDalDbNoActive}).  

\begin{figure}
    \centering
    \includegraphics[width=0.5\linewidth]{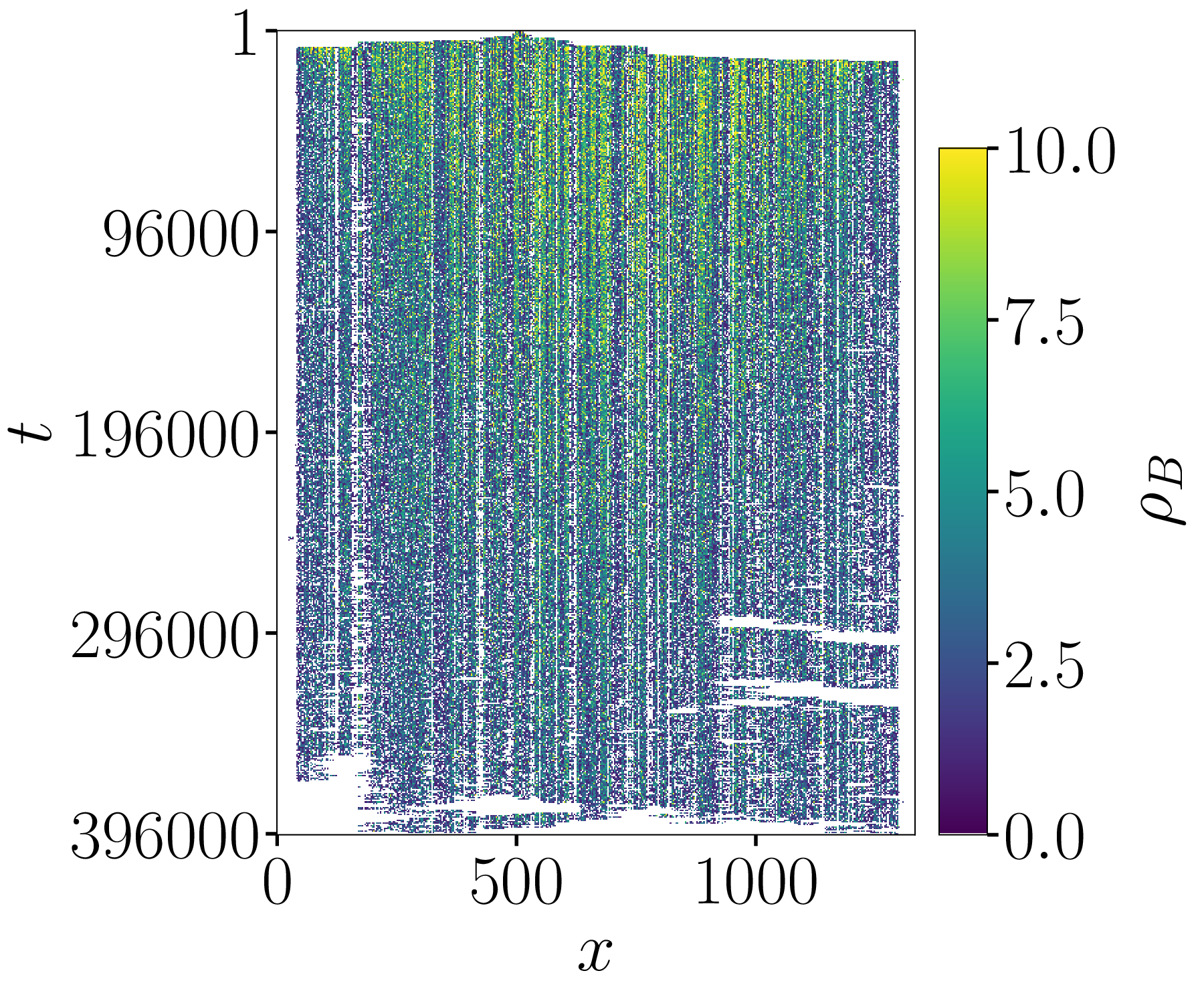}
    \caption{Temporal evolution (kymograph) of the infected-particle density $\rho_B$ for $\lambda=50$ and $D_A^{\mathrm{eff}}<D_B$, starting from a single infected seed. Following an initial rapid spreading driven by the large infection rate, propagation halts due to the presence of extended low-density regions. The spatially averaged density $\rho_B$ then decreases over time, consistent with flux conservation when $D_B>D_A^0$. At later times, spontaneous recoveries occur despite the large value of $\lambda$, as isolated infected particles increasingly occupy sites, ultimately leading to complete extinction of activity. Other parameters: $D_A^0=0.1$, $D_A^1=2000$, $p_A=0.5$, $D_B=1$, and $\rho=5$.}
    \label{fig:Sup_Kymographe}
\end{figure}

In Fig.~\ref{fig:Sup_Kymographe}, we show the kymograph from which Fig.~3(c) of the main text is extracted. It illustrates the competition between the time required for infected particles to reach the next dense region and the time needed for the local particle density to equilibrate.

In the limit $\lambda \gg 1$, flux balance on locally equilibrated sites imposes \(D_B \rho_B^{\infty} = D_A^0 \rho_A^{\infty},\)where $\rho_B^{\infty}$ ($\rho_A^{\infty}$) denotes the asymptotic density on equilibrated infected (healthy) sites. When $D_B > D_A^0$, this condition allows $\rho_B^{\infty}$ to become arbitrarily small. If spreading is halted long enough for local equilibration to occur, the system can therefore be driven toward extinction even at large $\lambda$.

Spontaneous recovery events can even occur earlier than expected due to a non-equilibrium mechanism involving correlations between neighboring sites. In the kymograph, this appears as long white bands inside the infected regions, \emph{e.g.} areas previously occupied by infected particles, emerging around $t \approx 3 \times 10^{5}$. The mechanism is as follows: when an infected site recovers, its diffusion coefficient switches from $D_B$ to $D_A^0$, abruptly reducing the outgoing particle flux. This induces a momentary but stronger-than-equilibrium density reduction on neighboring sites, before their densities relax back toward the new equilibrated value. Such transient density dips increase the likelihood of recovery of neighboring sites, potentially triggering a cascade of recoveries that propagates spatially. This avalanche-like process gives rise to the extended white bands observed in the kymograph.

\subsection{$D_B$ disorder}

As with \( D_A \)-disorder, we found \( D_B \)-disorder to be irrelevant when \( D_A < D_B^{\text{eff}} \). While we observed no change in behavior for stronger disorder, a detailed investigation was not conducted.  

\begin{figure}
    \centering
    \includegraphics[width=0.32\columnwidth]{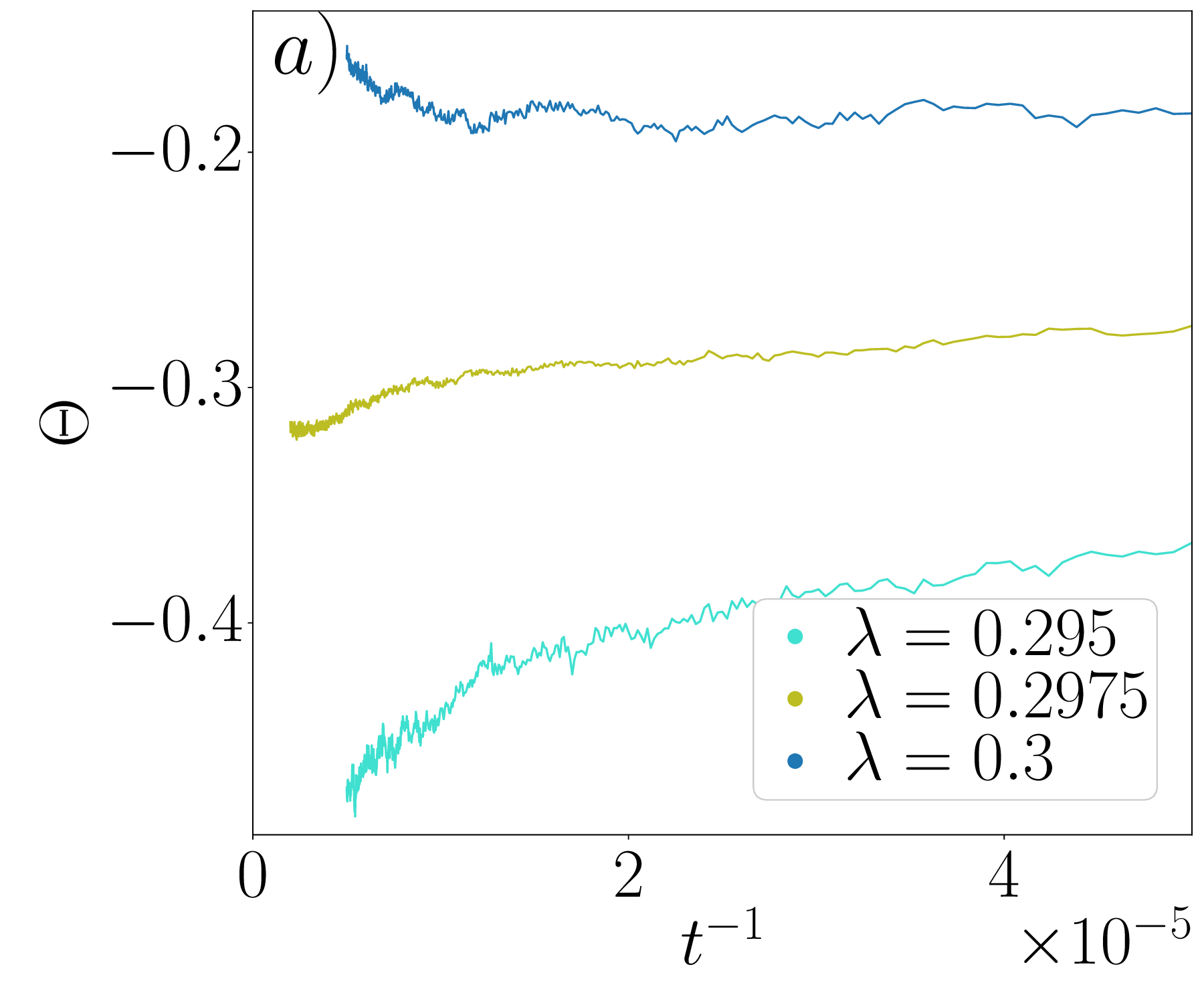}
    \includegraphics[width=0.32\columnwidth]{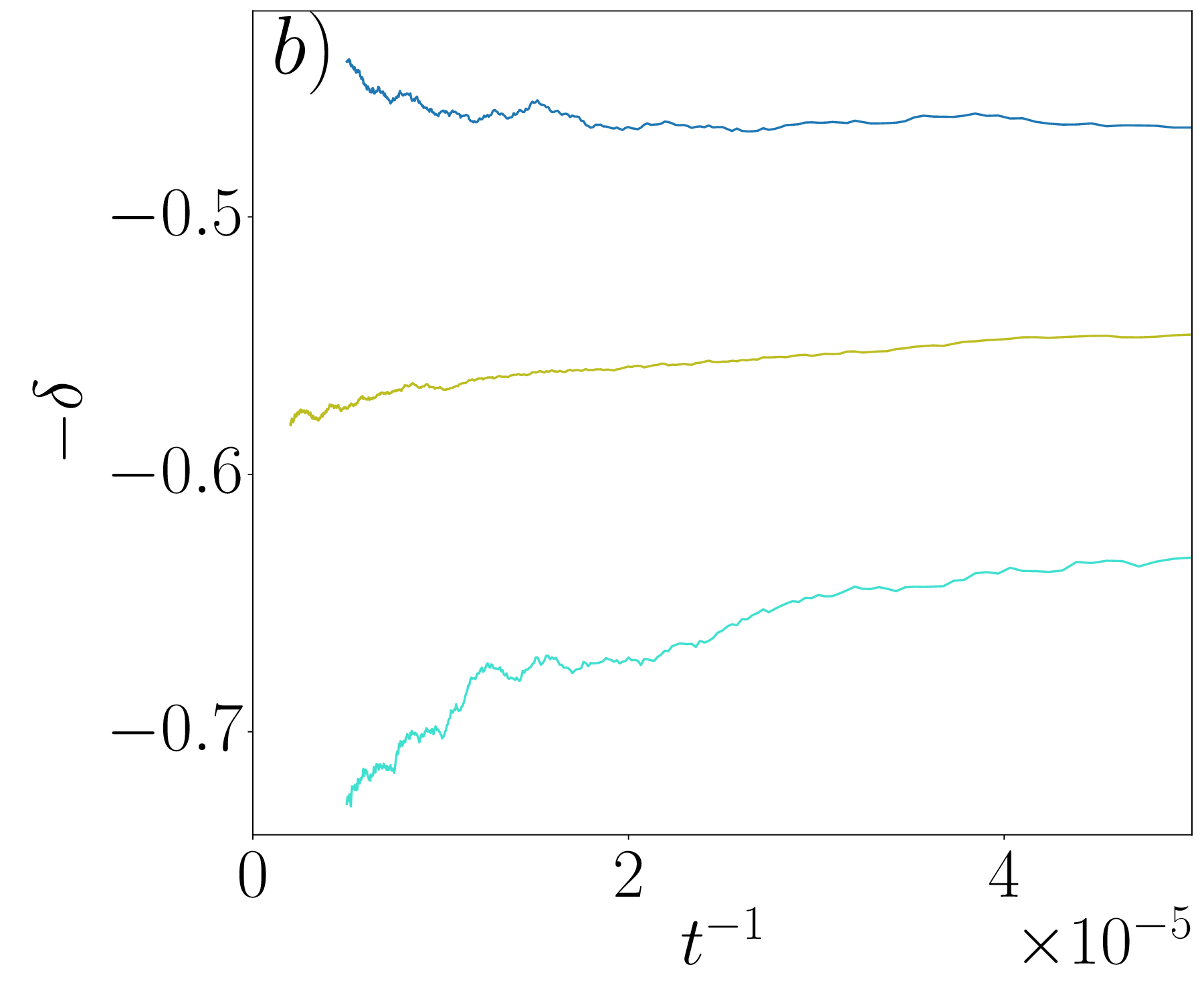}
    \includegraphics[width=0.32\columnwidth]{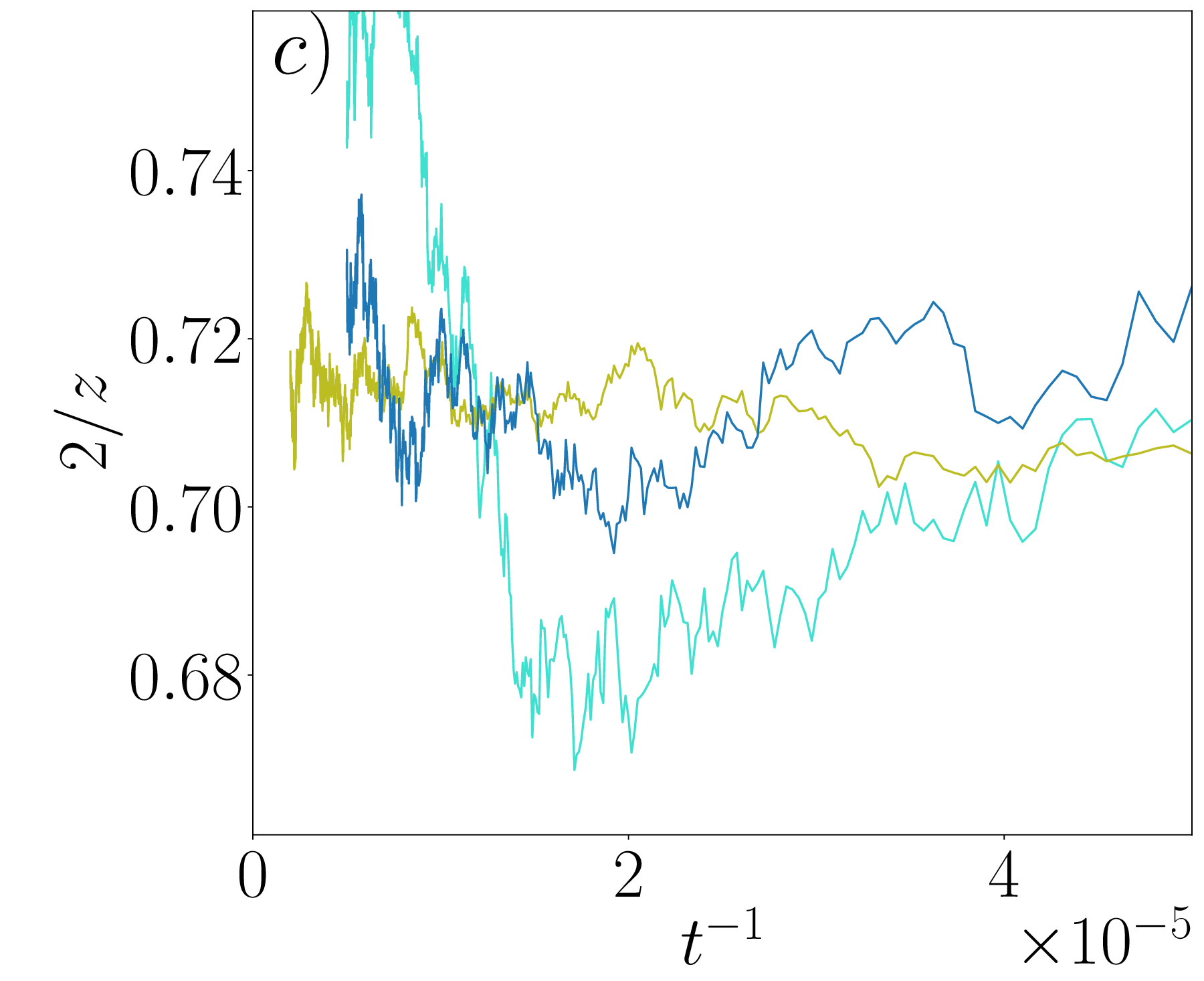}
    \includegraphics[width=0.32\columnwidth]{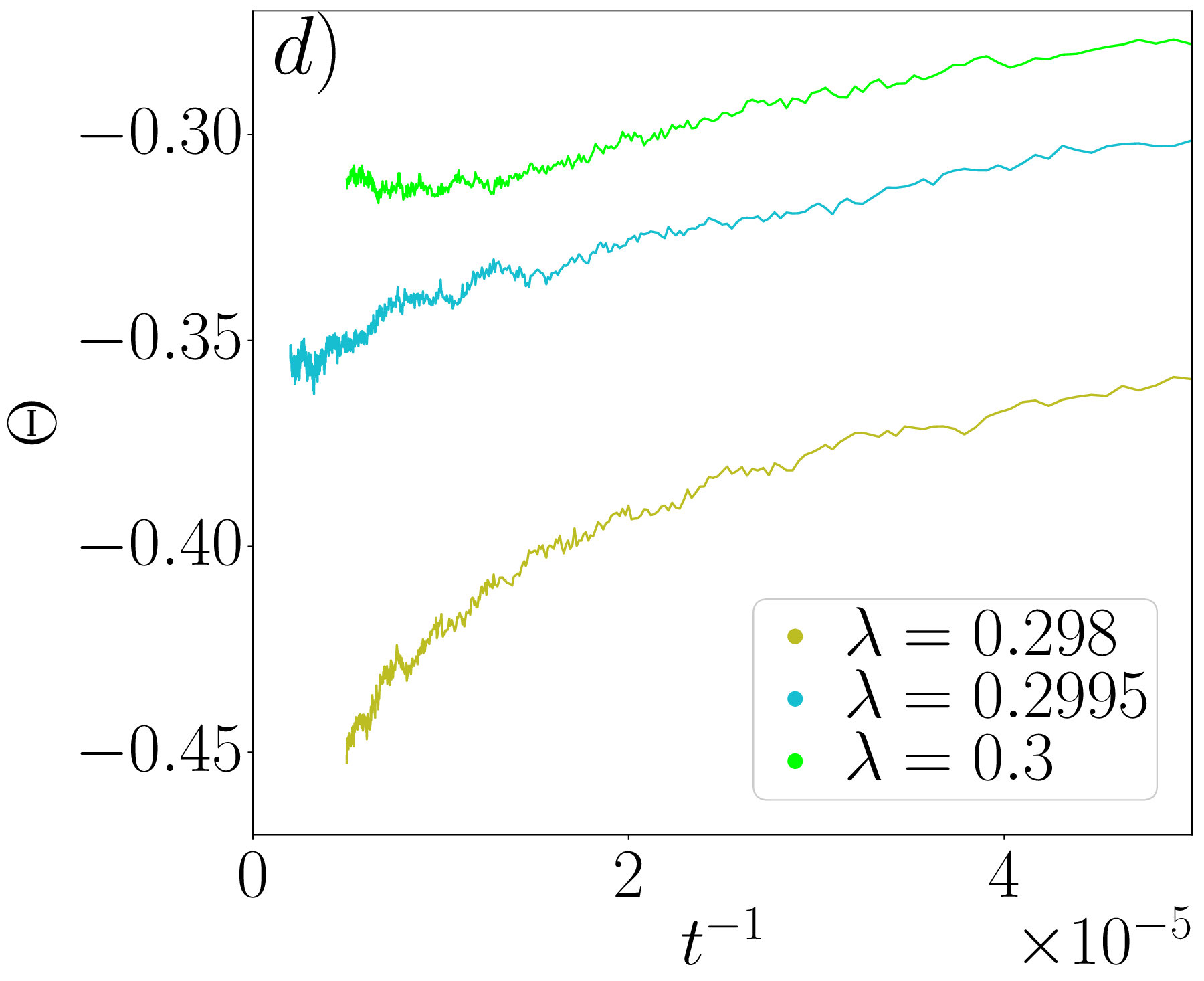}
    \includegraphics[width=0.32\columnwidth]{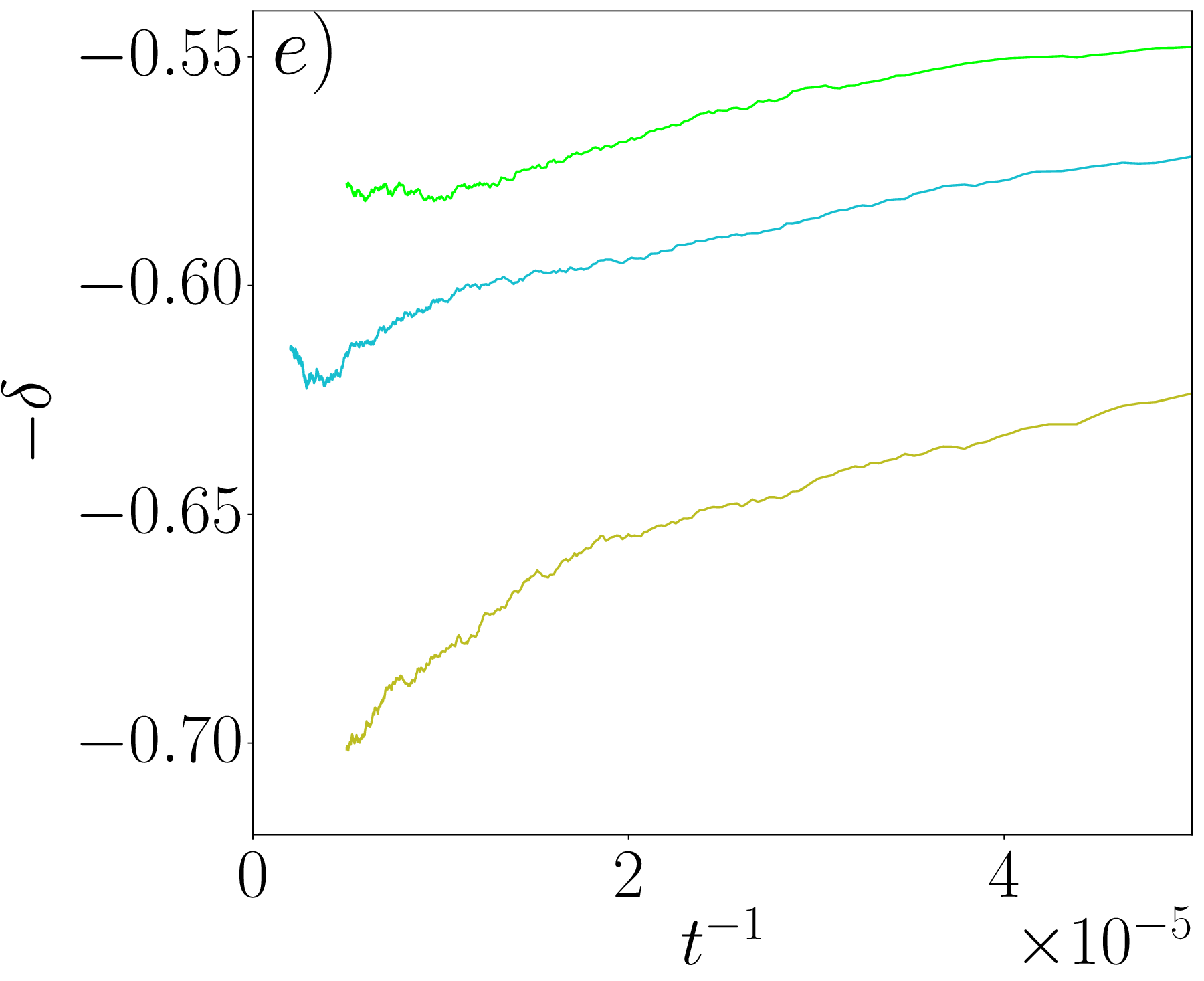}
    \includegraphics[width=0.32\columnwidth]{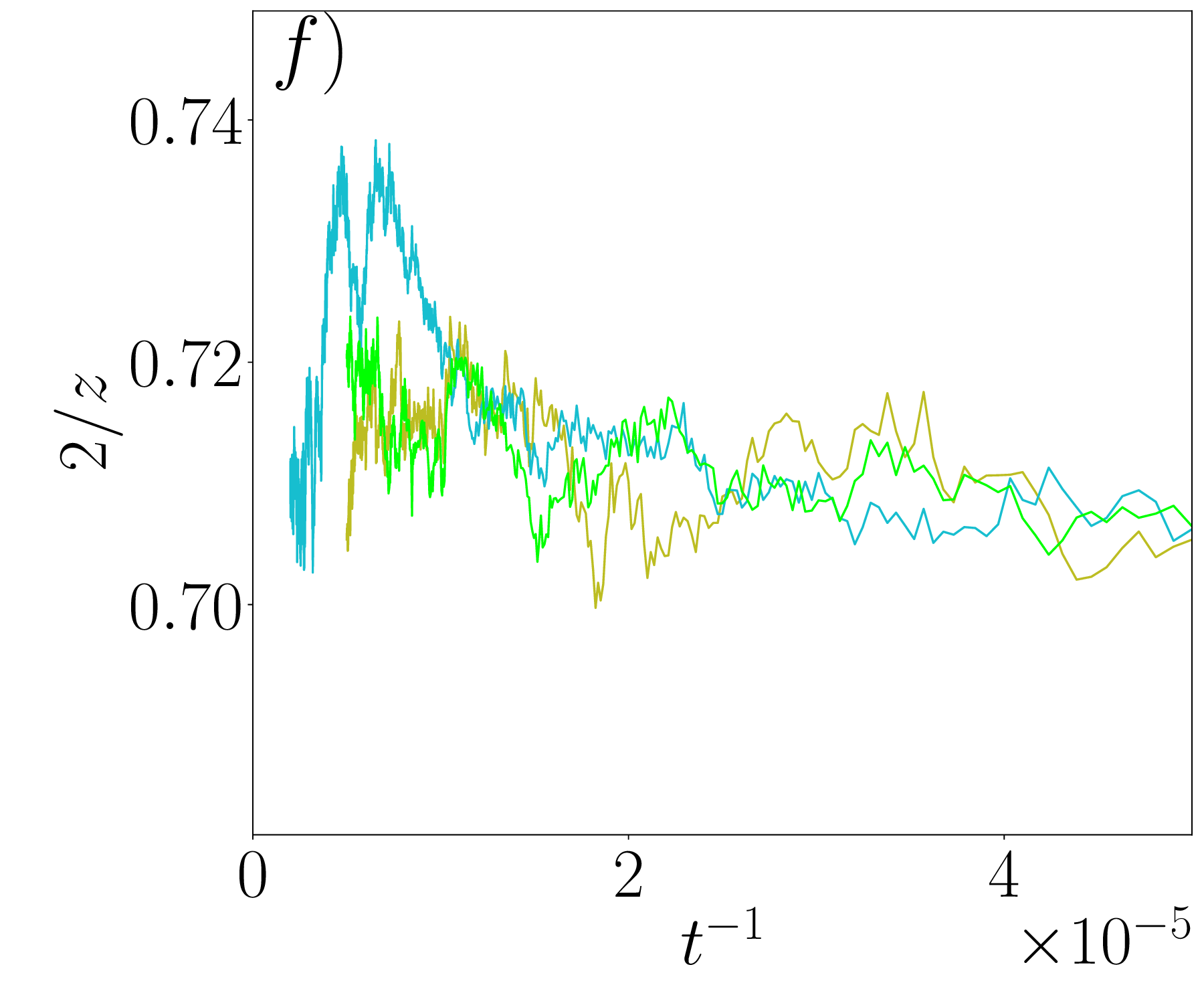}
    \caption{Critical properties of \(D_B\)-disordered DEP with \(0.376 \simeq D_B^{\text{eff}} < D_A = 1\).  (a-c) Using \(D_B^0 = 0.25, D_B^1=6\) and \(p_{B} = 0.65\). Effective critical exponents yielding the estimates \(\lambda_c = 0.2975(25)\), \(\Theta = -0.32(3)\), \(\delta = 0.58(3)\), and \(2/z = 0.72(2)\). $10^5$ disorder realization were used for $\lambda=0.2975$. 
    (d-f) Same as (a-c) using \(D_B^0 = 0.25, D_B^1=1, p_B = 0.55\) yielding the estimates  \(\lambda_c = 0.2995(20)\), \(\Theta = -0.36(3)\), \(\delta = 0.62(3)\), and \(2/z = 0.72(2)\).  $10^5$ disorder realization were used for $\lambda=0.2995$. }
    \label{fig:DbDisorderDblDa}
\end{figure}

\begin{figure}
    \centering
    \includegraphics[width=0.32\columnwidth]{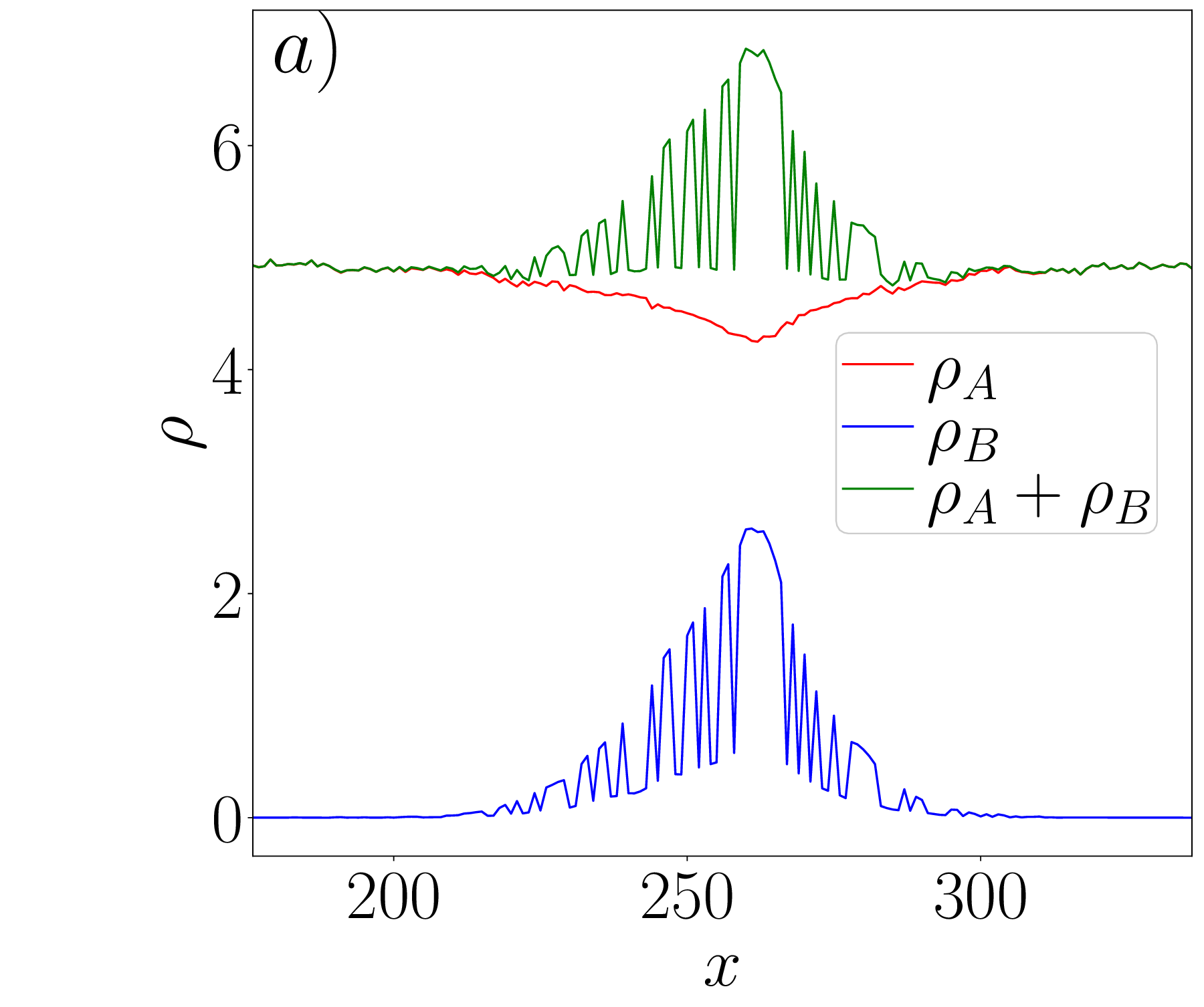}
    \includegraphics[width=0.32\columnwidth]{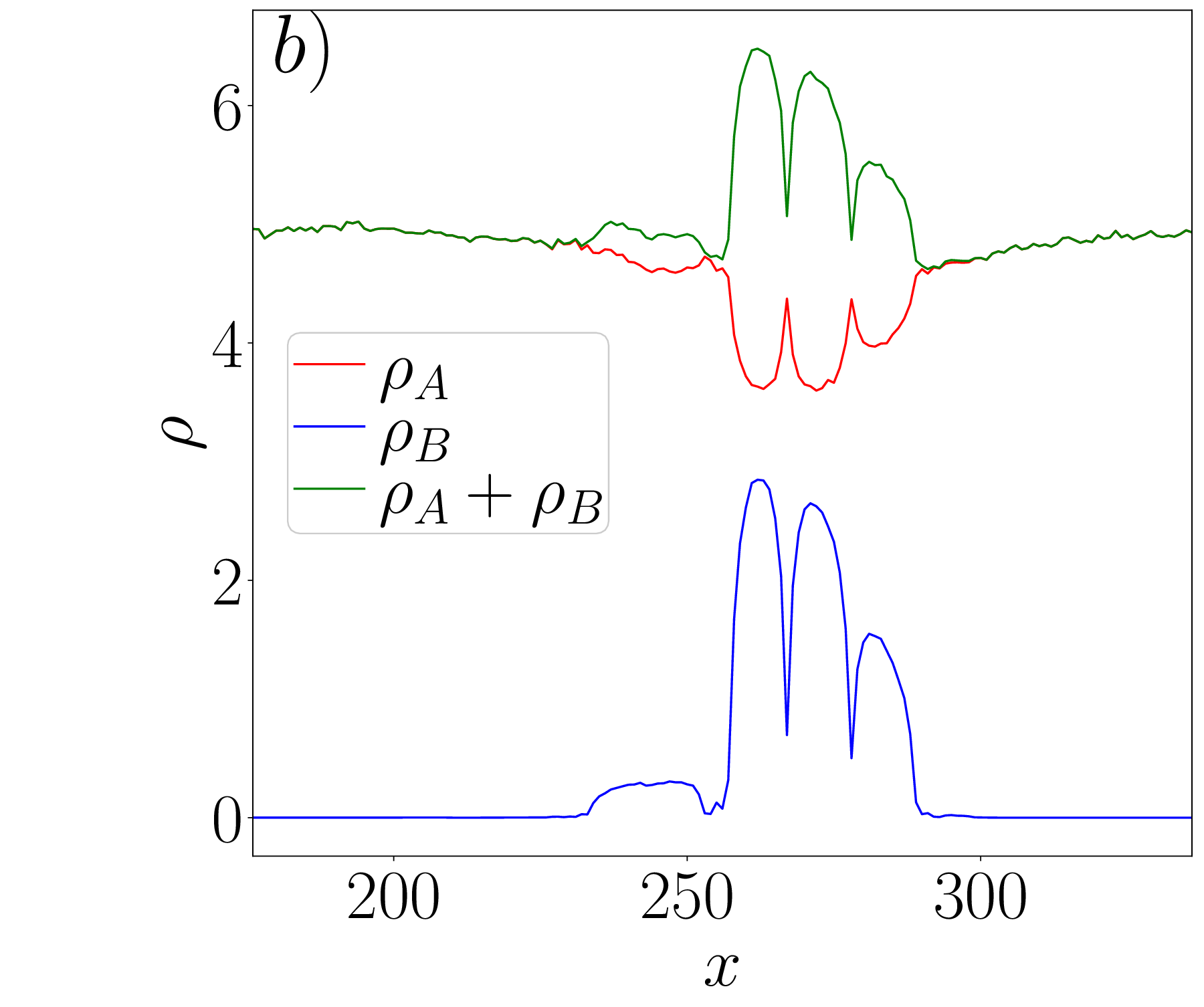}
    \includegraphics[width=0.32\columnwidth]{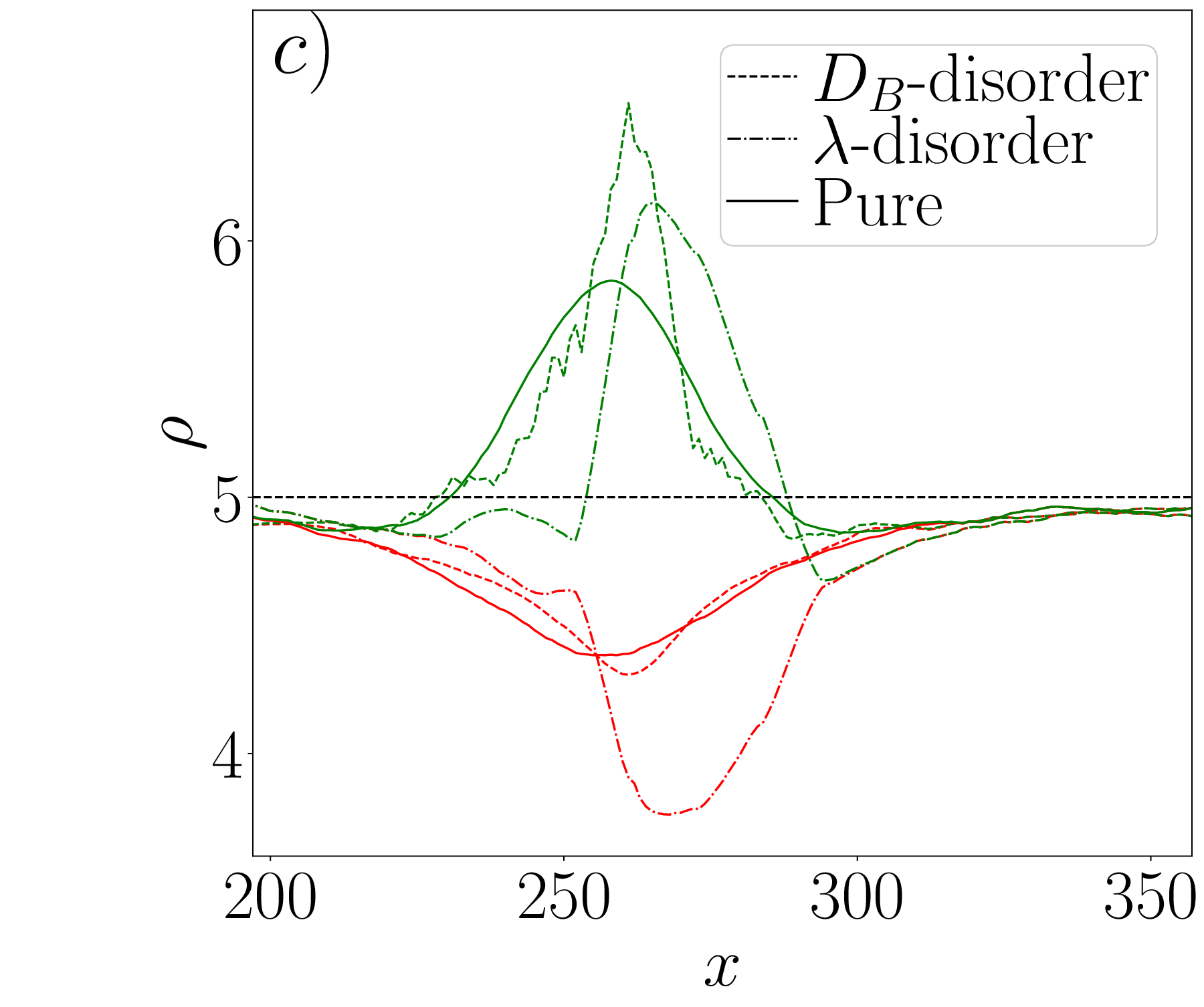}
    \caption{
    (a) \( D_B \)-disorder: Spatial density profiles of \( A \) particles (red), \( B \) particles (blue), and total density (green) at \( t = 1000 \), starting from a seed at \( x = 256 \). Averages over 50,000 independent runs with the same disorder configuration. Parameters: \( D_B^0 = 0.25 \), \( D_B^1 = 1 \), \( p_B = 0.55 \), \( \lambda = \lambda_c = 0.29 \), \( D_A = 1 \).  
    (b) Same as (a) but for \( \lambda \)-disorder. Parameters: \( D_B = 0.375 \), \( D_A = 1 \), \( c = 0.2 \), \( p_{\lambda} = 0.3 \), \( \lambda = \lambda_c = 0.37 \).  
    (c) Comparison of \( D_B \)-disorder, \( \lambda \)-disorder, and the pure case. Disordered curves are smoothed using a central moving average with a window size of 11. All cases share \( D_A = 1 \) and \( D_B^{\text{eff}} \approx 0.375 \). The density profiles for \( D_B \)-disorder and the pure case exhibit similar shapes, while for \( \lambda \)-disorder, the peak shifts and becomes more localized, with a sharper drop in total density surrounding it.}
    \label{fig:comparisonDensityDisorder}
\end{figure}

\begin{figure}
    \centering
    \includegraphics[width=0.32\linewidth]{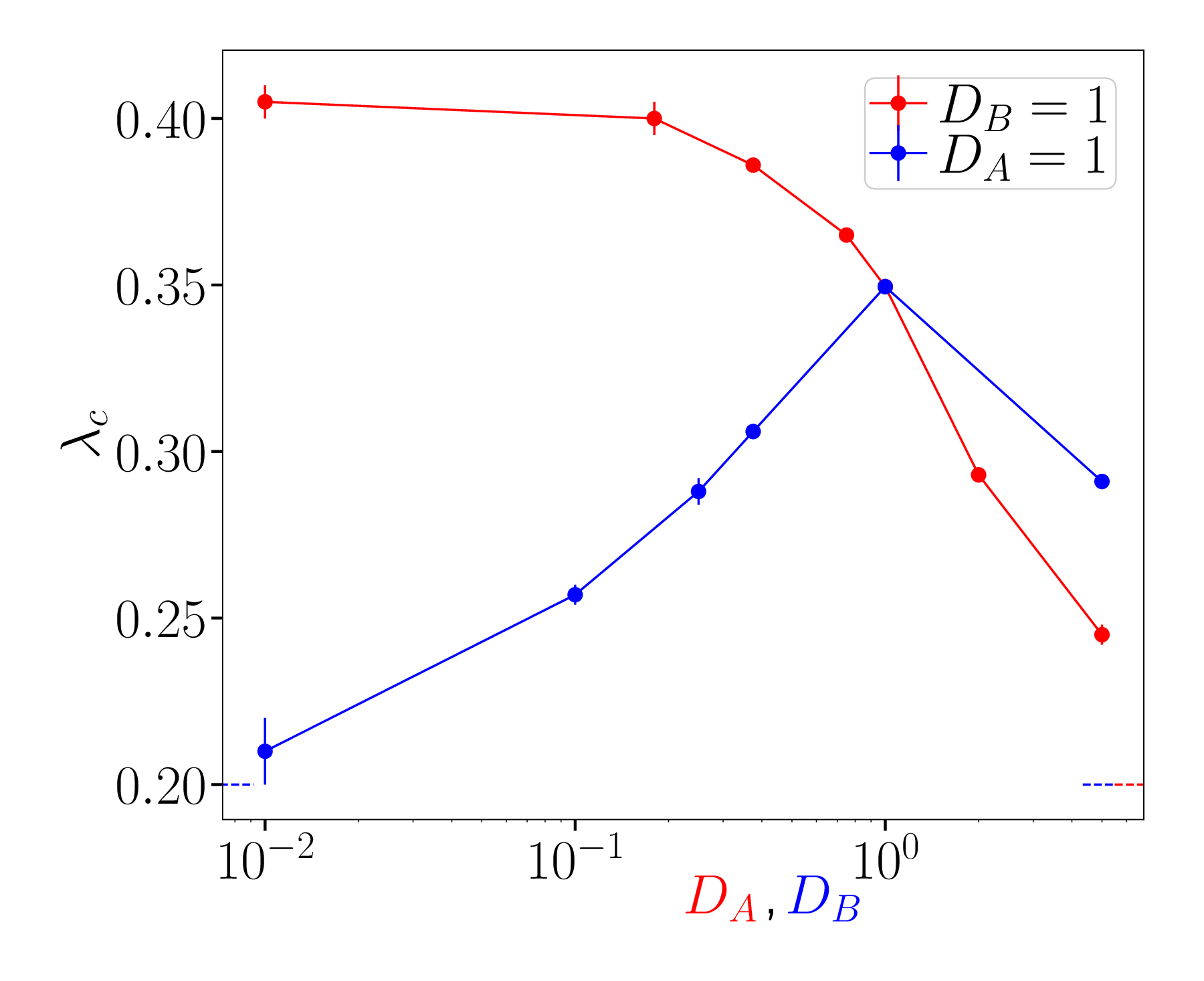}
    \centering
    \includegraphics[width=0.32\linewidth]{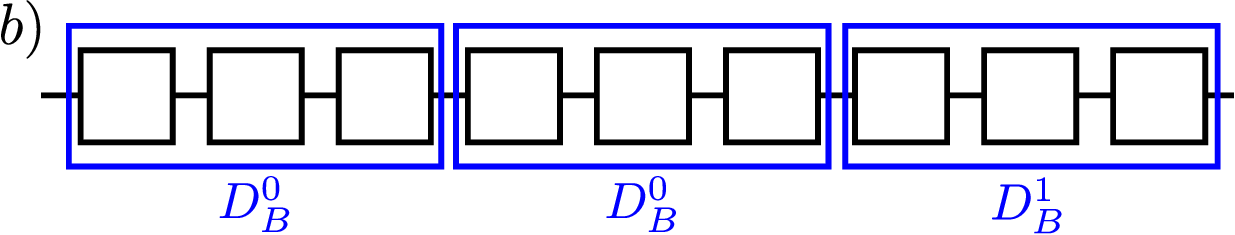}
    \caption{a) Critical contamination rate $\lambda_c$ in the pure DEP function of $D_B$ with $D_A=1$ fixed (blue) an function of $D_A$ with $D_B=1$ fixed (red). Dash dotted line correspond to the expected $\lambda_c$ with $D_B\to 0$, $D_B\to\infty$, $D_A \to \infty$. b) Illustration of the block disorder. Each site in the block of size $X$ (here $X=3$) has the same diffusion rate, allowing having larger regions with the same diffusion rate without changing the probability $p_B$.}
    \label{fig:Sup_PurePhaseDiagram_fct_Da_or_Db}
\end{figure}

\begin{figure}
    \centering
    \includegraphics[width=0.32\linewidth]{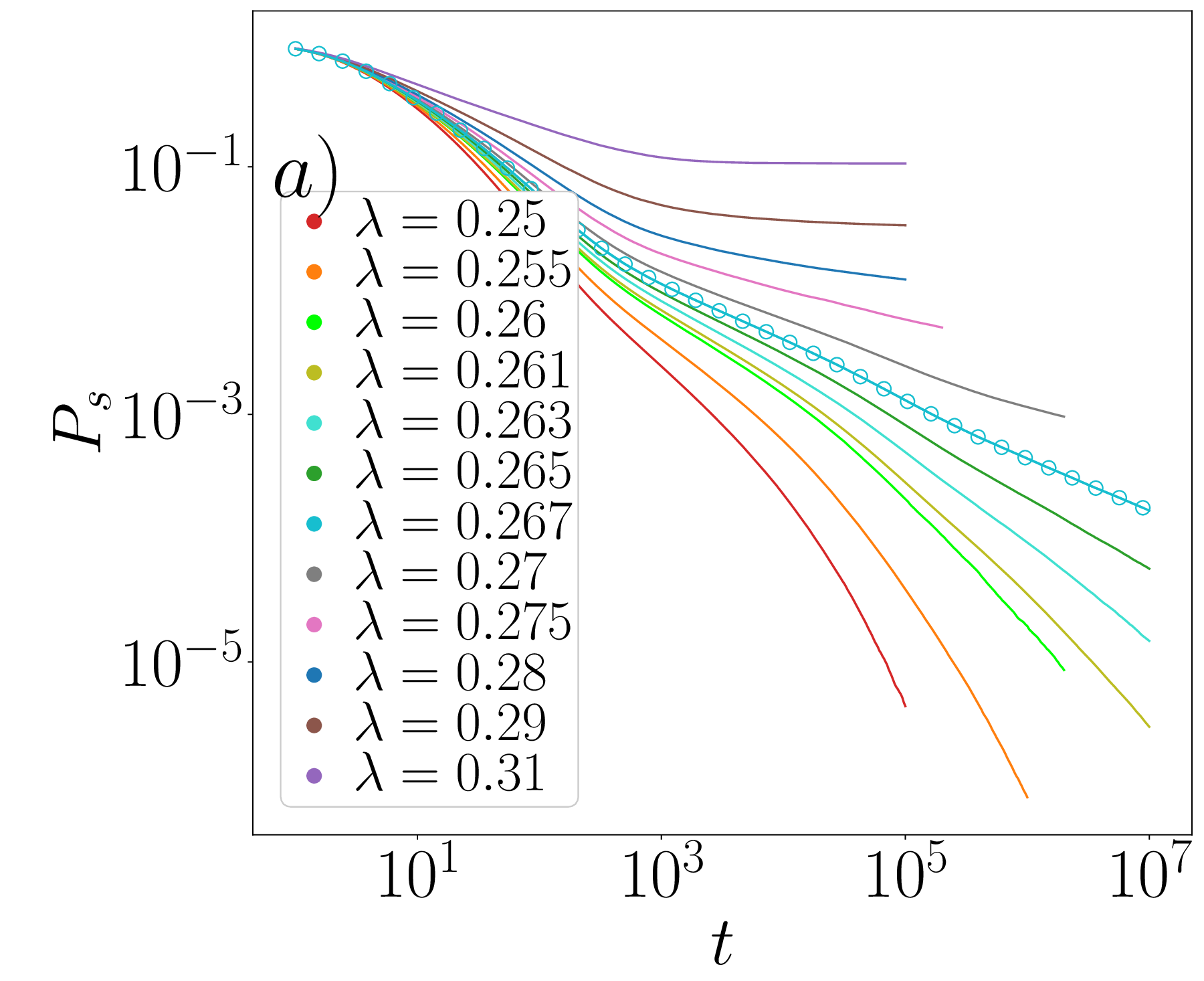}
    \includegraphics[width=0.32\linewidth]{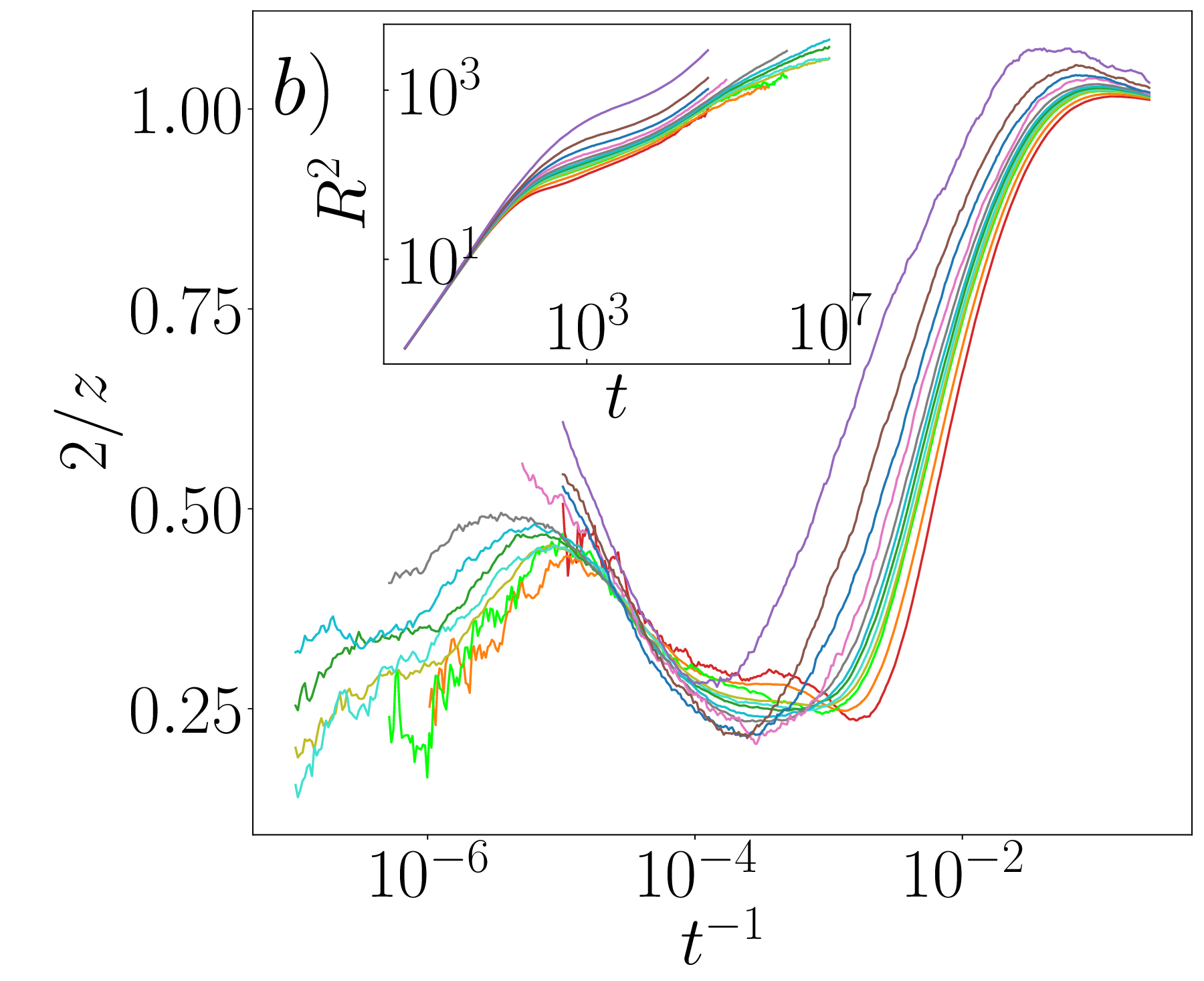}
    \includegraphics[width=0.32\linewidth]{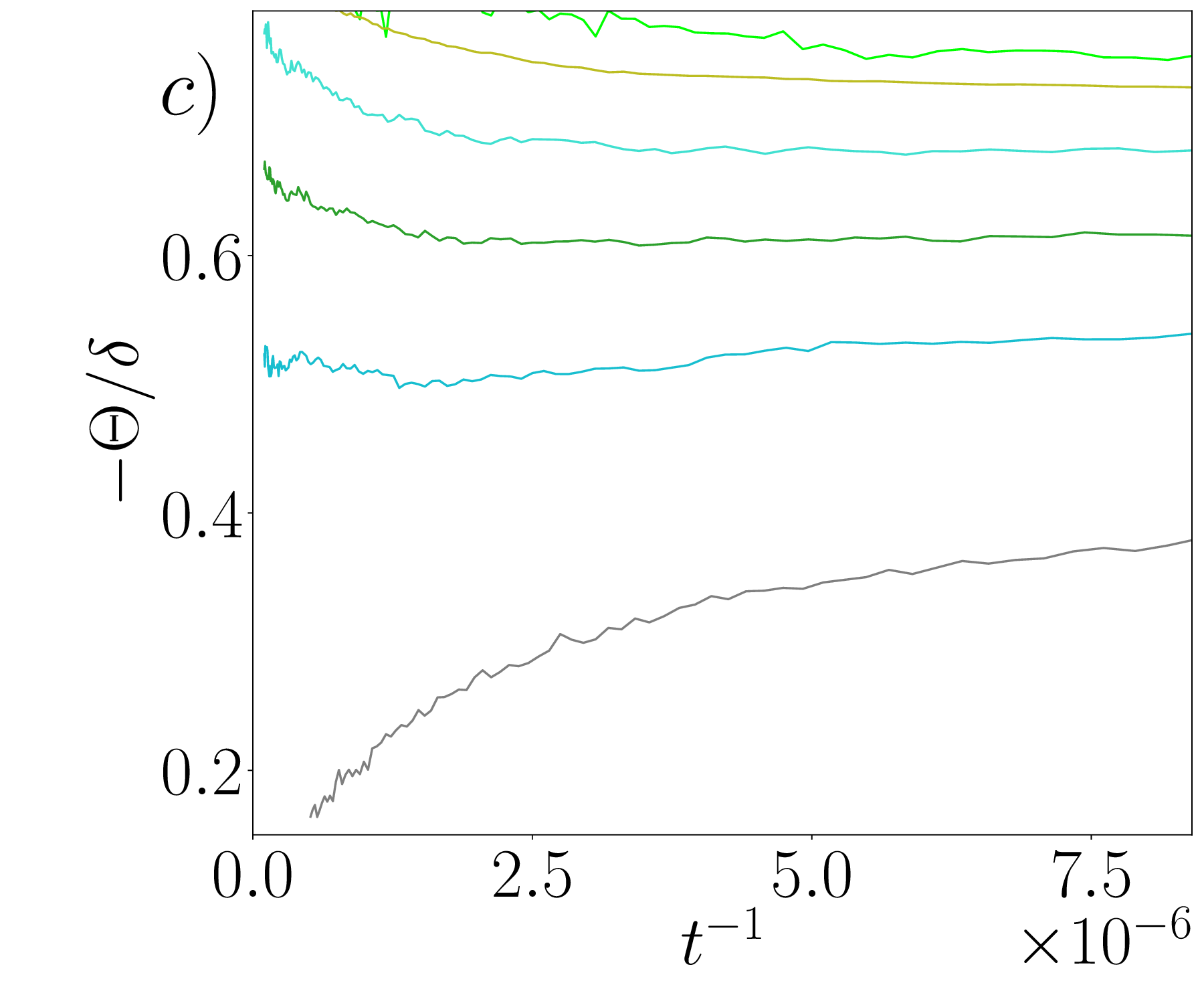}
    \caption{$D_B$-block disorder with $X=10$, $D_B^0=0.1$, $D_B^1=1$, and $p_B=0.2$. 
    (a) For contamination rates below the estimated critical point (circle) but larger than $\lambda=0.26$, the late-time decay of the survival probability $P_s$ is compatible with a power law. 
    (b) The effective dynamical exponent $2/z$ shows a complex time dependence, indicating the emergence of a mesoscopic timescale induced by the block disorder. Near the critical point, both $2/z(t)$ and its linear extrapolation (not shown) to $t\!\to\!\infty$ are smaller than the pure value. Inset: $R^2$ vs $t$.
    (c) The critical point is estimated from the effective exponent ratio $-\Theta/\delta$, yielding $\lambda_c = 0.267(2)$. }
    \label{fig:Sup_BlockDbDis_DblDa_Variation_ln_Ps_with_ln_t}
\end{figure}

What is unexpected is the similarity in critical properties between the disordered and pure cases when \( D_A > D_B^{\text{eff}} \), despite the Harris criterion predicting the relevance of disorder. We tested various disorder distributions with \( D_B^0, D_B^1 \leq D_A \) and \( D_B^0 < D_A < D_B^1 \), consistently obtaining the same result. In Fig.~\ref{fig:DbDisorderDblDa}, the estimated disordered critical exponents closely match the pure case (see Table~\ref{tab:exponents1D}). Two possible explanations arise:

\begin{itemize}
    \item \textbf{Weaker Effective Disorder for \( D_B \)-Disorder and large transient}:
    Spatial quenched disorder in \( D_B \) appears to have a weaker impact on the dynamics compared to \( D_A \)- or \( \lambda \)-disorder. Figure~\ref{fig:comparisonDensityDisorder}c) compares density profiles at \( t = 500 \) for \( D_B \)-disorder, \( \lambda \)-disorder, and the pure case, all with similar diffusion rates (\( D_A = 1, D_B^{\text{eff}} \approx 0.375 \)). For \( \lambda \)-disorder, infected particles are highly localized: the peak shifts relative to the pure case, narrows, and is surrounded by a lower total density. In contrast, the density profile for \( D_B \)-disorder remains closer to the pure case, indicating a weaker effect. This is further illustrated in Fig.~\ref{fig:comparisonDensityDisorder}a), where the \( A \)-particle density shows minimal spatial variation, whereas the \( B \)-particle density follows the disorder distribution. In contrast, Fig.~\ref{fig:comparisonDensityDisorder}b) reveals high-contamination regions clearly distinguishable in both \( A \)- and \( B \)-density profiles.  

    One possible reason for the effective disorder to be weaker is related to the non-monotonic critical contamination rate with \( D_B \), peaking around \( D_A \), Fig.~\ref{fig:Sup_PurePhaseDiagram_fct_Da_or_Db}a) in the pure case. The reduced critical contamination rate for \( D_B \ll D_A \) or \( D_B \gg D_A \) arises respectively from the mechanisms: (i) dense clusters of infected particles and (ii) sufficient mobility to reduce fluctuations in particle encounters, approaching the mean-field critical value as \( D_B \to \infty \). Both involve a characteristic length scale (cluster size or visited region size), meaning disorder effects must be considered over comparable regions. Consequently, fluctuations in the effective diffusion—averaged over these regions (typically spanning up to hundreds of sites, depending on parameters)—are significantly weaker than site-to-site fluctuations, making \( D_B \)-disorder less impactful than \( D_A \)-disorder and \(\lambda\)-disorder.  

    As a result, the time required to reach the dirty critical point could be significantly longer for \( D_B \)-disorder than for \( D_A \)- or \( \lambda \)-disorder, explaining why pure critical exponents persist even at large times.  

    To enhance the effect of disorder in \( D_B \), we introduce spatial correlations by grouping neighboring sites into blocks sharing the same diffusion rate,  Fig.~\ref{fig:Sup_PurePhaseDiagram_fct_Da_or_Db}b). The value of each block follows the same binary distribution as before, the block size is a free parameter that can affect the disorder strength without modifying \( p_B \). We use \( p_B = 0.2 \), \( D_B^0 = 0.1 \), \( D_B^1 = 1 \) (yielding \( D_B^{\text{eff}} \approx 0.36 \)), a block size of ten sites, and \( D_A = 1 \). This block disorder introduces additional length and time scales, producing distinct dynamical regimes in the observables which complicates the interpretation of the data (Fig.~\ref{fig:Sup_BlockDbDis_DblDa_Variation_ln_Ps_with_ln_t}). In particular, it limits the accessible late-time window for precise critical analysis. Up to the maximum time used, $t=10^7$, we found the position of the critical point to be \( \lambda_c \approx 0.267(2) \) from the evolution of $P_s$ vs $N_B$ whose effective exponent is shown in Fig.~\ref{fig:Sup_BlockDbDis_DblDa_Variation_ln_Ps_with_ln_t}c). Below the estimated critical point, and particularly for $\lambda=0.265$, a power-law decay of the survival probability \( P_s(t) \) is observed for almost the two last decades. The mean-squared displacement grows with an effective exponent \( 2/z(t) \) much below the pure value ($<0.5$) at late time and seems to continuously decay. These signatures indicate deviation from the pure critical properties and larger maximum time would be needed to check if they could be compatible with an infinite-disorder fixed point (IDFP) or if it could be due to long transient dynamics. 
    Examining the effective exponents $\Theta(t)$ and $\delta(t)$, we find that they remain close to those of the pure fixed point at $\lambda = 0.265$ (not shown), i.e., slightly below the expected disordered critical point. This suggests that the actual critical point may lie near $\lambda = 0.265$ instead, and, as in the case of $\lambda$-disorder with $D_A = D_B$, the fixed-point properties may depend on the disorder strength primarily through the dynamical exponent $z$.

    Our results therefore suggest that quenched disorder in \( D_B \) becomes relevant when \( D_B^{\text{eff}} < D_A \). The change in the critical properties only when using block disorder is consistent with an intrinsically weak \( D_B \) disorder. Nevertheless, we cannot exclude the possibility of long-lived transients. Whether weak \( D_B \) disorder can ultimately destabilize the pure critical point, as theoretically expected, and whether the resulting critical behavior coincides with that induced by \( D_A \) disorder remain open questions.


    \item \textbf{Possible Breakdown of Theoretical Predictions:}  
    Another possibility is a deviation from established theoretical expectations, such as the Harris criterion \cite{harris_effect_1974} or the Chayes inequality \cite{chayes_finite-size_1986}. Since both $D_A$ and $D_B$ influence the location of the critical point without altering the symmetries of the system, it should be possible to use the Harris criterion to predict the relevance of introducing heterogeneities into $D_A$ or $D_B$. This is expected to hold at least when, for all sites $i$, $D_A^i$ ($D_B^i$) is larger (smaller) than $D_B$ ($D_A$). However, it remains unclear why $D_A$-disorder conforms to this criterion while $D_B$-disorder appears not to, without resorting to block disorder. Although a slight deviation in the critical exponents from their pure-system values could still be compatible with the Harris criterion, our current estimates lack the precision required to confirm this. And even in such case, the strong difference (IDFP vs conventional FP) between $D_A$-disorder and $D_B$-disorder remains puzzling.
\end{itemize}


\section{Two-species correlated diffusion rate disorder}

We consider only perfectly correlated disorder of the diffusion rates implying $D_{A,i} = \mu D_{B,i}$ with $\mu > 0$. 

\subsection{$\mu=1$}
\begin{figure}
    \centering
    \includegraphics[width=0.32\columnwidth]{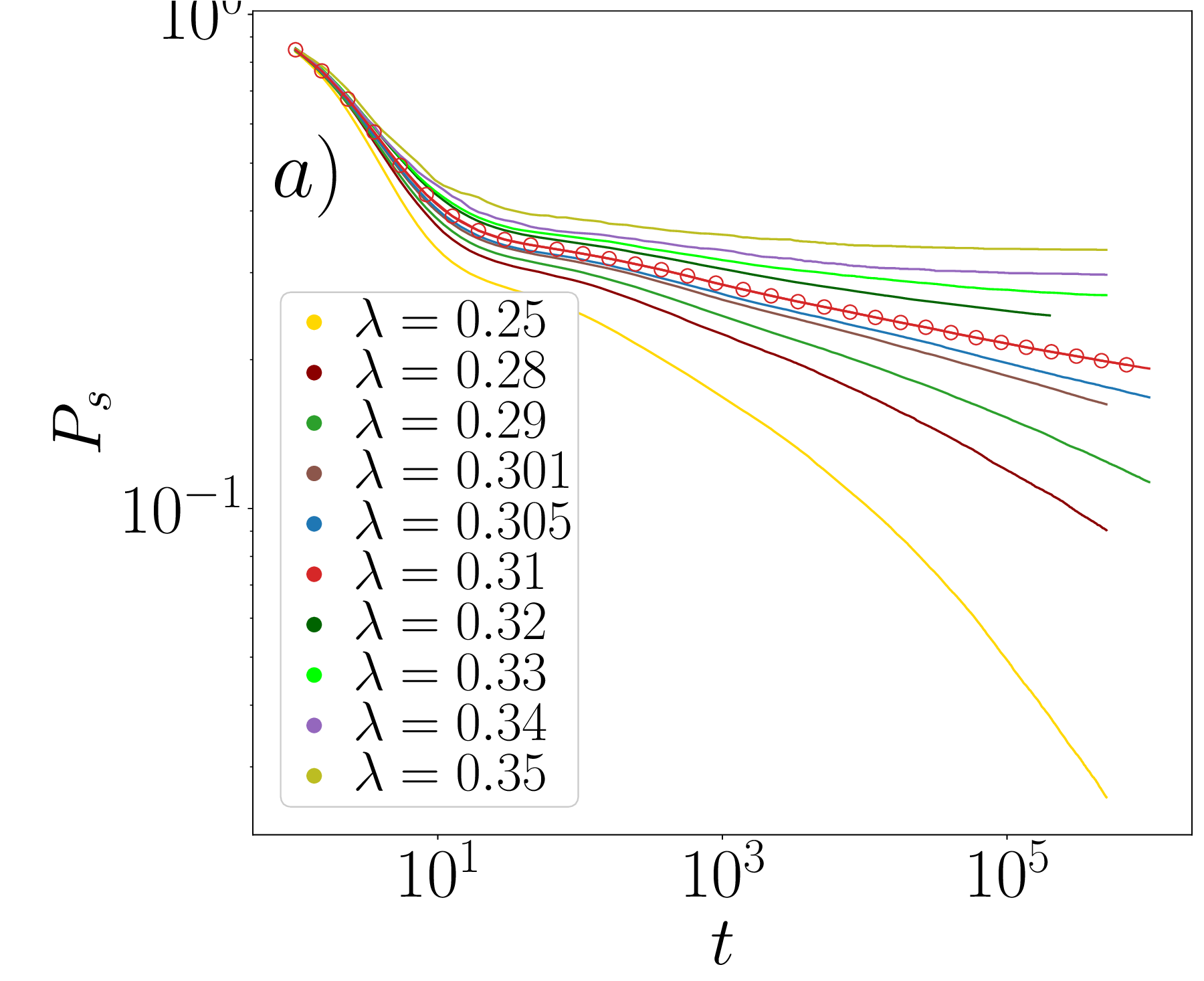}
    \includegraphics[width=0.32\columnwidth]{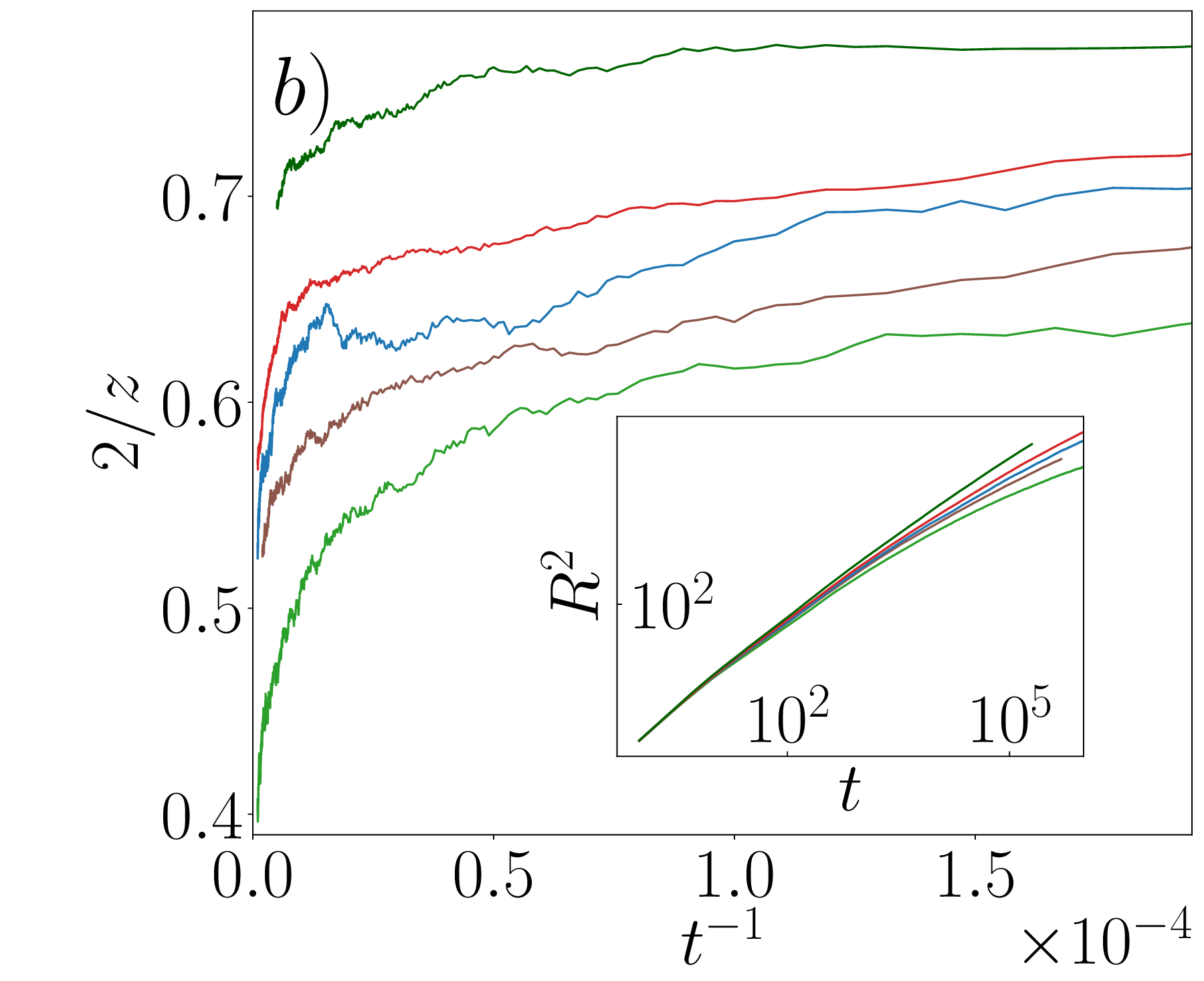}
    \includegraphics[width=0.32\columnwidth]{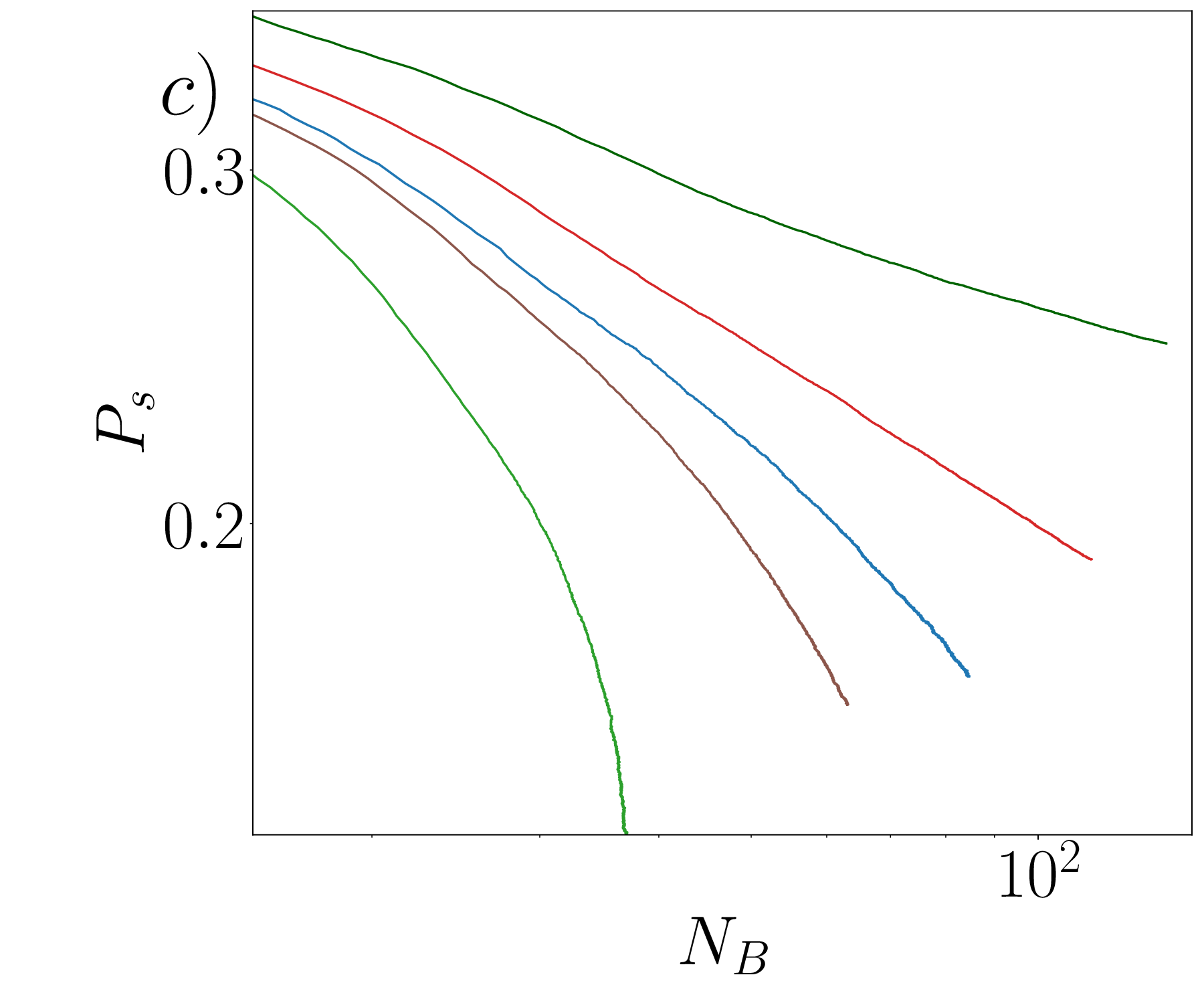}
    \caption{Correlated disorder with \( D^0=0.25 \), \( D^1=1 \), \( p=0.1 \), \( D_B=1 \).  
    (a) Time evolution of the survival probability \( P_s \) on a log-log scale. At the estimated critical point (\(\lambda_c \approx 0.31\), circle), a clear power-law behavior is not observed, whereas a relatively well-defined one appears at a lower contamination rate. This suggests the presence of an Infinite Disorder Fixed Point with a Griffiths phase.  
    (b) The effective dynamical exponent \( 2/z \) decreases to abnormally low values near and below the critical point, further supporting an Infinite Disorder Fixed Point. This corresponds to a slower-than-power-law growth of the mean-square displacement \( R^2 \) (inset).  
    (c) The survival probability \( P_s \) is plotted against the number of particles \( N_B \) on a log-log scale to estimate the critical point at \( \lambda_c \approx 0.31 \). A total of \( 3 \times 10^3 \) disorder realizations were used for \( \lambda = 0.31 \).  }
    \label{fig:CorrDaDbDis_DbeqDa}
\end{figure}

In addition to the data presented in the main text, we provide detailed figures on slow dynamics consistent with an IDFP scenario with \( \mu = 1 \) (see Fig.~\ref{fig:CorrDaDbDis_DbeqDa}). However, as with \( \lambda \)-disorder when \( D_A = D_B \), the critical properties may depend on disorder strength, with the dynamical exponent \( z \) increasing up to be formally infinite at the IDFP. This aligns with our simplified picture, which requires sufficiently strong disorder. Consequently, at large \( p \) or for weak differences between \( D^0 \) and \( D^1 \), critical properties may still be governed by a conventional fixed point.

\subsection{$\mu\neq1$}

For \( \mu \neq 1 \), the simple picture relating to the contact process no longer holds. When \( \mu \neq 1 \), infected and healthy sites can be distinguished by their density, causing effective contamination rates between dense regions to become time-dependent. For instance, when \( \mu < 1 \) (\( D_{A,i} > D_{B,i} \)), the density of infected sites decreases as the number of infected sites increases due to particle conservation. We provide numerical data for both \( \mu > 1 \) and \( \mu < 1 \), confirming the expected behavior in each case.

\begin{figure}
    \centering
    \includegraphics[width=0.34\columnwidth]{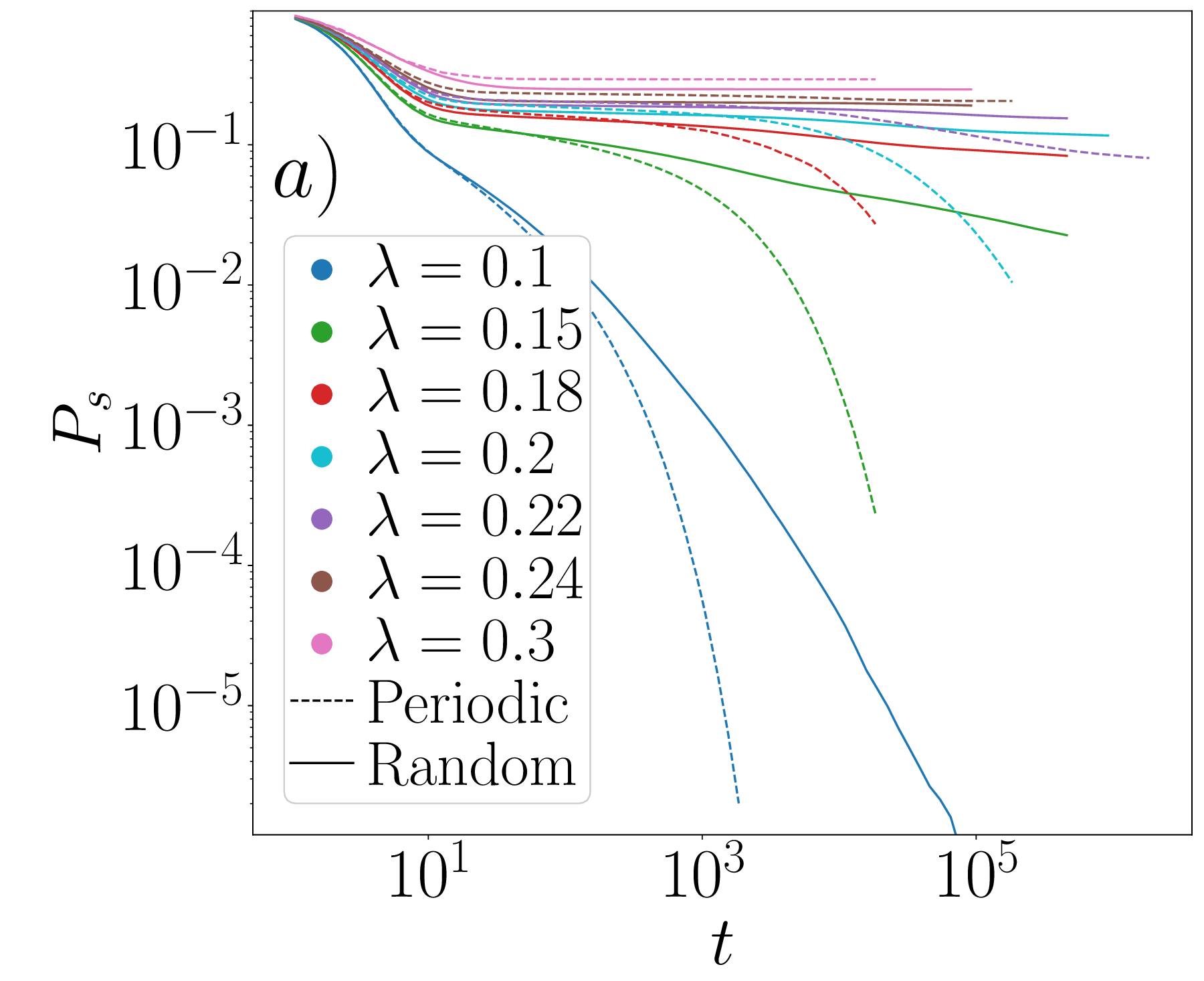}
    \includegraphics[width=0.34\columnwidth]{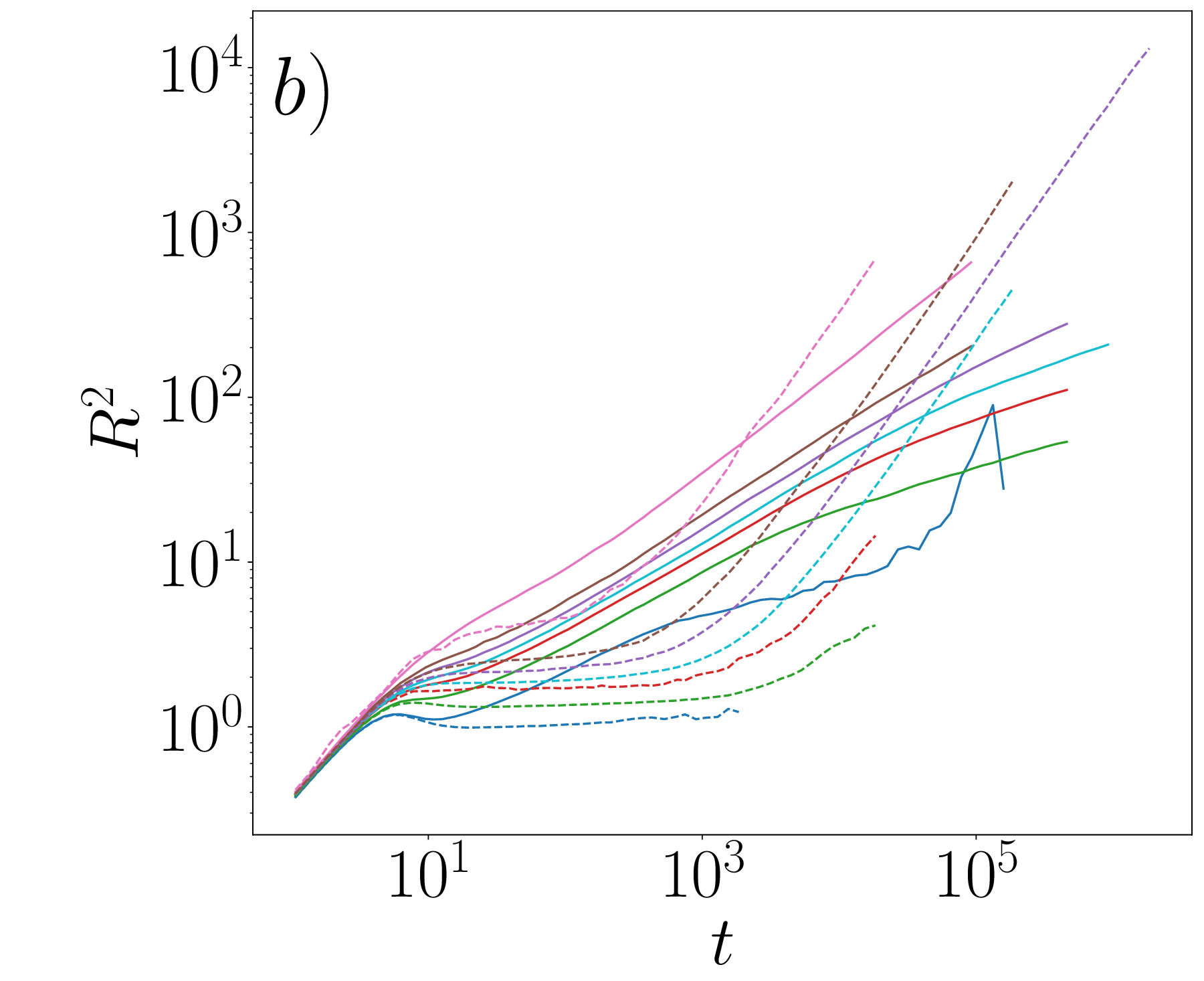}
    \includegraphics[width=0.34\columnwidth]{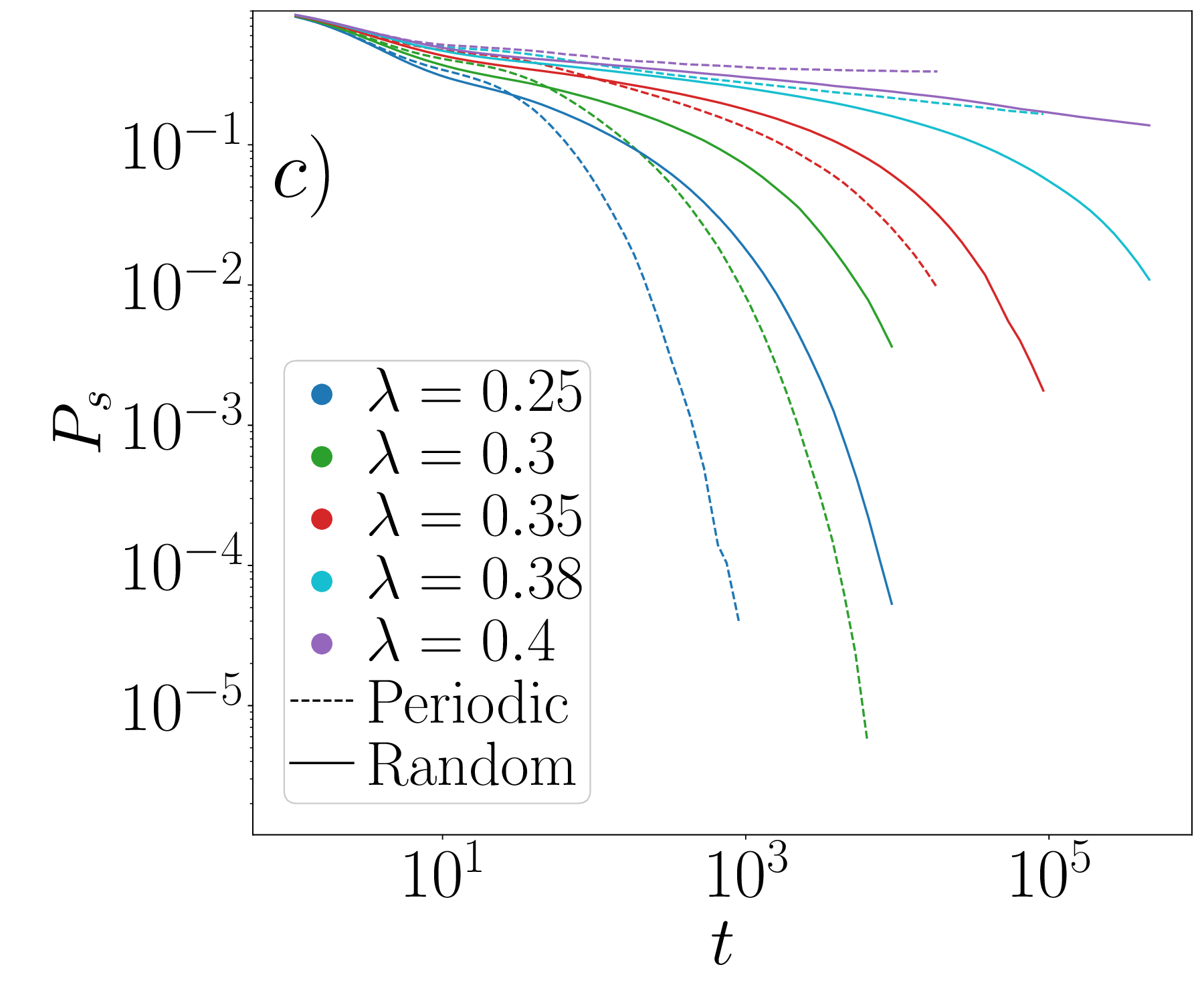}
    \includegraphics[width=0.34\columnwidth]{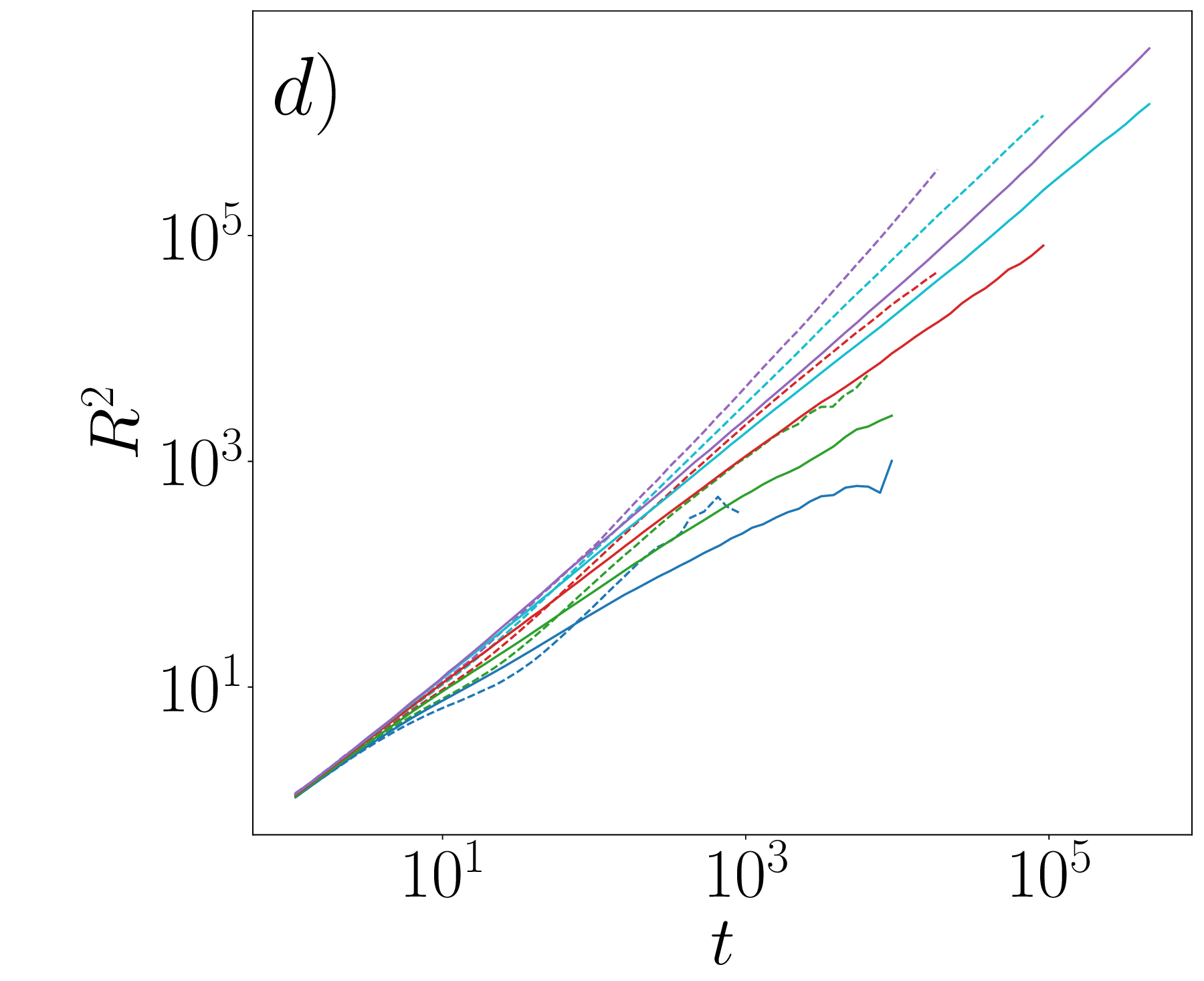}
    \caption{Correlated disorder with \( D^0=0.25 \), \( D^1=1 \), \( p=0.1 \), \( D_B=1 \).  
    \textbf{Upper panel:}\( \mu = 0.375 \), with \( D_{A, i} = D^{0/1} \).  
    (a) Comparison of the survival probability \( P_s \) evolution for periodic (dotted lines) and random (solid lines) disorder. In the periodic case, \( P_s \) decays exponentially in the absorbing phase, whereas in the random case, it does not within the maximum time considered.  
    (b) Comparison of the mean-square displacement \( R^2 \). For periodic disorder, an initial plateau is followed by a power-law growth. In contrast, for random disorder, the growth is significantly slower.  
    \textbf{Lower panel:} Same as the upper panel but with \( \mu = 2.67 \) and \( D_{B, i} = D^{0/1} \). (c) \( P_s \) decays exponentially for both disorder. (d) The growth of \( R^2 \) lacks an initial plateau in the periodic case. At the estimated critical points, \( \lambda_c \approx 0.38 \) (periodic, dotted light blue) and \( \lambda_c \approx 0.4 \) (random, solid purple), both $R^2(t)$ curves grow with similar exponents. }
    \label{fig:CorrDaDbDis_DblgDa}
\end{figure}

We first consider \( \mu = 0.375 \) (\( D_{A,i} > D_{B,i} \)), where quenched disorder is expected to be relevant. For \( D^0 = 0.25, D^1 = 1, p = 0.1 \), with periodic (dotted) disorder, the time required to infect a neighboring slow site is on the order of 500 seconds, as indicated by the early plateau in the mean squared displacement \( R^2 \) (Fig.~\ref{fig:CorrDaDbDis_DblgDa}b). This relatively large time scale means that simulations need much longer durations to capture critical properties. We estimate \( \lambda_c \approx 0.217 \) with periodic disorder (data not shown).As in the \( \mu = 1 \) case, \( P_s(t) \) and \( R^2(t) \) behave differently for periodic (dotted) and random (solid) disorder: \( P_s \) decays and \( R^2 \) grows more slowly in the random disorder case. These results are consistent with an IDFP scenario up to the maximum considered time.

In contrast, for \( \mu = 2.67 \) (\( D_{A,i} < D_{B,i} \)), the behavior differs significantly (Fig.~\ref{fig:CorrDaDbDis_DblgDa}(c-d)). Though the decay of \( P_s \) is slower in the random case than in the periodic case, both decay similarly in the absorbing phase. From the survival probability decays, we estimate \( \lambda_c \approx 0.38 \) for periodic disorder and \( \lambda_c \approx 0.4 \) for random disorder. The growth of \( R^2 \) at these critical points is similar at late times. Although a full analysis of the critical properties hasn't been performed, both cases align with pure critical properties as expected.

\bibliography{references3} 